\documentclass[a4paper,11pt]{article}
\pdfoutput=1 % if your are submitting a pdflatex (i.e. if you have
\usepackage{jheppub} % for details on the use of the package, please
\usepackage{slashed} 
\usepackage{mathtools} 
\usepackage{bbm}
\usepackage{float}
\usepackage[normalem]{ulem}
\usepackage{subcaption}
\usepackage{adjustbox}
\usepackage[T1]{fontenc} % if needed
\usepackage[utf8]{inputenc}

\newcommand{\pd}{\partial}

\def\eps{\epsilon}
\def\mmu2om2{\left (\frac{\mu^2}{m^2} \right )^{\!\eps}}

\def\spa#1.#2{\langle #1 #2\rangle}
\def\spb#1.#2{[ #1 #2]}
\def\spab#1.#2.#3{\langle #1 |#2| #3] }

\title{Master Integrals for Electroweak corrections to $gg\rightarrow \gamma\gamma$ - Light quark contributions}

\author{Gabriele Fiore,}   
\author{Ciaran Williams}    

\affiliation{Department of Physics,\\University at Buffalo, The State University of New York, Buffalo
14260 USA}

\emailAdd{gfiore@buffalo.edu}
\emailAdd{ciaranwi@buffalo.edu}

\begin{abstract}{
		We present a calculation of the master integrals (MI's) required for the calculation of the Electroweak corrections to $gg\rightarrow \gamma\gamma$ production in which the process contains a light quark loop. 
		The integrals can be broken down into five categories based on the flow of the heavy vector bosons throughout the loop. Three of the families are planar, and two are non-planar. We determine a canonical basis for each family which allows an efficient solution of the resulting differential equations via iterated integrals. We calculate the families in relevant physical kinematics and obtain an efficient numerical evaluation  based on an implementation of Chen-iterated integrals. 
	}
\end{abstract}

\begin{document} 
\maketitle
\flushbottom
%%% Until the content of the paper is finished, use clearpage to flush the
% positioning of figures in chapters

\section{Introduction}

The production of prompt photon pairs has long been a staple process at hadron colliders with many observations ranging back over the last four decades~\cite{Bonvin:1988yu,Alitti:1992hn,Abe:1992cy,Abazov:2010ah,Aaltonen:2012jd,Chatrchyan:2011qt,Aad:2012tba,Aad:2013zba,Aaltonen:2011vk,Chatrchyan:2013mwa,Chatrchyan:2014fsa,Marini:2016zjr,CMS:2018qao,ALICE:2019rtd,ATLAS:2019buk,ATLAS:2019iaa,CMS:2019jlq,ATLAS:2021mbt,ATLAS:2023yrt}. A centerpiece of this work and culmination of many years of effort was the discovery of the Higgs boson in this channel in 2012~\cite{Aad:2012tfa,Chatrchyan:2012ufa}. Going forward into the era of the High Luminosity LHC (HL-LHC) the process will continue to be studied in ever greater experimental
precision, mandating a continued theoretical effort to provide increasingly accurate predictions for this channel.

At the lowest order in perturbation theory the production of photon pairs proceeds through a $q\overline{q}$ annihilation channel. 
For many years the theoretical state of the art corresponded to Next-to-Leading Order (NLO) QCD predictions, implemented into the
Monte Carlo code Diphox~\cite{Binoth:1999qq}. However over the last decade significant theoretical progress has been made, including Next-to-Next-to Leading Order (NNLO) QCD predictions~\cite{Campbell:2016yrh,Catani:2011qz} and $q_T$ resummed predictions at order N$^3$LL matched to NNLO~\cite{Neumann:2021zkb}. Further, the theoretical predictions in association with one or more jets is well under control~\cite{DelDuca:2003uz,Gehrmann:2013bga,Badger:2013ava,Chawdhry:2021hkp}.
These results, in combination with the recent computation of the three-loop amplitudes for $q\overline{q} \rightarrow\gamma\gamma$~\cite{Caola:2020dfu}\footnote{and similarly for inclusive photon production~\cite{Bargiela:2022lxz}.} suggests that it is not unreasonable to assume the availability of N$^3$LO predictions over the course of the HL-LHC.

At $\mathcal{O}(\alpha_s^2)$ (NNLO in QCD) the gluon-initiated $gg\rightarrow \gamma\gamma$ partonic configuration becomes accessible for the first time.  
For LHC kinematics there is a large flux of gluons, which somewhat compensates the suppression from the increased powers of the coupling. Thus for a long time~\cite{Glover:1988fe} 
the $gg$ initiated production of vector boson pairs has received special attention in the theoretical literature. 
The NLO QCD corrections to this process (which corresponds to two-loop diagrams) for massless quarks were presented around twenty years ago in refs.~\cite{Bern:2001df,Bern:2002jx} and more recently massive top loops have also been included at this order~\cite{Chen:2019fla,Maltoni:2018zvp}. 
The three-loop $gg\rightarrow \gamma\gamma$  calculation~\cite{Bargiela:2021wuy} and two-loop $gg\rightarrow g\gamma\gamma$ amplitudes (for light quarks) has further extended the known theoretical structure of this process in QCD.

One of the reasons the $gg$ initiated process is so fascinating is its role in the continued study of the Higgs boson. 
The $gg\rightarrow \gamma\gamma$ amplitude interferes with the amplitude for the production of a Higgs boson through a top loop and 
subsequent decay to photons. This interference can lead to changes in the measured properties of the Higgs~\cite{Martin:2012xc,Dixon:2003yb,Dixon:2013haa} when compared to a signal only 
theoretical predictions. These initial studies have generated significant interest, leading to calculations over a wider range of production modes~\cite{deFlorian:2013psa,Coradeschi:2015tna,Campbell:2017rke}.
In particular, it has been noted that the interference, and hence width extraction, is sensitive to higher order corrections in QCD~\cite{Cieri:2017kpq,Bargiela:2022dla}.  
This motivates the study of the electroweak corrections to the interference pattern, of which the starting point is provided in this paper by the calculation of two-loop integrals required. 

Over the last decade significant progress has been made in the computation of multi-loop Feynman integrals due to the advances related to solution via differential equations.
Using differential equations to simplify and solve Feynman integrals is a well-established technique~\cite{Kotikov:1990kg,Remiddi:1997ny,Gehrmann:1999as,Gehrmann:2000xj,Argeri:2007up}, however, the field experienced a major advance when, in ref.~\cite{Henn:2013pwa}, a refined basis approach was proposed. By modifying the basis from a traditional one, derived for instance via reduction by integration by parts identities, ref.~\cite{Henn:2013pwa} proposed a new canonical basis which obeyed a simpler differential equation which factorized the regulating parameter from the kinematics.  If such a basis can be found, solutions to the differential equation can be more readily extracted.   Since its introduction, the canonical differential equation technique has led to the calculation of many two-loop process for $2\rightarrow 2$ and even $2\rightarrow 3$ kinematics; for an incomplete list see e.g. \cite{Gehrmann:2001ck,Bonciani:2016ypc,DiVita:2017xlr,Heller:2019gkq,Hasan:2020vwn,Bonetti:2020hqh,Behring:2020cqi,Canko:2020ylt,Duhr:2021fhk,Abreu:2021smk,Bonciani:2021zzf,Kardos:2022tpo,Abreu:2022vei,Badger:2022hno,Armadillo:2022bgm,Bonetti:2022lrk,Buccioni:2022kgy}. 

One of the negative features of calculations beyond one-loop is the lack of common basis. This means that typically each new calculation often requires a dedicated calculation of the relevant MIs for the process, further since different authors use different normalizing factors and conventions, it is often inconvenient or impractical to combine results from different MI calculations. 
Our aim is to calculate the electroweak corrections to $gg\rightarrow \gamma\gamma$, in this first paper we present the MIs for the simplest set of Feynman diagrams, namely those which proceed through a loop of quarks which can be taken to be massless. Although more complicated MI calculations have been performed in the literature, we were unable to find specific MIs for several of our topologies (relating to diagrams involving three massive $W$ bosons radiating photons). We therefore believe it is worthwhile to discuss these integrals in one place with a consistent framework. We postpone the discussion of the third generation quark loops, which contain many new MIs, to a future publication. 

This paper proceeds as follows, in Section~\ref{sec:Overview} we present a discussion of the canonical basis, and the solution of the differential equation via Chen-iterated integrals. Section~\ref{sec:Plan} then presents the calculation of the planar topologies and section~\ref{sec:nonplan} presents the calculation of the non-planar topologies, finally we draw our conclusions in section~\ref{sec:conc}. Electronic files containing solutions obtained in this paper are included as ancillary files with this paper.

%=====================================================================

%============================= OVERVIEW 

%=====================================================================
\section{Overview}
\label{sec:Overview}

The primary focus of this paper is the calculation of the Master Integrals (MI's) required for the evaluation of the electroweak (EW) corrections to the $gg\rightarrow \gamma\gamma$ process which proceeds through the following kinematic reaction
\begin{equation}\label{eq:process}
0 \rightarrow g(k_1) + g(k_2)+ \gamma(k_3) + \gamma(k_4),
\end{equation}
where we consider all external momenta to be outgoing, and light-like ($k_i^2 =0$).  It is convenient to parameterize the scattering in terms of the following Mandelstam invariants
\begin{align}\label{eq:mandelstams}
s\! =\! (k_1\! +\! k_2)^2 = 2 k_1 \cdot k_2, \nonumber\\\; 
t  \! =\! (k_2 \!+ \!k_3)^2 =2 k_2 \cdot k_3,\nonumber\\\;
u \!=\! (k_1\! +\! k_3)^2 = 2 k_1 \cdot k_3.
\end{align}
Momentum conservation further constrains the sum of the Mandelstam invariants $s+t+u=0$, and for physical scatterings $s > 0$ and $u, t <0$. We employ the 't Hooft-Veltman scheme in which loop momenta are evaluated in $d=4-2\epsilon$ dimensions and polarization vectors $(\epsilon_i)$ and momenta attributed to the final state bosons are taken in four dimensions. We also note that when computing loop topologies later in this paper we introduce the momenta set $p_1,p_2$ and $p_3$, where the assigned signs of $p_i \cdot p_j$ can vary based on which of the $k_i$ we assign to each $p_j$. 

We define the  $\ell$-loop color stripped amplitude as follows
\begin{equation}\label{eq:physical_amplitude}
{A}_{\ell}^{a_1 a_2} = 16 \pi^2 \delta^{a_1a_2}\alpha^0_{s} \alpha^0 \, \mathcal{A}_{\ell}(s,t),\;
\end{equation}
where $\alpha^0_{s}$ represents the bare strong and $\alpha^0$ denotes the bare electromagnetic coupling, $\delta^{a_1 a_2}$ defines the leading order QCD color factor. 
In our calculation we will decompose the amplitude into a series of tensor structures 
following the procedure outlined in refs.~\cite{Peraro:2019cjj,Peraro:2020sfm}.  In total there are eight relevant tensor structures
\begin{equation}\label{eq:tensor_decomp}
\mathcal{A}(s,t) = \sum_{i=1}^{8} {F}_i \: T_i.
\end{equation}
Here the tensors $T_i$ are defined as:
\begin{align} \label{eq:Tensors}
&T_1 = \epsilon_1 \! \cdot \! k_3\; \epsilon_2 \! \cdot \! k_1\; \epsilon_3 \! \cdot \! k_1\; \epsilon_4 \! \cdot \! k_2 \;, \nonumber \\
&T_2 = \epsilon_1 \! \cdot \! k_3\; \epsilon_2 \! \cdot \! k_1\; \epsilon_3 \! \cdot \! \epsilon_4 , \quad
T_3 = \epsilon_1 \! \cdot \! k_3\; \epsilon_3 \! \cdot \! k_1\; \epsilon_2 \! \cdot \! \epsilon_4 , \nonumber \\
&T_4 = \epsilon_1 \! \cdot \! k_3\; \epsilon_4 \! \cdot \! k_2\; \epsilon_2 \! \cdot \! \epsilon_3, \quad T_5 = \epsilon_2 \! \cdot \! k_1\; \epsilon_3 \! \cdot \! k_1\; \epsilon_1 \! \cdot \! \epsilon_4 , \nonumber \\
&T_6 = \epsilon_2 \! \cdot \! k_1\; \epsilon_4 \! \cdot \! k_2\; \epsilon_1 \! \cdot \! \epsilon_3 , \quad
T_7 = \epsilon_3 \! \cdot \! k_1\; \epsilon_4 \! \cdot \! k_2\; \epsilon_1 \! \cdot \! \epsilon_2 , \nonumber \\
&T_8 = \epsilon_1 \! \cdot \! \epsilon_2\;  \epsilon_3 \! \cdot \! \epsilon_4+ \epsilon_1 \! \cdot \! \epsilon_4\;  \epsilon_2 \! \cdot \! \epsilon_3 + \epsilon_1 \! \cdot \! \epsilon_3\;  \epsilon_2 \! \cdot \! \epsilon_4 \;.
\end{align}
Each tensor has a corresponding form-factor ${F}_i$. In defining the tensor decomposition we make the cyclic gauge choice $\epsilon_i  \! \cdot \! k_{i+1} = 0$, with $k_5 = k_1$ and impose the transversality condition $\epsilon_i \! \cdot \! k_i = 0$.
We use projector operators in order to isolate individual form factors entering into the expansion in eq.~\ref{eq:tensor_decomp}. Each projector is defined through an orthogonality condition with the corresponding tensor such that $\sum_{pol} P_i T_j= \delta_{ij}$.  Since the resulting projectors are the same as those appearing in the literature ~\cite{Peraro:2019cjj,Peraro:2020sfm,Bargiela:2021wuy} we do not list them explicitly here for the sake of brevity. 

\begin{figure}[h]
	\centering
	\begin{subfigure}{0.32\linewidth}
		\includegraphics[width=\linewidth]{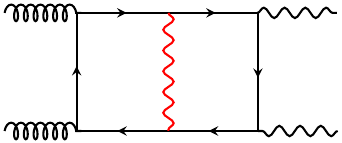}
		\caption{Topology P$_\text{I}$}
		\label{fig:PI_feyn}
	\end{subfigure}\hspace{0.1cm}
	\begin{subfigure}{0.32\linewidth}
		\includegraphics[width=\linewidth]{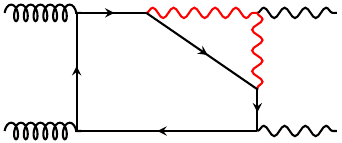}
		\caption{Topology P$_\text{II}$}
		\label{fig:PII_feyn}
	\end{subfigure}\hspace{0.1cm}
	\begin{subfigure}{0.32\linewidth}
		\includegraphics[width=\linewidth]{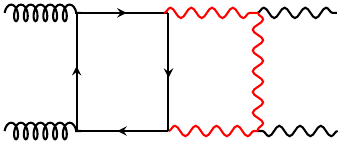}
		\caption{Topology P$_\text{III}$}
		\label{fig:PIII_feyn}
	\end{subfigure}\vspace{0.5cm}

	\begin{subfigure}{0.32\linewidth}
		\includegraphics[width=\linewidth]{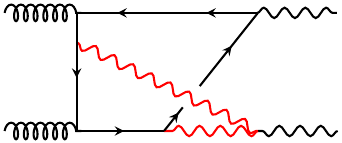}
		\caption{Topology N$_\text{I}$}
		\label{fig:NI_feyn}
	\end{subfigure}\hspace{0.5cm}
	\begin{subfigure}{0.32\linewidth}
		\includegraphics[width=\linewidth]{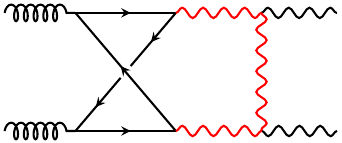}
		\caption{Topology N$_\text{II}$}
		\label{fig:NII_feyn}
	\end{subfigure}
	\caption{Feynman diagrams associated with  the relevant topologies. Propagators marked in red are massive. }
	\label{fig:feynAll}
\end{figure}
The lowest order $gg \rightarrow \gamma\gamma$ amplitude occurs at $\mathcal{O}(\alpha_s \alpha)$ and proceeds through a closed loop of quarks. Here we are interested in EW corrections to the lowest order amplitude and are therefore calculating two-loop topologies. As a step towards the full $\mathcal{O}(\alpha)$ corrections we present in this work the MIs associated with quark loops arising from the first two generations ($q= u,d,c,s$) for which the quark mass can be set to zero. We postpone analysis of the third generation to a future publication. 
We work in the t'Hooft-Feynman gauge with $\xi=1$. 
The resulting diagrams can be classified in terms of five-master topologies, three planar and two non-planar for which representative Feynman diagrams are shown in 
Fig.\ref{fig:feynAll}. We note that four of the topologies $\text{P}_\text{I-III}$ and $\text{N}_\text{I}$ are also relevant for the mixed QCD-EW corrections to $q\overline{q}\to\gamma\gamma$, whereas $\text{N}_\text{II}$ can only occur with a closed loop of fermions. We note that, in addition to the topologies listed above which all contain at least one massive boson, there are contributions from a photon exchange. However, these contributions are well understood from the studies of $gg\rightarrow gg$ (see e.g.~\cite{Henn:2013pwa} for a canonical basis for these integrals) and we do not consider them explicitly in this paper (although we have checked our results obtained for these diagrams against the literature, finding perfect agreement). 

After computing the two-loop diagrams and applying the projectors described above the resulting the form-factors  consist of thousands of unique scalar integrals. 
However, through the application of well-known reduction techniques the amplitude can be written in terms of a much smaller set of Master Integrals (MI's). 
The choice of MIs is not unique, however, and a judicious choice is a set which satisfy differential equations of the following form~\cite{Henn:2013pwa}
\begin{eqnarray}
 \frac{\pd}{\pd x_i}\vec{\mathcal{G}}(\{x_i\}) = \epsilon \mathcal{A}(\{x_i\}) \,  \vec{\mathcal{G}}(\{x_i\}),
\end{eqnarray}
here $\vec{\mathcal{G}}(\{{x}_i\})$ denotes the vector constructed from the MI's which depends on the external scales $x_i$ (typically chosen to be dimensionless ratios). 
$\mathcal{A}(\{x_i\})$ is a matrix which depends on these external scales, but crucially not on the dimensional regulating parameter $\epsilon$. In this case the kinematics have factorized from the $\epsilon$-series expansion and the integrals are said to be in canonical form. If such a basis can be found, the solution to the above differential equation can be readily obtained, and the MIs can be written as the following Chen iterated integrals~\cite{Chen:1977oja,Caron-Huot:2014lda}
\begin{equation}
\label{eq:chenSol}
	\vec{\mathcal{G}}(\{x_i\})  = \left(\mathbbm{1}+\epsilon \int_\gamma d\mathcal{A}+\epsilon^2\int_\gamma d\mathcal{A}\,d\mathcal{A}+\dots  \right)\vec{\mathcal{G}}(\{x^0_i\}).
\end{equation}
Here $\gamma$ is defined as a smooth map $\gamma:[0,1] \to D \forall p\in D$ such that $\gamma(0) = \{x^0_i\}$ and $\gamma(1) = \{x_i\}$  and $\vec{\mathcal{G}}(\{x^0_i\})$ denotes a boundary vector which must be determined by independent means. If such a boundary vector is known, then the solution is obtained at a given weight by successive integrations over the integration kernel $d\mathcal{A}$. The value of the integral does not depend on the chosen path $\gamma$, provided no singular points (or branch cuts) are crossed. 
By writing the vector of MIs as a series expansion in $\epsilon$ through $\epsilon^4$,
\begin{eqnarray}
\vec{\mathcal{G}}(\{x_i\}) = \sum_{j=0}^{4} \epsilon^j \vec{\mathcal{G}}^{(j)}(\{x_i\}),
\end{eqnarray}
and comparing to eq.~\ref{eq:chenSol}, we can quickly define each coefficient: 
\begin{align}
	& \vec{\mathcal{G}}^{(1)}(\{x_i\}) = \int_\gamma d\mathcal{A}\,\vec{\mathcal{G}}^{(0)}(\{x^0_i\})+\vec{\mathcal{G}}^{(1)}(\{x^0_i\}),\\
	& \vec{\mathcal{G}}^{(2)}(\{x_i\}) = \int_\gamma d\mathcal{A}\,d\mathcal{A}\,\vec{\mathcal{G}}^{(0)}(\{x^0_i\})+ \int_\gamma d\mathcal{A}\,\vec{\mathcal{G}}^{(1)}(\{x^0_i\})+\vec{\mathcal{G}}^{(2)}(\{x^0_i\}),\\
	& \vec{\mathcal{G}}^{(3)}(\{x_i\}) = \int_\gamma d\mathcal{A}\,d\mathcal{A}\,d\mathcal{A}\,\vec{\mathcal{G}}^{(0)}(\{x^0_i\})+ \int_\gamma d\mathcal{A}\,d\mathcal{A}\,\vec{\mathcal{G}}^{(1)}(\{x^0_i\})+\int_\gamma d\mathcal{A}\,\vec{\mathcal{G}}^{(2)}(\{x^0_i\})\nonumber\\
	&\hspace{45pt}+\vec{\mathcal{G}}^{(3)}(\{x^0_i\}),\\
	& \vec{\mathcal{G}}^{(4)}(\{x_i\}) = \int_\gamma d\mathcal{A}\,d\mathcal{A}\,d\mathcal{A}\,d\mathcal{A}\,\vec{\mathcal{G}}^{(0)}(\{x^0_i\})+ \int_\gamma d\mathcal{A}\,d\mathcal{A}\,d\mathcal{A}\,\vec{\mathcal{G}}^{(1)}(\{x^0_i\})+\int_\gamma d\mathcal{A}\,d\mathcal{A}\,\vec{\mathcal{G}}^{(2)}(\{x^0_i\})\nonumber\\	
	&\hspace{45pt} +\int_\gamma d\mathcal{A}\,\vec{\mathcal{G}}^{(3)}(\{x^0_i\}) +\vec{\mathcal{G}}^{(4)}(\{x^0_i\}).
\end{align}
It is convenient to write the integration kernel in a $dlog$ form as follows
\begin{equation}
	d \mathcal{A} = \sum_{k}\mathcal{M}_k \; d\log{\eta_k},
\end{equation}
where $\eta_k$ defines a letter, $\mathcal{M}_k$ is the coefficient matrix relative to said letter and the set of all letters $\{\eta_k\}$ defines the alphabet of the problem.
Following the notation in ref.~\cite{Bonciani:2016ypc} we see that, when expanded to a given weight,  each term which enters the solution is of the form
\begin{eqnarray}
\int_{\gamma} d\log{\eta_{k_j}} \dots d\log{\eta_{k_1}}   = \int_{0 \le t_1 \le \dots \le t_j \le 1} g^{\gamma}_{k_j}(t_j) \dots g^{\gamma}_{k_1}(t_1) dt_1 \dots d t_j,
\label{eq:chenlett}
\end{eqnarray}
where
\begin{eqnarray}
g_k^{\gamma}(t) = \frac{d}{dt} \log{\eta_k(\gamma(t))}, 
\end{eqnarray}
defines the integration of the letter along the chosen path. Following the notation of ref.~\cite{Bonciani:2016ypc} once more, we can more compactly define
\begin{eqnarray}
C^{[\gamma]}_{k_j\dots k_i} \equiv \int_{\gamma} d\log{\eta_{k_j}} \dots d\log{\eta_{k_i}}.
\end{eqnarray}
Our solution is then made up of many terms with different orderings of the letters $k_j\dots k_i$. If all of the letters are rational then the term may be written in terms of the well-known Goncharov polylogarithms:
\begin{align}
	G(a;x_0) & = \int_0^{x_0} \frac{dt}{t-a},\\
	G(a_n,\dots,a_1;x_0) & = \int_0^{x_0} G(a_{n-1},\dots,a_1;x_0) \frac{dt}{t-a_n},\\
	G(\vec{0}_n;x_0) & = \frac{1}{n!}\log{(x_0)}^n.
\end{align}
However, we will frequently encounter non-rational letters which must be directly integrated through eq.~\ref{eq:chenlett}. Nominally one would expect to perform $k$-integrations to obtain a weight-$k$ contribution, however Chen-iterated integrals have several pleasing properties which can systematically reduce the complexity of the problem. We refer interested readers to the more complete overviews which can be found, for instance, in ref.~\cite{Bonciani:2016ypc}. Here we note only the most important identities required in our calculation. 
Firstly we note that at weight one, path independence renders the integration trivial, one must simply evaluate the integral at its end-points.\newpage Then at weight two one can  use this to perform the inner most integration and then perform the following integration by parts identity writing
\begin{eqnarray}
C^{[\gamma]}_{ab} =\int_0^1 g_b(t) (\log{\eta_a({\bf{x}}(t)}) - \log{\eta_a({{\bf{x}}_0)} )dt},
\end{eqnarray}
such that all weight two contributions can be written as single parameter integrals. Once all weight two functions are obtained one can systematically reduce the number of integrations, by using the following generalized integration by parts identity
\begin{eqnarray}
C^{[\gamma]}_{k_j\dots k_i}  = \log{\eta_{k_j}({\bf{x}}}) \;C^{[\gamma]}_{k_{j-1}\dots k_i}   - \int  \log{\eta_{k_j}({\bf{x}}(t))} \;g_{k_{j-1}}(t) \;C^{[\gamma_t]}_{k_{j-2}\dots k_i}  dt,
\end{eqnarray}
where $ [ \gamma_t ] $ is  a modified path which integrates over the domain $[0,t ]$ (and the complete path is restored as $t\rightarrow 1$). 
Up to weight four order, we therefore are able to write weight three contributions as one parameter integrals, and weight four contributions as integrals over two parameters. 
Finally, situations frequently occur when the inner most letter vanishes at the boundary of integration, which introduces potential singular contributions. 
In order to extract and remove theses contributions we employ a shuffle algebra to systematically extract singular logarithms (and cancel them with the corresponding terms arising at the boundary)
\begin{eqnarray}
C^{[\gamma]}_{{\pmb{k}}}C^{[\gamma]}_{{\pmb{j}}} =  \sum_{\rm{shuffles} \;\; \sigma} C^{[\gamma]}_{\sigma(k_k),\dots \sigma(k_1),\sigma(j_j),\dots,\sigma(j_1)},
\end{eqnarray}
where ${\pmb{k}}$ and ${\pmb{j}}$ are the vectors of letters with length $k$ and $j$ and the sum runs over all permutations which preserve the relative order of ${\pmb{k}}$ and ${\pmb{j}}$. 

Our general setup is as follows: starting from the relevant Feynman diagrams for each topology we project out the form factors obtaining scalar integrals. 
These are then reduced using {\tt{Kira 2.0}}~\cite{Klappert:2020nbg,Maierhofer:2017gsa} to a minimal set of MIs. We then promote the obtained basis to 
canonical form using the Magnus-Dyson approach outlined in ref.~\cite{Argeri:2014qva}. We then use the physical properties of the Feynman integrals to 
determine boundaries at special kinematic points. Our final answers are then obtained via path integration away from these boundary vectors. In all instances 
we are able to check our results numerically using the package {\tt{AMFlow}}~\cite{Liu:2022chg} (we have additionally checked with {\tt{pySecDec}}~\cite{Borowka:2015mxa} where possible too).
During our calculations we make frequent use of the software  {\tt{PolyLogTools}}~\cite{Duhr:2019tlz} and {\tt{HandyG}}~\cite{Naterop:2019xaf}. 
The next sections describe these calculations in detail.

%=====================================================================

%================================== PLANAR RESULTS 

%=====================================================================

\section{Planar Topologies} 
\label{sec:Plan}

\subsection{Planar topology P$_\text{I}$}

%\subsubsection{Setup}
\begin{figure}[H]
	\centering
	\includegraphics[scale=1.5]{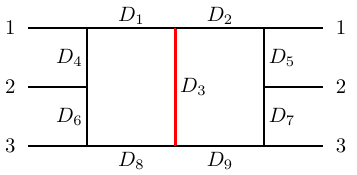}
	\caption{The auxiliary topology diagram for P$_\text{I}$ as defined in Eq.~\ref{eq:topo_PI}. Propagators marked in red are massive.}
	\label{fig:aux_PI}
\end{figure}
\begin{figure}[H]
	\centering
	\includegraphics[scale=0.62]{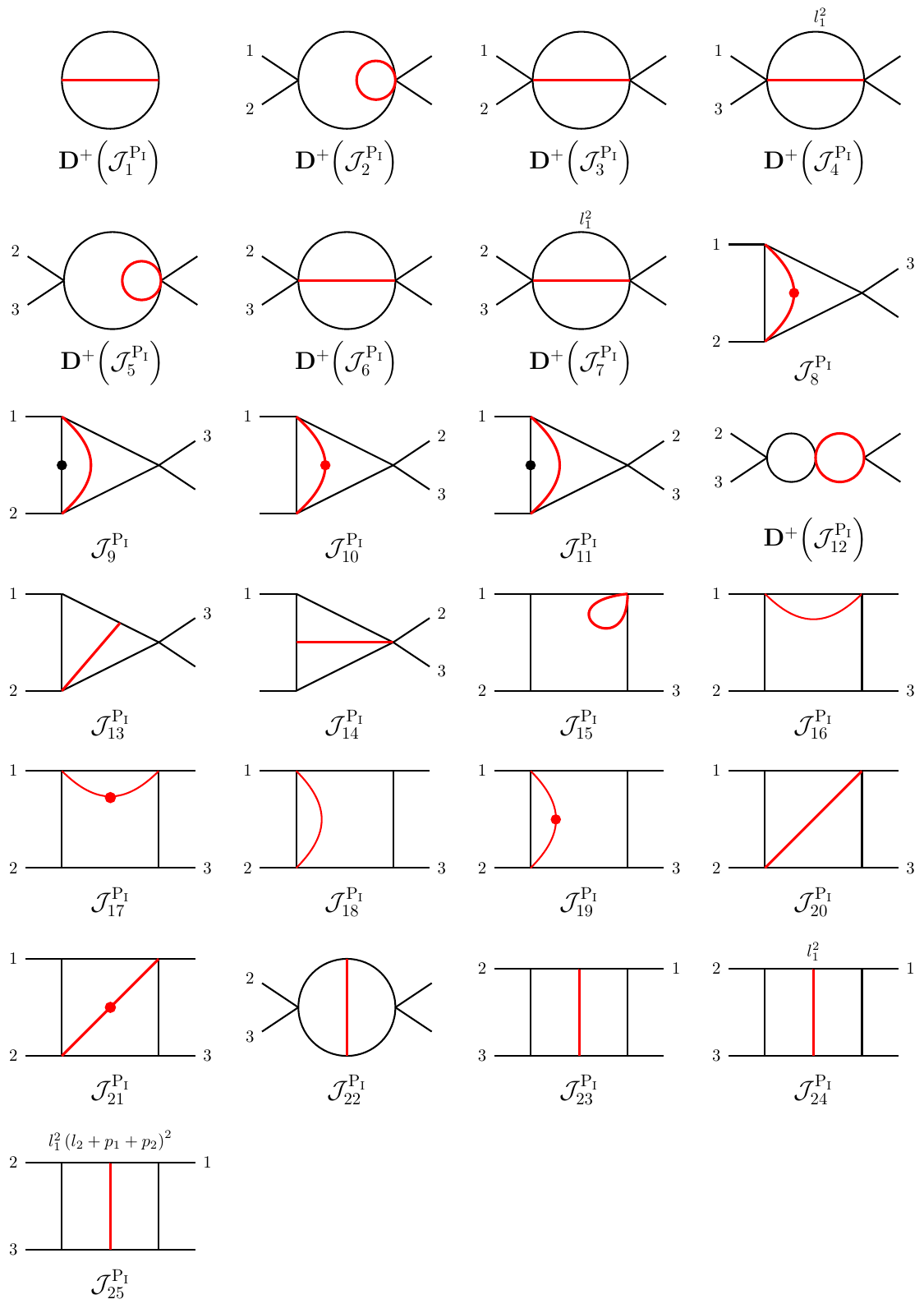}
	\caption{Master integrals for the P$_\text{I}$ topology. Dotted propagators have higher power of the respective denominator, while red propagators are massive.}
	\label{fig:MI_PI}
\end{figure}

The first planar topology we encounter involves the exchange of a single massive boson. The associated propagator structure has the following form
\begin{eqnarray}
	I^{\text{P}_{\text{I}}}_{a_1\dots a_9} =\left(\frac{M_W^{2\epsilon}}{\Gamma({\epsilon})}\right)^2 \int \frac{d^d \ell_1}{(2\pi)^d}\frac{d^d \ell_2}{(2\pi)^d} 
	\frac{1}{D_1^{a_1}D_2^{a_2}D_3^{a_3}D_4^{a_4}D_5^{a_5}D_6^{a_6}D_7^{a_7}D_8^{a_8}D_9^{a_9}},
\end{eqnarray}
with
\begin{align}
	\label{eq:topo_PI}
	& D_1=\ell_1^2,  
	&& D_{2}=\ell_2^2,
	&& D_{3}=(\ell_1 + \ell_2)^2- M_W^2, \nonumber\\
	& D_{4}= (\ell_1 - p_1)^2,
	&& D_{5}= (\ell_2 + p_1)^2, 
	&& D_{6}= (\ell_1 - p_1 - p_2)^2,\nonumber\\
	& D_{7}= (\ell_2 + p_1 + p_2)^2, 
	&&  D_{8}= (\ell_1 - p_1 - p_2 - p_3)^2,
	&&  D_{9}= (\ell_2 + p_1 + p_2 + p_3)^2.
\end{align}
These propagators form an auxiliary topology which is presented in Fig.~\ref{fig:aux_PI}. We note that, since the external momentum $\{k_i\}$ can be permuted into any ordering, we need to evaluate this topology in three distinct regions defined by the signs of $p_i\cdot p_j$.
All the integrals that appear in the physical amplitude can be expressed, via integration by parts identities (IBPs), in terms of a minimal basis of Master Integrals (MIs). We used {\tt{Kira}}~\cite{Klappert:2020nbg} to obtain such set of identities and we found that for the sum over all Feynman diagrams with a single massive boson with a fixed order of the final state momentum we need a total of 25 MIs. 
Therefore, in order to obtain a basis in canonical form we start with the following 25 integrals:
\begin{align}
	\mathcal{J}_{1} & = {\bf{D}}^-(I^{\text{P}_{\text{I}}}_{111000000}),
	&\mathcal{J}_{2} & = {\bf{D}}^-(I^{\text{P}_{\text{I}}}_{101001000}),
	&\mathcal{J}_{3} & = {\bf{D}}^-(I^{\text{P}_{\text{I}}}_{011001000}),\nonumber\\
	\mathcal{J}_{4} & = {\bf{D}}^-(I^{\text{P}_{\text{I}}}_{(-1)11001000}),
	&\mathcal{J}_{5} & = {\bf{D}}^-(I^{\text{P}_{\text{I}}}_{001100010}),
	&\mathcal{J}_{6} & = {\bf{D}}^-(I^{\text{P}_{\text{I}}}_{001010010}),\nonumber\\
	\mathcal{J}_{7} & = {\bf{D}}^-(I^{\text{P}_{\text{I}}}_{(-1)01010010}),
	&\mathcal{J}_{8} & = I^{\text{P}_{\text{I}}}_{102011000},
	&\mathcal{J}_{9} & = I^{\text{P}_{\text{I}}}_{101021000},\nonumber\\
	\mathcal{J}_{10} & = I^{\text{P}_{\text{I}}}_{012100010},
	&\mathcal{J}_{11} & = I^{\text{P}_{\text{I}}}_{021100010},
	&\mathcal{J}_{12} & = {\bf{D}}^-(I^{\text{P}_{\text{I}}}_{000110011}),\nonumber\\
	\mathcal{J}_{13} & = I^{\text{P}_{\text{I}}}_{111011000},
	&\mathcal{J}_{14} & = I^{\text{P}_{\text{I}}}_{111010010},
	&\mathcal{J}_{15} & = I^{\text{P}_{\text{I}}}_{101101010},\nonumber\\
	\mathcal{J}_{16} & = I^{\text{P}_{\text{I}}}_{011101010},
	&\mathcal{J}_{17} & = I^{\text{P}_{\text{I}}}_{012101010},
	&\mathcal{J}_{18} & = I^{\text{P}_{\text{I}}}_{101011010},\nonumber\\
	\mathcal{J}_{19} & = I^{\text{P}_{\text{I}}}_{102011010},
	&\mathcal{J}_{20} & = I^{\text{P}_{\text{I}}}_{011011010},
	&\mathcal{J}_{21} & = I^{\text{P}_{\text{I}}}_{012011010},\nonumber\\
	\mathcal{J}_{22} & = I^{\text{P}_{\text{I}}}_{001110011},
	&\mathcal{J}_{23} & = I^{\text{P}_{\text{I}}}_{011111011},
	&\mathcal{J}_{24} & = I^{\text{P}_{\text{I}}}_{(-1)11111011},\nonumber\\
	\mathcal{J}_{25} & = I^{\text{P}_{\text{I}}}_{(-1)11111(-1)11},
\end{align}
which are presented in Fig.~\ref{fig:MI_PI}. It is convenient to write some of these integrals using the dimension shifting operator
\begin{eqnarray}
	{\bf{D}}^-(I^{\text{P}_\text{I}}_{a_1\dots a_9})&=&\bigg((a_1 \mathbf{1}^+ + a_4 \mathbf{4}^++ a_6  \mathbf{6}^++ a_8 \mathbf{8}^+)(a_2 \mathbf{2}^+ + a_5 \mathbf{5}^+ + a_7 \mathbf{7}^+ + a_9 \mathbf{9}^+) 
	\nonumber\\&&+ a_3 \mathbf{3}^+ (\sum_{j \ne 3} \,a_j {\mathbf{j}^+}) \bigg)I^{\text{P}_\text{I}}_{a_1\dots a_9}. \nonumber\\
\end{eqnarray}
Where $\mathbf{j}^+$ increases the power of the $j^{\rm{th}}$ propagator by one.  
In order to write our basis in canonical form, we introduce the following finite and dimensionless integrals
\begin{eqnarray}
	\mathcal{F}_{X}= \epsilon^2 \; (M_W)^{\alpha} \mathcal{J}_{X},
\end{eqnarray}
where $\alpha$ is defined for $ \mathcal{J}_{X}=  f(I^{\text{P}_{\text{I}}}_{a_1\dots a_9})$ as $(\sum {a_i}) -4$, and we utilize the techniques of the Magus-Dyson approach outlined in ref.~\cite{Argeri:2014qva} to obtain the following canonical basis:
\begin{align}
	&\mathcal{G}_{1}^{\text{P}_\text{I}}=\mathcal{F}_{1},
	&&\mathcal{G}_{2}^{\text{P}_\text{I}}=x \mathcal{F}_{2},
	&&\mathcal{G}_{3}^{\text{P}_\text{I}}=(1+x) \mathcal{F}_{3},\nonumber\\[5pt]
	&\mathcal{G}_{4}^{\text{P}_\text{I}}=\mathcal{F}_{4}-\mathcal{F}_{3},
	&&\mathcal{G}_{5}^{\text{P}_\text{I}}=y \mathcal{F}_{5},
	&&\mathcal{G}_{6}^{\text{P}_\text{I}}=(1+y) \mathcal{F}_{6},\nonumber\\[5pt]
	&\mathcal{G}_{7}^{\text{P}_\text{I}}=\mathcal{F}_{7},
	&&\mathcal{G}_{8}^{\text{P}_\text{I}}=\epsilon x  \mathcal{F}_{8},
	&&\mathcal{G}_{9}^{\text{P}_\text{I}}=\epsilon x \mathcal{F}_{9},\nonumber\\[5pt]
	&\mathcal{G}_{10}^{\text{P}_\text{I}}=\epsilon y\mathcal{F}_{10},
	&&\mathcal{G}_{11}^{\text{P}_\text{I}}=\epsilon y\mathcal{F}_{11},
	&&\mathcal{G}_{12}^{\text{P}_\text{I}}=y^2 \mathcal{F}_{12},\nonumber\\[5pt]
	&\mathcal{G}_{13}^{\text{P}_\text{I}}=\epsilon^2 x  \mathcal{F}_{13},
	&&\mathcal{G}_{14}^{\text{P}_\text{I}}= \epsilon ^2 y \mathcal{F}_{14},
	&&\mathcal{G}_{15}^{\text{P}_\text{I}}=\epsilon(\epsilon-1 )x y  \mathcal{F}_{15},\nonumber\\[5pt]
	&\mathcal{G}_{16}^{\text{P}_\text{I}}= \epsilon y ((2 \epsilon-1 )\mathcal{F}_{16}+\mathcal{F}_{17}),
	&&\mathcal{G}_{17}^{\text{P}_\text{I}}= \epsilon x y \mathcal{F}_{17},
	&&\mathcal{G}_{18}^{\text{P}_\text{I}}= \epsilon  ((2 \epsilon-1 )x \mathcal{F}_{18}+\mathcal{F}_{19}),\nonumber\\[5pt]
	&\mathcal{G}_{19}^{\text{P}_\text{I}}= \epsilon x y\mathcal{F}_{19},
	&&\mathcal{G}_{20}^{\text{P}_\text{I}}=\epsilon^2(x+y) \mathcal{F}_{20},
	&&\mathcal{G}_{21}^{\text{P}_\text{I}}=-\epsilon(x+y+x y) \mathcal{F}_{21},\nonumber\\[5pt]
	&\mathcal{G}_{22}^{\text{P}_\text{I}}= \epsilon(1-2 \epsilon )y \mathcal{F}_{22},
	&&\mathcal{G}_{23}^{\text{P}_\text{I}}=\epsilon^2 (1+x) y^2 \mathcal{F}_{23},
	&&\mathcal{G}_{24}^{\text{P}_\text{I}}=\epsilon ^2 y^2 (\mathcal{F}_{24}-\mathcal{F}_{23}),\nonumber\\[5pt]
	&\mathcal{G}_{25}^{\text{P}_\text{I}}= \mathrlap{x \left(\frac{1}{4} y\mathcal{F}_{12}+2 \epsilon ^2 \mathcal{F}_{20}+y \epsilon ^2\mathcal{F}_{23}\right)+ \epsilon^2y(\mathcal{F}_{25}-\mathcal{F}_{22}).}
\end{align}
Here the differential equation is expressed in terms of the following dimensionless ratios,
\begin{eqnarray}
	\label{eq:PI_vars}
	x =  -\frac{2p_1 \cdot p_2 }{M_W^2}   {\qquad\text{and}\qquad}    y = -\frac{2p_2 \cdot p_3}{M_W^2}.
\end{eqnarray}
This basis is indeed in canonical form, and satisfies the following differential equations
\begin{eqnarray}
	\frac{\partial \mathcal{G}^{\text{P}_{\text{I}}}_j}{\partial x} = \epsilon (A_x  \mathcal{G}^{\text{P}_{\text{I}}})_j  \qquad\text{and}\qquad  \frac{\partial \mathcal{G}^{\text{P}_{\text{I}}}_j}{\partial y} = \epsilon (A_y   \mathcal{G}^{\text{P}_{\text{I}}})_j.
\end{eqnarray}
For this topology the MIs depend only on rational functions of $x$ and $y$. As a result, the solutions of the differential equation can be expressed in terms of Goncharov polylogarithms (GPLs) and efficiently evaluated using {\tt{HandyG}}~\cite{Naterop:2019xaf} at every point in the phase space. In particular, the kinematic regions of interest, in terms of the variables in Eq.\ref{eq:PI_vars}, are given by 
\begin{align}
	\label{eq:PI_regionA}
	&{\rm{Region \; \; A}}: && (p_1\cdot p_2 > 0,\; p_1\cdot p_3 < 0,\; p_2\cdot p_3 < 0) \quad \to &&  x < 0,\; y > 0,\; \lvert x \lvert>\lvert y \lvert, \\[5pt]
	\label{eq:PI_regionB}
	&{\rm{Region \; \; B}}: && (p_1\cdot p_2 < 0,\; p_1\cdot p_3 > 0,\; p_2\cdot p_3 < 0) \quad \to &&  x > 0,\; y > 0 ,\\[5pt]
	\label{eq:PI_regionC}
	&{\rm{Region \; \; C}}: && (p_1\cdot p_2 < 0,\; p_1\cdot p_3 < 0,\; p_2\cdot p_3 > 0) \quad \to &&  x > 0,\; y < 0,\; \lvert y \lvert>\lvert x \lvert .
\end{align}
The boundary conditions can be determined for these integrals as follows:
\begin{itemize}
\item $\mathcal{G}_{3}^{\text{P}_\text{I}}$ and $\mathcal{G}_{4}^{\text{P}_\text{I}}$ are fixed since they are finite as $x\rightarrow 0$, and in this limit $\mathcal{G}_{3}^{\text{P}_\text{I}} = \mathcal{G}_{1}^{\text{P}_\text{I}}$.
\item Similarly  $\mathcal{G}_{6}^{\text{P}_\text{I}}$ and  $\mathcal{G}_{7}^{\text{P}_\text{I}}$  are finite as $y\rightarrow 0$, and in this limit $\mathcal{G}_{6}^{\text{P}_\text{I}} = \mathcal{G}_{1}^{\text{P}_\text{I}}$.
\item $\mathcal{G}_{8}^{\text{P}_\text{I}}$ and $\mathcal{G}_{9}^{\text{P}_\text{I}}$ can be determined in terms of $\mathcal{G}_{2}^{\text{P}_\text{I}}$ by demanding that these integrals are finite as $x\rightarrow0$ and noting that  $\mathcal{G}_{8}^{\text{P}_\text{I}}$ vanishes in this limit. Further $\mathcal{G}_{2}^{\text{P}_\text{I}}$ can then be constrained in terms of $\mathcal{G}_{1}^{\text{P}_\text{I}}$ by demanding that $\mathcal{G}_{9}^{\text{P}_\text{I}}$ is finite as $x\rightarrow 1$. In a similar fashion integrals 10 and 11 and 5 are fixed with $x\leftrightarrow y$ in terms of  $\mathcal{G}_{1}^{\text{P}_\text{I}}$.
\item  $\mathcal{G}_{12}^{\text{P}_\text{I}}$ is the product of two trivial one-loop integrals and can be determined quickly via direct integration. 
\item  $\mathcal{G}_{13}^{\text{P}_\text{I}}$ and  \{$\mathcal{G}_{14}^{\text{P}_\text{I}},\mathcal{G}_{22}^{\text{P}_\text{I}}\}$ are fixed since they must vanish as $x\rightarrow 0$ and $y\rightarrow 0$ respectively. 
\item $\mathcal{G}_{15-21}^{\text{P}_\text{I}}$,  and $\mathcal{G}_{23-25}^{\text{P}_\text{I}}$ are fixed by demanding finiteness in the limit $x\rightarrow -y$.  
\end{itemize}
Our integrals are then completely determined in terms of the overall normalizing integral $\mathcal{G}_{1}^{\text{P}_\text{I}}$ which is easily computed by direct integration, to our required order we have
\begin{eqnarray}
 \mathcal{G}_{1}^{\text{P}_\text{I}}=-1 - 2\zeta_2 \epsilon^2+2\zeta_3\epsilon^3-9\zeta_4 \epsilon^4 + \mathcal{O}(\epsilon^5).
\end{eqnarray}
The evaluation of these integrals for a sample point for each kinematic region is presented in Appendix~\ref{appendix:PI}.  We have checked our results against the numerical package {\tt{AMFlow}}~\cite{Liu:2022chg}, finding perfect agreement.  

\subsection{Planar topology P$_\text{II}$}
%\subsubsection{Setup}
\begin{figure}[H]
	\centering
	\includegraphics[scale=1.5]{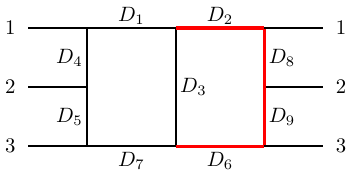}
	\caption{The auxiliary topology diagram for P$_\text{II}$ as defined in Eq.~\ref{eq:topo_PII}. Propagators marked in red are massive.}
	\label{fig:aux_PII}
\end{figure}
\begin{figure}[H]
	\centering
	\includegraphics[scale=0.75]{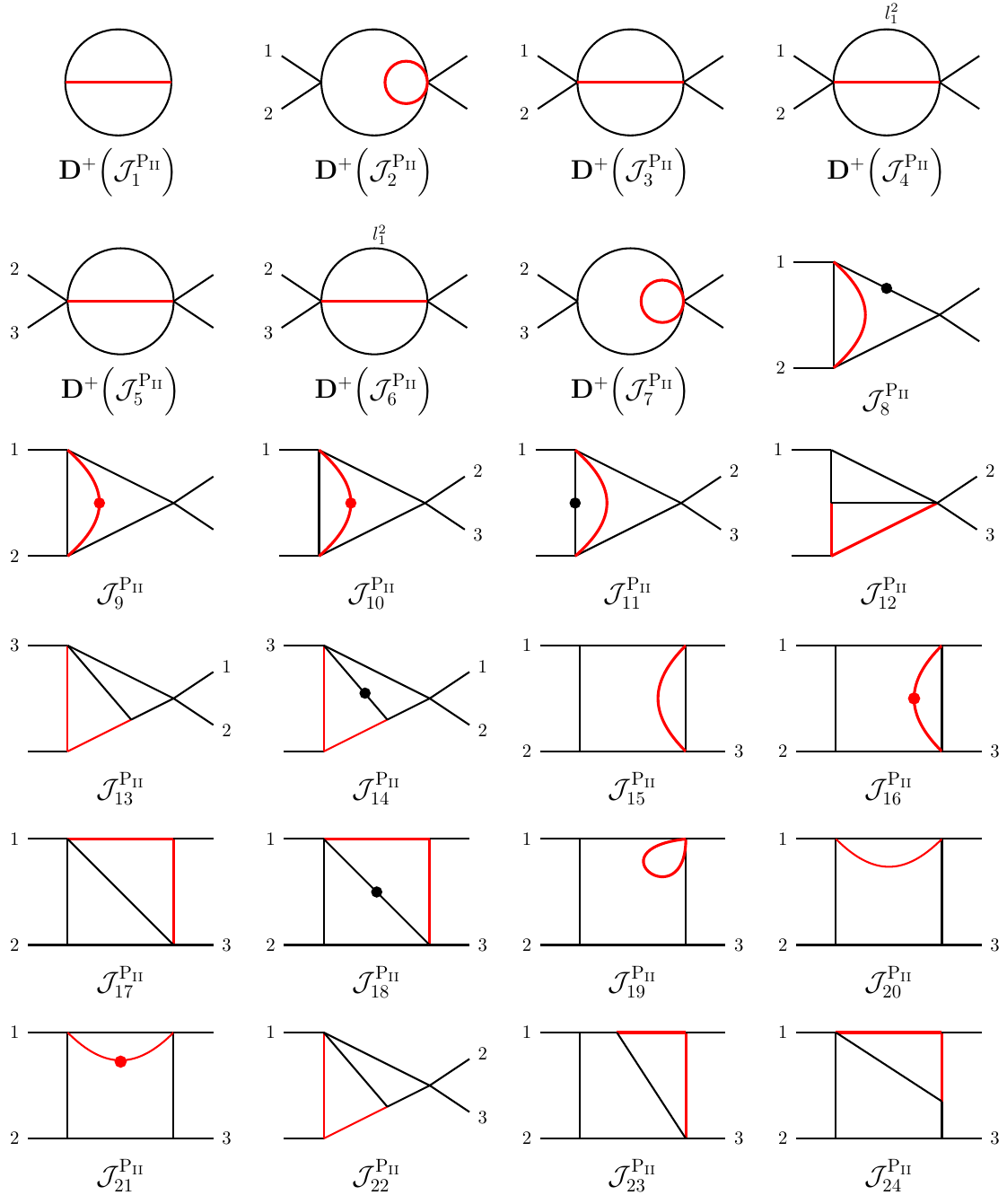}
	\caption{Master integrals for topology  P$_\text{II}$. Dotted propagators have higher power of the respective denominator, while red propagators are massive.}
	\label{fig:MI_PII}
\end{figure}

The second planar topology we will consider arises when one external photon is radiated from a $W$-boson and the other is radiated from the quark loop. 
The various permutations of photons and gluons can be written as functions of the following auxiliary Feynman integral
\begin{eqnarray}
	I^{\text{P}_{\text{II} }}_{a_1\dots a_9} = \left(\frac{M_W^{2\epsilon}}{\Gamma({\epsilon})}\right)^2\int \frac{d^d \ell_1}{(2\pi)^d}\frac{d^d \ell_2}{(2\pi)^d} 
	\frac{1}{D_1^{a_1}D_2^{a_2}D_3^{a_3}D_4^{a_4}D_5^{a_5}D_6^{a_6}D_7^{a_7}D_8^{a_8}D_9^{a_9}},
\end{eqnarray}
where:
\begin{align}
	\label{eq:topo_PII}
	& D_{1}=\ell_1^2, 
	&& D_{2}=\ell_2^2- M_W^2, 
	&& D_{3}=(\ell_1 + \ell_2)^2,\nonumber\\
	& D_{4}= (\ell_1  - p_1)^2,
	&&D_{5}= (\ell_1 - p_1 - p_2)^2,
	&& D_{6}= (\ell_2 + p_1 + p_2+p_3)^2- M_W^2 \nonumber\\
	&D_{7}= (\ell_1 - p_1-p_2-p_3)^2,
	&&D_{8}= (\ell_2 + p_1)^2-M_W^2, 
	&&D_{9}= (\ell_2 + p_1 + p_2)^2-M_W^2. 
\end{align}
This topology is shown in Fig.~\ref{fig:aux_PII}. For contributions to $gg\rightarrow \gamma\gamma$ considered here, $D_8$ and $D_9$ do not
appear with positive power. 
As before, by including the various permutations of physical momenta $\{k_1,k_2,k_3\}$ we find that we must calculate this auxiliary topology in two distinct regions, depending on the sign of the associated $p_i \cdot p_j$.
Following the same procedure as before, we reduce all the integrals appearing in the physical amplitude using {\tt{Kira}}~\cite{Klappert:2020nbg} finding that for this topology we need a total of 24 MIs. In order to derive a canonical basis we start with the following set of integrals:
\begin{align}
	&\mathcal{J}_{1} = {\bf{D}}^-(I^{\text{P}_{\text{II}}}_{111000000}),\quad
	&&\mathcal{J}_{2} = {\bf{D}}^-(I^{\text{P}_{\text{II}}}_{110010000}),\quad
	&&\mathcal{J}_{3} = {\bf{D}}^-(I^{\text{P}_{\text{II}}}_{011010000}),\nonumber\\
	&\mathcal{J}_{4} = {\bf{D}}^-(I^{\text{P}_{\text{II}}}_{(-1)11010000}),\quad
	&&\mathcal{J}_{5} = {\bf{D}}^-(I^{\text{P}_{\text{II}}}_{001101000}),\quad
	&&\mathcal{J}_{6} = {\bf{D}}^-(I^{\text{P}_{\text{II}}}_{(-1)01101000}),\nonumber\\
	&\mathcal{J}_{7} = {\bf{D}}^-(I^{\text{P}_{\text{II}}}_{010100100}),\quad
	&&\mathcal{J}_{8} = I^{\text{P}_{\text{II}}}_{201011000},\quad
	&&\mathcal{J}_{9} = I^{\text{P}_{\text{II}}}_{101012000},\nonumber\\
	&\mathcal{J}_{10} = I^{\text{P}_{\text{II}}}_{021100100},\quad
	&&\mathcal{J}_{11} = I^{\text{P}_{\text{II}}}_{012100100},\quad
	&&\mathcal{J}_{12} = I^{\text{P}_{\text{II}}}_{111101000},\nonumber\\
	&\mathcal{J}_{13} = I^{\text{P}_{\text{II}}}_{111011000},\quad
	&&\mathcal{J}_{14} = I^{\text{P}_{\text{II}}}_{112011000},\quad
	&&\mathcal{J}_{15} = I^{\text{P}_{\text{II}}}_{101111000},\nonumber\\
	&\mathcal{J}_{16} = I^{\text{P}_{\text{II}}}_{101112000},\quad
	&&\mathcal{J}_{17} = I^{\text{P}_{\text{II}}}_{011111000},\quad
	&&\mathcal{J}_{18} = I^{\text{P}_{\text{II}}}_{012111000},\nonumber\\
	&\mathcal{J}_{19} = I^{\text{P}_{\text{II}}}_{110110100},\quad
	&&\mathcal{J}_{20} = I^{\text{P}_{\text{II}}}_{011110100},\quad
	&&\mathcal{J}_{21} = I^{\text{P}_{\text{II}}}_{021110100},\nonumber\\
	&\mathcal{J}_{22} = I^{\text{P}_{\text{II}}}_{011101100},\quad
	&&\mathcal{J}_{23} = I^{\text{P}_{\text{II}}}_{111111000},\quad
	&&\mathcal{J}_{24} = I^{\text{P}_{\text{II}}}_{011111100},\nonumber\\
\end{align}
these integrals are also represented in Fig.~\ref{fig:MI_PII}. The three propagator integrals are written using the dimension shifting operator, which for this topology is defined as
\begin{eqnarray}
	{\bf{D}}^-(I^{\text{P}_{\text{II}}}_{a_1\dots a_9})&=&\bigg((a_1 \mathbf{1}^+ + a_4  \mathbf{4}^++ a_5  \mathbf{5}^++ a_7 \mathbf{7}^+)(a_2 \mathbf{2}^+ + a_6 \mathbf{6}^+ a_8 \mathbf{8}^+ + a_9 \mathbf{9}^+) 
	\nonumber\\&& +\; a_3 \mathbf{3}^+ \sum_{j \ne 3} \,a_j {\mathbf{j}^+} \bigg)I^{\text{P}_{\text{II} }}_{a_1\dots a_9}.
\end{eqnarray}
We repeat the steps outlined in the previous section to rewrite our basis in canonical form by first rendering the integrals finite and dimensionless
\begin{eqnarray}
	\mathcal{F}_{X}= \epsilon^2 \; (M_W)^{\alpha} \mathcal{J}_{X},
\end{eqnarray}
and ultimately we obtain the following canonical basis:
\begin{align}
	\label{eq:PII_can}
	&\mathcal{G}_{1}^{\text{P}_\text{II}}=\mathcal{F}_{1},
	&&\mathcal{G}_{2}^{\text{P}_\text{II}}=x \mathcal{F}_{2},\hspace{50pt}
	&&\mathcal{G}_{3}^{\text{P}_\text{II}}=(1+x) \mathcal{F}_{3},\nonumber\\[5pt]
	&\mathcal{G}_{4}^{\text{P}_\text{II}}=\mathcal{F}_{4}-\mathcal{F}_{3},
	&&\mathcal{G}_{5}^{\text{P}_\text{II}}=(1+y) \mathcal{F}_{5},
	&&\mathcal{G}_{6}^{\text{P}_\text{II}}=\mathcal{F}_{6},\nonumber\\[5pt]
	&\mathcal{G}_{7}^{\text{P}_\text{II}}=y \mathcal{F}_{7},\nonumber\\[5pt]
	&\mathcal{G}_{8}^{\text{P}_\text{II}}=\mathrlap{\frac{2x^2(2 \epsilon -1)\mathcal{F}_8+2\epsilon x (\mathcal{F}_1-\mathcal{F}_9)  +x\mathcal{F}_2}{2 (x-1)},}
	&&&&\mathcal{G}_{9}^{\text{P}_\text{II}}=\epsilon x   \mathcal{F}_{9},\nonumber\\[5pt]
	&\mathcal{G}_{10}^{\text{P}_\text{II}}=\epsilon y  \mathcal{F}_{10},
	&&\mathcal{G}_{11}^{\text{P}_\text{II}}=\epsilon y \mathcal{F}_{11},
	&&\mathcal{G}_{12}^{\text{P}_\text{II}}=\epsilon^2 y \mathcal{F}_{12},\nonumber\\[5pt]
	&\mathcal{G}_{13}^{\text{P}_\text{II}}=\epsilon^2 x  \mathcal{F}_{13},
	&&\mathcal{G}_{14}^{\text{P}_\text{II}}=\epsilon  x  \mathcal{F}_{14},
	&&\mathcal{G}_{15}^{\text{P}_\text{II}}=\epsilon x((2 \epsilon-1	) \mathcal{F}_{15}+\mathcal{F}_{16})\nonumber\\[5pt],
	&\mathcal{G}_{16}^{\text{P}_\text{II}}=\epsilon x y \mathcal{F}_{16},
	&&\mathcal{G}_{17}^{\text{P}_\text{II}}=\epsilon^2 (x+y) \mathcal{F}_{17},
	&&\mathcal{G}_{18}^{\text{P}_\text{II}}=\epsilon(x+y+x y) \mathcal{F}_{18},\nonumber\\[5pt]
	&\mathcal{G}_{19}^{\text{P}_\text{II}}=\epsilon(\epsilon-1 ) x y \mathcal{F}_{19},
	&&\mathcal{G}_{20}^{\text{P}_\text{II}}=\epsilon y   ((2 \epsilon-1) \mathcal{F}_{20}+\mathcal{F}_{21}),
	&&\mathcal{G}_{21}^{\text{P}_\text{II}}=\epsilon x y \mathcal{F}_{21},\nonumber\\[5pt]
	&\mathcal{G}_{22}^{\text{P}_\text{II}}=\epsilon^2 y  \mathcal{F}_{22},
	&&\mathcal{G}_{23}^{\text{P}_\text{II}}=\epsilon^2 x y \mathcal{F}_{23},
	&&\mathcal{G}_{24}^{\text{P}_\text{II}}=\epsilon^2 x y \mathcal{F}_{24},\nonumber\\
\end{align}
where the differential equation variables are defined to the same as those in the previous topology
\begin{eqnarray}
x =-\frac{2p_1\cdot p_2}{M_W^2} \quad {\rm{and}} \quad y =-\frac{2p_2\cdot p_3}{M_W^2}.
\end{eqnarray}
When considering all possible assignments of physical external momenta, we find we have to consider two kinematic regions, which we define as:
\begin{align}
	\label{eq:PII_regionA}
	&{\rm{Region \; \; A}}: && (p_1\cdot p_2 > 0,\; p_1\cdot p_3 < 0,\; p_2\cdot p_3 < 0) \quad \to &&  x < 0,\; y > 0,\; \lvert x \lvert>\lvert y \lvert, \\[5pt]
	\label{eq:PII_regionB}
	&{\rm{Region \; \; B}}: && (p_1\cdot p_2 < 0,\; p_1\cdot p_3 > 0,\; p_2\cdot p_3 < 0) \quad \to &&  x > 0,\; y > 0 .
\end{align}
%Writing the differential equations in $dlog$ form 
%\begin{eqnarray}
%d \mathcal{\vec{G}} =   \sum_{k}\mathcal{M}_k \; d\log{\eta_k}  \;  \vec{\mathcal{G}}
%\end{eqnarray}
%we find that we draw letters from the following alphabet 
%\begin{eqnarray}
%\{x,x\pm1, y,y\pm 1, x+y,x + y + x y\}.
%\end{eqnarray}
The basis in Eq.~\ref{eq:PII_can} is in canonical form, and satisfies the differential equations
\begin{eqnarray}
	\frac{\partial \mathcal{G}^{\text{P}_{\text{II}}}_j}{\partial x} = \epsilon (A_x  \mathcal{G}^{\text{P}_{\text{II}}})_j  \qquad\text{and}\qquad  \frac{\partial \mathcal{G}^{\text{P}_{\text{II}}}_j}{\partial y} = \epsilon (A_y   \mathcal{G}^{\text{P}_{\text{II}}})_j, 
\end{eqnarray}
where $A_x$ and $A_y$ have no dependence on $\epsilon$. Additionally, since the coefficient matrices depend only on rational functions of the variables $x$ and $y$, the solution can be written entirely in-terms of Goncharov polylogarithms and can be readily determined, once appropriate boundary conditions are known.
The boundary constants can be determined as follows, 
\begin{itemize}
\item 
$\mathcal{G}^{\text{P}_\text{II}}_3$ and $\mathcal{G}^{\text{P}_\text{II}}_4$ are finite as $x\rightarrow 0$, and further $\mathcal{G}^{\text{P}_\text{II}}_3 = \mathcal{G}^{\text{P}_\text{II}}_1$ when $x=0$. 
\item 
The finiteness of $\mathcal{G}^{\text{P}_\text{II}}_5$ at $y=0$ fixes $\mathcal{G}^{\text{P}_\text{II}}_6$ and the relation $\mathcal{G}^{\text{P}_\text{II}}_5=\mathcal{G}^{\text{P}_\text{II}}_1$ at $y=0$ fixes $\mathcal{G}^{\text{P}_\text{II}}_5$. 
\item 
The finiteness of $\mathcal{G}^{\text{P}_\text{II}}_9$ at $x=0$ fixes $\mathcal{G}^{\text{P}_\text{II}}_{8}$ and the vanishing of $\mathcal{G}^{\text{P}_\text{II}}_9$ at $y=0$ fixes $\mathcal{G}^{\text{P}_\text{II}}_9$, such that $\mathcal{G}^{\text{P}_\text{II}}_8$ and $\mathcal{G}^{\text{P}_\text{II}}_9$ are determined in terms of $\mathcal{G}^{\text{P}_\text{II}}_{2}$. $\mathcal{G}^{\text{P}_\text{II}}_2$ can be further constrained by requiring that $\mathcal{G}^{\text{P}_\text{II}}_8$ is finite as $x\rightarrow 1$. 
\item 
The finiteness of $\mathcal{G}^{\text{P}_\text{II}}_{10}$ at $y=0$ and $\mathcal{G}^{\text{P}_\text{II}}_{11}$ at $y=1$ fixes $\mathcal{G}^{\text{P}_\text{II}}_{11}$ and $\mathcal{G}^{\text{P}_\text{II}}_{7}$. $\mathcal{G}^{\text{P}_\text{II}}_{10}$ can be further constrained since the integral vanishes at $y=0$. 
\item 
$\mathcal{G}^{\text{P}_\text{II}}_{12}$ vanishes as $y\rightarrow 0$.
\item 
$\mathcal{G}^{\text{P}_\text{II}}_{13}$ and $\mathcal{G}^{\text{P}_\text{II}}_{14}$ vanish as $x\rightarrow 0$.
\item 
$\mathcal{G}^{\text{P}_\text{II}}_{15}$ and $\mathcal{G}^{\text{P}_\text{II}}_{16}$, and $\mathcal{G}^{\text{P}_\text{II}}_{22}$  are finite as $y\rightarrow 0$.
\item 
$\mathcal{G}^{\text{P}_\text{II}}_{17}$ and $\mathcal{G}^{\text{P}_\text{II}}_{18}$ are finite as $x\rightarrow -y$ and $x\rightarrow 0$. 
\item 
$\mathcal{G}^{\text{P}_\text{II}}_{19}$, $\mathcal{G}^{\text{P}_\text{II}}_{20}$, $\mathcal{G}^{\text{P}_\text{II}}_{21}$,  $\mathcal{G}^{\text{P}_\text{II}}_{23}$ and $\mathcal{G}^{\text{P}_\text{II}}_{24}$ are finite as $x\rightarrow -y$. 
\end{itemize}
This fixes all of our integrals in terms of the overall normalizing integral $\mathcal{G}^{\text{P}_\text{II}}_1$, which can be readily determined by direct integration: 
\begin{eqnarray}
\mathcal{G}^{\text{P}_\text{II}}_{1} = -1 -2\zeta_2 \epsilon^2+2\zeta_3 \epsilon^3- 9\zeta_4 \epsilon^4 + \mathcal{O}(\epsilon^5).  
\end{eqnarray}
We note that our normalizing integral is the same as that of the $\text{P}_\text{I}$ topology.
In obtaining the boundary conditions we made use of the {\tt{PolyLogTools}}~\cite{Duhr:2019tlz} package and it's interface to {\tt{GiNaC}}~\cite{Naterop:2019xaf} to write combinations of boundary GPLs as simple functions of $\zeta_i$. 
We present results for these integrals in both regions of interest in Appendix~\ref{appendix:PII}, which we obtained using {\tt{HandyG}}~\cite{Naterop:2019xaf} to numerically evaluate the GPLs. We have checked our results against the numerical package {\tt{AMFlow}}~\cite{Liu:2022chg}, finding perfect agreement.  

\subsection{Planar topology P$_\text{III}$}
\label{sec:W3PA}
%\subsubsection{Setup}
\begin{figure}[H]
	\centering
	\includegraphics[scale=1.5]{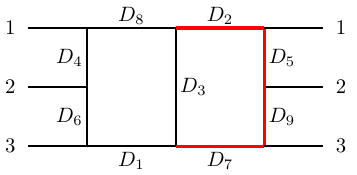}
	\caption{The auxiliary topology diagram for P$_\text{III}$ as defined in Eq.~\ref{eq:topo_PIII}. Propagators marked in red are massive.}
	\label{fig:aux_PIII}
\end{figure}
\begin{figure}[H]
	\centering
	\includegraphics[scale=0.72]{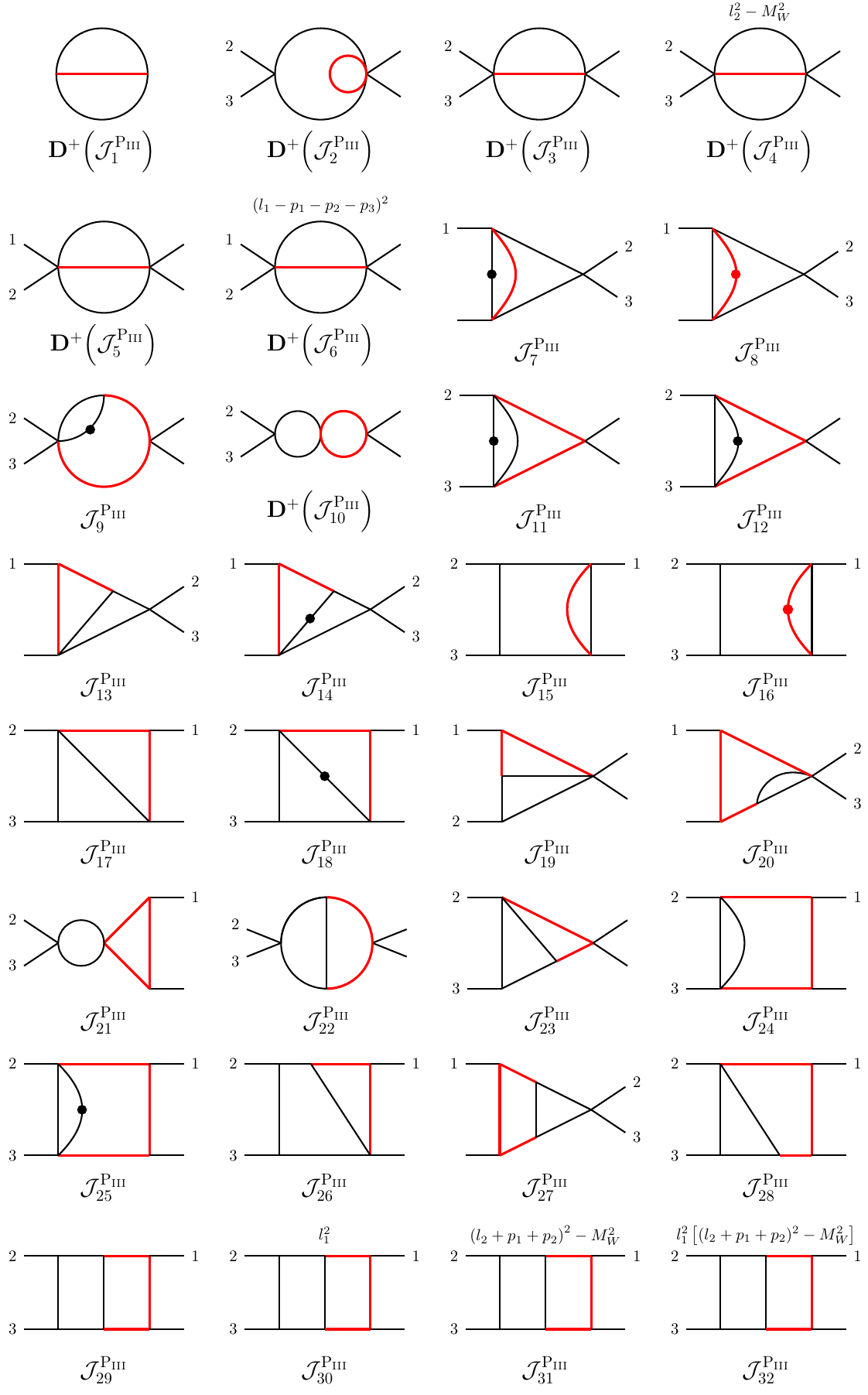}
	\caption{Master integrals for topology P$_\text{III}$. Dotted propagators have higher power of the respective denominator, while red propagators are massive.  }
	\label{fig:MI_PIII}
\end{figure}

The final, and most complicated, planar topology arises from Feynman diagrams which contain three massive vector bosons. 
The contributions to the form factors ${F}_i$ arising from the planar topology $P_\text{III}$ can be written in terms of the following scalar loop integral
\begin{eqnarray}
	I^{\text{P}_{\text{III} }}_{a_1\dots a_9} = \left(\frac{M_W^{2\epsilon}}{\Gamma({\epsilon})}\right)^2\int \frac{d^d \ell_1}{(2\pi)^d}\frac{d^d \ell_2}{(2\pi)^d} 
	\frac{1}{D_1^{a_1}D_2^{a_2}D_3^{a_3}D_4^{a_4}D_5^{a_5}D_6^{a_6}D_7^{a_7}D_8^{a_8}D_9^{a_9}}.
\end{eqnarray}
%where the overall normalization is defined as 
%\begin{eqnarray}
%\mathcal{C}(\epsilon,M_W^2)  = \left(\frac{M_W^{2\epsilon}}{\Gamma({\epsilon})}\right)^2.
%\end{eqnarray}
The propagators entering the loop are defined as follows:  
\begin{align}
	\label{eq:topo_PIII}
	&D_{1}= (\ell_1 - p_1-p_2-p_3)^2,
	&& D_{2}=\ell_2^2- M_W^2, 
	&& D_{3}=(\ell_1 + \ell_2)^2,\nonumber\\
	& D_{4}= (\ell_1  - p_1)^2,
	&&D_{5}= (\ell_2 + p_1)^2-M_W^2, 
	&&D_{6}= (\ell_1 - p_1 - p_2)^2,\nonumber\\
	&D_{7}= (\ell_2 + p_1 + p_2+p_3)^2- M_W^2
	&& D_{8}=\ell_1^2, 
	&&D_{9}= (\ell_2 + p_1 + p_2)^2-M_W^2. 
\end{align}
These propagators form an auxiliary topology which is shown in Fig.~\ref{fig:aux_PIII}. We note that this auxiliary topology is equivalent (up-to re-labeling) as that appearing 
in  P$_\text{II}$.
For the physical scattering process, propagators $D_8$ and $D_9$ do not appear with positive powers ($a_8 < 0$ and $a_9 < 0$). 
We have constructed the topology in the same notation as topology P$_\text{II}$, which minimizes the number of unique MIs which must be computed, however this means that in order to correctly assign the  physical momenta $k_i$ to the topology momenta $p_i$, mappings such as $\{p_1,p_2,p_3\} \rightarrow \{k_3,k_1,k_2\}$ must occur. For physical scattering processes we must therefore evaluate the auxiliary topology with
 $p_1\cdot p_2 < 0$ and $p_2 \cdot p_3 > 0$. 
The set of scalar integrals entering the form factors from this topology can be reduced to 32 MI's, (for which we used {\tt{Kira}}~\cite{Klappert:2020nbg}).
In order to obtain a canonical basis we begin with the following set of integrals: 
\begin{align}
	&\mathcal{J}_{1} = {\bf{D}}^-(I^{\text{P}_{\text{III}}}_{111000000}),\quad
	&&\mathcal{J}_{2} = {\bf{D}}^-(I^{\text{P}_{\text{III}}}_{110100000}),\quad
	&&\mathcal{J}_{3} = {\bf{D}}^-(I^{\text{P}_{\text{III}}}_{101010000}),\nonumber\\
	&\mathcal{J}_{4} = {\bf{D}}^-(I^{\text{P}_{\text{III}}}_{1(-1)1010000}),\quad
	&&\mathcal{J}_{5} = {\bf{D}}^-(I^{\text{P}_{\text{III}}}_{011001000}),\quad
	&&\mathcal{J}_{6} = {\bf{D}}^-(I^{\text{P}_{\text{III}}}_{(-1)11001000}),\nonumber\\
	&\mathcal{J}_{7} = I^{\text{P}_{\text{III}}}_{112100000},\quad
	&&\mathcal{J}_{8} = I^{\text{P}_{\text{III}}}_{121100000},\quad
	&&\mathcal{J}_{9} = I^{\text{P}_{\text{III}}}_{102010100},\nonumber\\
	&\mathcal{J}_{10} = {\bf{D}}^-(I^{\text{P}_{\text{III}}}_{100110100}),\quad
	&&\mathcal{J}_{11} = I^{\text{P}_{\text{III}}}_{002011100},\quad
	&&\mathcal{J}_{12} = I^{\text{P}_{\text{III}}}_{001021100},\nonumber\\
	&\mathcal{J}_{13} = I^{\text{P}_{\text{III}}}_{111110000},\quad
	&&\mathcal{J}_{14} = I^{\text{P}_{\text{III}}}_{112110000},\quad
	&&\mathcal{J}_{15} = I^{\text{P}_{\text{III}}}_{111101000},\nonumber\\
	&\mathcal{J}_{16} = I^{\text{P}_{\text{III}}}_{121101000},\quad
	&&\mathcal{J}_{17} = I^{\text{P}_{\text{III}}}_{111011000},\quad
	&&\mathcal{J}_{18} = I^{\text{P}_{\text{III}}}_{112011000},\nonumber\\
	&\mathcal{J}_{19} = I^{\text{P}_{\text{III}}}_{011111000},\quad
	&&\mathcal{J}_{20} = I^{\text{P}_{\text{III}}}_{111010100},\quad
	&&\mathcal{J}_{21} = I^{\text{P}_{\text{III}}}_{110110100},\nonumber\\
	&\mathcal{J}_{22} = I^{\text{P}_{\text{III}}}_{101110100},\quad
	&&\mathcal{J}_{23} = I^{\text{P}_{\text{III}}}_{101011100},\quad
	&&\mathcal{J}_{24} = I^{\text{P}_{\text{III}}}_{011011100},\nonumber\\
	&\mathcal{J}_{25} = I^{\text{P}_{\text{III}}}_{012011100},\quad
	&&\mathcal{J}_{26} = I^{\text{P}_{\text{III}}}_{111111000},\quad
	&&\mathcal{J}_{27} = I^{\text{P}_{\text{III}}}_{111110100},\nonumber\\
	&\mathcal{J}_{28} = I^{\text{P}_{\text{III}}}_{111011100},\quad
	&&\mathcal{J}_{29} = I^{\text{P}_{\text{III}}}_{111111100},\quad
	&&\mathcal{J}_{30} = I^{\text{P}_{\text{III}}}_{1111111(-1)0},\nonumber\\
	&\mathcal{J}_{31} = I^{\text{P}_{\text{III}}}_{11111110(-1)},\quad
	&&\mathcal{J}_{32} = I^{\text{P}_{\text{III}}}_{1111111(-1)(-1)},\quad
\end{align}
these integrals are presented in Fig.~\ref{fig:MI_PIII}. Here the dimension shifting operator is defined as
\begin{eqnarray}
	{\bf{D}}^-(I^{\text{P}_{\text{III}}}_{a_1\dots a_9})&=&\bigg((a_1 \mathbf{1}^+ + a_4  \mathbf{4}^++ a_6  \mathbf{6}^++ a_8 \mathbf{8}^+)(a_2 \mathbf{2}^+ + a_5 \mathbf{5}^+ a_7 \mathbf{7}^+ + a_9 \mathbf{9}^+) 
	\nonumber\\&& +\; a_3 \mathbf{3}^+ \sum_{j \ne 3} \,a_j {\mathbf{j}^+} \bigg)I^{\text{P}_{\text{III} }}_{a_1\dots a_9}.
\end{eqnarray}
We then utilize the same strategy as the previous sections by introducing the following finite and dimensionless integrals
\begin{eqnarray}
\mathcal{F}_{X}=\epsilon^2 \, (M_W^2)^{\alpha} \mathcal{J}_{X},
\end{eqnarray}
where $\alpha$ is defined for $\mathcal{J}_{X} =f(I^{\text{P}_{\text{III}}}_{a_1\dots a_9})$ as $(\sum a_i)-4$. 
We can introduce the  resulting basis: 
\begingroup
\allowdisplaybreaks
\begin{align}
	&\mathcal{G}_{1}^{\text{P}_\text{III}}=\mathcal{F}_{1},
	&&\mathcal{G}_{2}^{\text{P}_\text{III}}=y \mathcal{F}_{2},
	&&\mathcal{G}_{3}^{\text{P}_\text{III}}=(1+y) \mathcal{F}_{3},\nonumber\\[5pt]
	&\mathcal{G}_{4}^{\text{P}_\text{III}}=\mathcal{F}_{3}+\mathcal{F}_{4},
	&&\mathcal{G}_{5}^{\text{P}_\text{III}}=(1+x) \mathcal{F}_{5},
	&&\mathcal{G}_{6}^{\text{P}_\text{III}}=\mathcal{F}_{6},\nonumber\\[5pt]
	&\mathcal{G}_{7}^{\text{P}_\text{III}}= \epsilon y \mathcal{F}_{7},
	&&\mathcal{G}_{8}^{\text{P}_\text{III}}=\epsilon y  \mathcal{F}_{8},\nonumber\\[5pt]
	&\mathcal{G}_{9}^{\text{P}_\text{III}}=\mathrlap{\frac{1}{4 \sqrt{r_1}}\left( y^2 \mathcal{F}_{3}-2\mathcal{F}_{4}-y (\mathcal{F}_{1}-3 \mathcal{F}_{3}-4\mathcal{F}_{9}+8 \epsilon  \mathcal{F}_{9})\right),}
	&&&&\mathcal{G}_{10}^{\text{P}_\text{III}}=y\sqrt{r_1}\mathcal{F}_{10},\nonumber\\[5pt]
	&\mathcal{G}_{11}^{\text{P}_\text{III}}=\epsilon  y \mathcal{F}_{11},
	&&\mathcal{G}_{12}^{\text{P}_\text{III}}=\frac{\sqrt{r_1}}{2} (\epsilon  \mathcal{F}_{11}+2 (2 \epsilon -1)	\mathcal{F}_{12}),
	&&\mathcal{G}_{13}^{\text{P}_\text{III}}= \epsilon ^2 y \mathcal{F}_{13},\nonumber\\[5pt]
	&\mathcal{G}_{14}^{\text{P}_\text{III}}=\epsilon y \mathcal{F}_{14},
	&&\mathcal{G}_{15}^{\text{P}_\text{III}}=\epsilon y ((2 \epsilon-1 )\mathcal{F}_{15}+\mathcal{F}_{16}),
	&&\mathcal{G}_{16}^{\text{P}_\text{III}}= \epsilon x y \mathcal{F}_{16},\nonumber\\[5pt]
	&\mathcal{G}_{17}^{\text{P}_\text{III}}=\epsilon ^2(x+y) \mathcal{F}_{17},
	&&\mathcal{G}_{18}^{\text{P}_\text{III}}=\epsilon(x+y+x y) \mathcal{F}_{18},
	&&\mathcal{G}_{19}^{\text{P}_\text{III}}= \epsilon ^2 x \mathcal{F}_{19},\nonumber\\[5pt]
	&\mathcal{G}_{20}^{\text{P}_\text{III}}= \epsilon  (2 \epsilon-1 )y\mathcal{F}_{20},
	&&\mathcal{G}_{21}^{\text{P}_\text{III}}= \epsilon  (2 \epsilon -1)y\mathcal{F}_{21},
	&&\mathcal{G}_{22}^{\text{P}_\text{III}}= \epsilon  (2 \epsilon -1)y\mathcal{F}_{22},\nonumber\\[5pt]
	&\mathcal{G}_{23}^{\text{P}_\text{III}}= \epsilon ^2y \mathcal{F}_{23},
	&&\mathcal{G}_{24}^{\text{P}_\text{III}}= \epsilon  (2 \epsilon -1)y\mathcal{F}_{24},
	&&\mathcal{G}_{25}^{\text{P}_\text{III}}= \epsilon\sqrt{r_2}  \mathcal{F}_{25},\nonumber\\[5pt]
	&\mathcal{G}_{26}^{\text{P}_\text{III}}= \epsilon ^2x y\mathcal{F}_{26},
	&&\mathcal{G}_{27}^{\text{P}_\text{III}}=\epsilon^2y\sqrt{r_1}  \mathcal{F}_{27},
	&&\mathcal{G}_{28}^{\text{P}_\text{III}}= \epsilon ^2 \sqrt{r_3} \mathcal{F}_{28},\nonumber\\[5pt]
	&\mathcal{G}_{29}^{\text{P}_\text{III}}=\epsilon ^2y \sqrt{r_2}  \mathcal{F}_{29},
	&&\mathcal{G}_{30}^{\text{P}_\text{III}}=\epsilon ^2\sqrt{r_1}  (2x \mathcal{F}_{28}+x y \mathcal{F}_{29}+y \mathcal{F}_{30}),
	&&\mathcal{G}_{31}^{\text{P}_\text{III}}=\epsilon ^2y^2 (\mathcal{F}_{29}+\mathcal{F}_{31}),\nonumber\\[5pt]
	&\mathcal{G}_{32}^{\text{P}_\text{III}}= \mathrlap{-\frac{1}{2} \epsilon\bigg(2 x (1-2 \epsilon ) \mathcal{F}_{21}+y^2 \epsilon(-\mathcal{F}_{27}+x \mathcal{F}_{29}+\mathcal{F}_{30})}\nonumber\\
	&\mathrlap{\hspace{30pt}+2 y \epsilon(2 \mathcal{F}_{17}+\mathcal{F}_{22}+x \mathcal{F}_{28}+x\mathcal{F}_{29}-\mathcal{F}_{32})\bigg).}\nonumber\\
\end{align}
\endgroup
where we have introduced the two dimensionless ratios (which are the same as those used in the previous planar topologies)
\begin{eqnarray}
	x =  -\frac{2 p_1\cdot p_2}{M_W^2}   {\qquad\text{and}\qquad}    y = -\frac{2p_2\cdot p_3}{M_W^2}.
\end{eqnarray}
For physical kinematics, since both photons are radiated from the $W$ bosons, there is only one region to consider $x >0$ and $y <0$, and in the physical region $|x| \le |y|$. In the differential equations the following roots occur
\begin{align}
	\label{eq:rootsPIII}
	&r_1 = y (4+y),
	&&r_2= x y \; (4 y + x (4 + y)),
	&&r_3 = y\left(y + 2 x y + x^2 (4 + y)\right).
\end{align}
In principle it is possible to rationalize these roots simultaneously, however the resulting variables are polynomials of higher degree, which leads to a proliferation of letters entering the alphabet. In this paper we pursue an approach which rationalizes just $r_1$, and results in a smaller alphabet, but with more letters with non-rational components. We find that this leads to a more optimal final numerical implementation in our framework.
There are two sub-regions of interest, corresponding to whether or not $y$ is above the threshold at $y = -4$. 
We define Region A for $y \in [0,-4 ]$ and define
 \begin{eqnarray}
\label{eq:PIII_regionA}
  y = - \frac{4 z_A^2}{1 + z_A^2} \quad   {\rm{with}} \quad  0 \ge y \ge -4,
\end{eqnarray}
we take $z_A >0$ and therefore $x + x z_A^2 < 4 z_A^2$ ensures $  x \le -y$. 
Region B is defined as $y \in (-4,-\infty)$ and here we write
 \begin{eqnarray}
\label{eq:PIII_regionB} 	
y=-\frac{4}{1 - z_B^2}  \quad  {\rm{with}} \quad  y \le -4,
\end{eqnarray}
with $z_B \in (0,1)$.  In terms of these variables the differential equations take the following $dlog$-form
\begin{eqnarray}
d\mathcal{A} =  \sum_k \mathcal{M}_{k} \; d\log{\eta_k}
\end{eqnarray}
where the alphabet $\{\eta_k\}$ is made from twenty-two letters which may be inspected in the attached ancillary files. 
In region A, the remaining roots which enter the alphabet are a function of the polynomials
\begin{eqnarray}
	r_2=-\left(\frac{4z_A}{1+z_A^2} \right)^2 f^A_1 \quad {\rm{and}} \quad 
	r_3 = -\left(\frac{4z_A}{1+z_A^2} \right)^2 f^A_2,  
\end{eqnarray}
with
\begin{eqnarray}
f^A_1 = x-4z_A^2  \quad {\rm{and}} \quad f^A_2=x^2 - z_A^2 - 2 x z_A^2,
\end{eqnarray}
while in region B we have
\begin{eqnarray}
	r_2 = \left(\frac{4}{z_B^2-1}\right)^2f^B_1  \quad {\rm{and}} \quad  r_3=\left(\frac{4}{z_B^2-1}\right)^2 f_B^2,
\end{eqnarray}
with
\begin{eqnarray}
	f^B_1 = 4 + x z_B^2  \quad {\rm{and}} \quad  f^B_2= 1 + 2 x + x^2 z_B^2.
\end{eqnarray}
We have written our alphabet in a form which allows for clean extraction in the limit $x\rightarrow 0$. 
In terms of $(x,z_{A/B})$ variables, the following 25 MIs contain only rational letters and may be evaluated in terms of GPLs
\begin{eqnarray}
\mathcal{G}_i^{\text{P}_\text{III}}  \; {\rm{with}} \quad i \in \{1 - 23, 26 ,27\} .
\end{eqnarray}
Many of these integrals are recycled from the previous planar topologies, the boundary conditions for the new integrals can be quickly determined via appropriate finiteness of vanishing of the integral in the limits $x\rightarrow 0$, $y\rightarrow 0$ and $x\rightarrow -y$, of which the vanishing of the integrals as $y\rightarrow 0$ is particularly useful.  This procedure systematically relates MIs to one another (expanded through weight 5 to extract the pertinent limits), up to an overall normalizing integral which we again take to be $\mathcal{G}^{\text{P}_\text{III}}_1$. This integral can be readily computed via direct integration and in our normalization we have
\begin{eqnarray}
\mathcal{G}^{\text{P}_\text{III}}_{1} = -1 -  2\zeta_2 \epsilon^2 + 2\zeta_3 \epsilon^3 - 9 \zeta_{4} \epsilon^4  + \mathcal{O}(\epsilon^5),
\end{eqnarray}
which is the same as that used in the previous planar topologies. 
We evaluate GPLs using the {\tt{HandyG}}~\cite{Naterop:2019xaf} package, and have also made extensive use of the {\tt{PolyLogTools}}~\cite{Duhr:2019tlz} package and it's interface to {\tt{GiNaC}}~\cite{Naterop:2019xaf} to write combinations of boundary GPLs as a simple functions of $\zeta_i$ and $\log{2}$.

With the choice of variables in Eq.~\ref{eq:PIII_regionA} and Eq.~\ref{eq:PIII_regionB}, the remaining 7 integrals $\{\mathcal{G}_{24}^{\text{P}_\text{III}} ,\mathcal{G}_{25}^{\text{P}_\text{III}} ,\mathcal{G}_{28-32}^{\text{P}_\text{III}}\}$ contain non-rational letters and cannot be evaluated in terms of pure GPLs. As discussed in section~\ref{sec:Overview}, a convenient means to evaluate the solution is through Chen-iterated integrals. The pre-requisite for evaluation is the knowledge of the MIs at a boundary point, such that the MIs at any other point can be determined by path integration. For our calculation the remaining task is thus to determine appropriate boundary conditions to evaluate our Chen-iterated integral expressions. 

In order to solve for boundary points we note the following limit
\begin{eqnarray}
\lim_{x\rightarrow 0 } \left( \frac {\pd \mathcal{G}^{\text{P}_\text{III}}_j}{\pd z_A }  \right),
\end{eqnarray} \raggedbottom
is regular, and further is free of non-rational functions. We can therefore solve along the $x=0$ boundary entirely in terms of GPL solutions, with letters drawn from the set $\{0,\pm i, \pm {1}/{\sqrt{3}}, \pm{i}/{\sqrt{5}} \}$
in region A. The unknown boundary constants attributed to the 7 integrals with non-rational functions in the bulk space can be quickly determined through the vanishing of the integrals as $y\rightarrow 0$  or ($z_A \rightarrow 0$). The equivalent boundary GPLs in Region B can be obtained from these results via analytic continuation (using $z_A = i /z_B$) when $y <-4$. 
We have written our alphabet to allow smooth extraction of the $x\rightarrow 0$ limit, however in doing so we introduce the spurious letter $\eta_1 = 1- 2x$, which introduces a (spurious) singularity in region A. In order to avoid paths which cross the resulting branch cut, we also solve the following differential equations
\begin{eqnarray}
\lim_{x\rightarrow -y } \left( \frac {\pd \mathcal{G}^{\text{P}_\text{III}}_j}{\pd z_A }  \right),
\end{eqnarray} 
which again, is entirely made up of GPL solutions (with the same letters as $x\to0$ limit), and known boundary constants. For region B, the boundary solutions at $x=0$ are sufficient.

\begin{figure}[H]
	\centering
	\includegraphics[width=4.2cm]{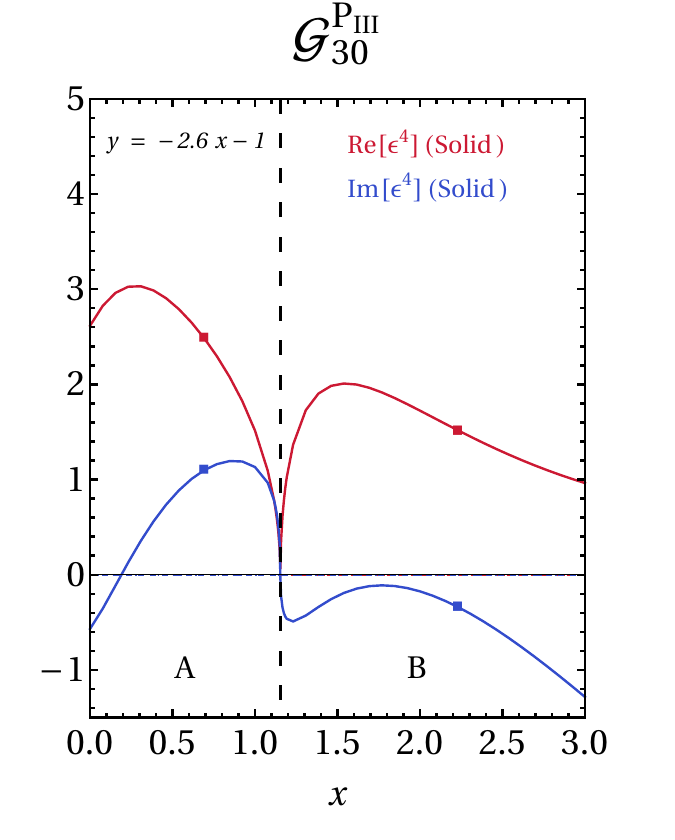}
	\includegraphics[width=4.2cm]{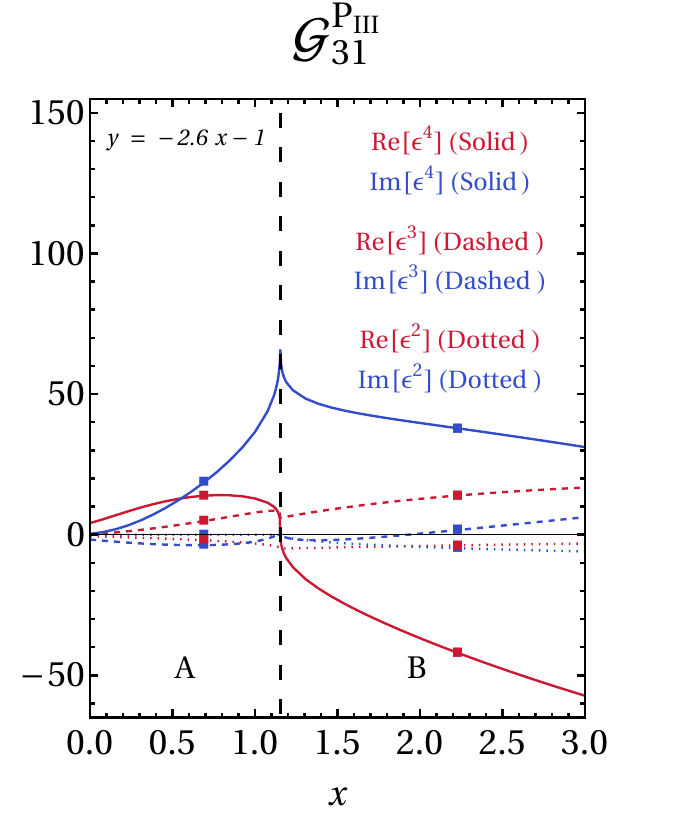}
	\includegraphics[width=4.2cm]{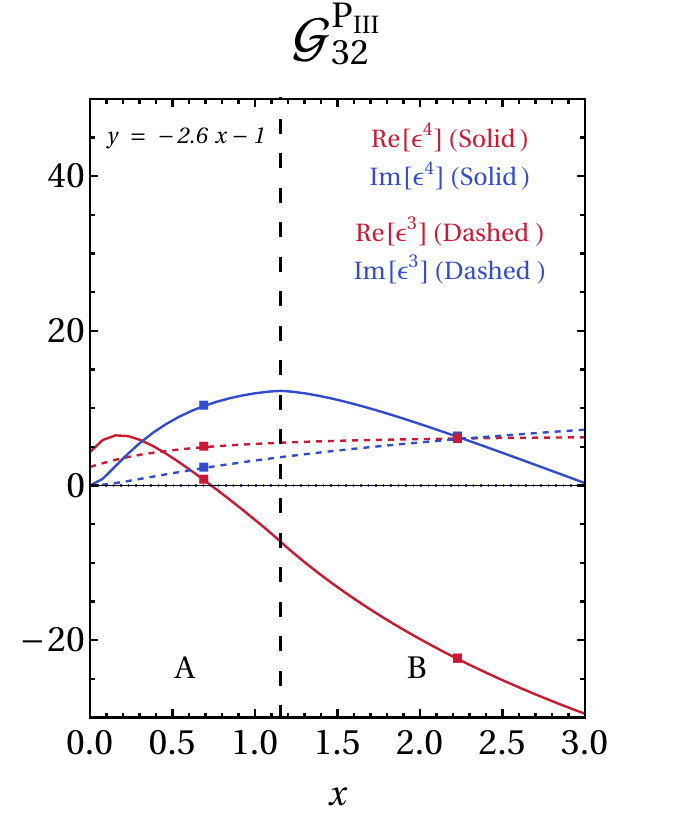}
	\caption{Coefficients of the $\epsilon^i$ terms (where $i=2,3,4$) in the expansion of the three 7 point integrals $\mathcal{G}^{\text{P}_{\text{III}}}_{30}$,	$\mathcal{G}^{\text{P}_{\text{III}}}_{31}$, and $\mathcal{G}^{\text{P}_{\text{III}}}_{32}$. Here $x \in [0,3]$ and $y$ is varied along the line $y = -2.6x-1$ between Region A and Region B, as defined in Eq.\ref{eq:PIII_regionA} and Eq.\ref{eq:PIII_regionB} respectively. Red and blue squares represent the real and imaginary part, respectively, of independent evaluation using {\tt{AMFlow}}.}
	\label{fig:PII_plot}
\end{figure}

With our boundary constants in hand we are therefore able to evaluate the remaining 7 integrals via path integration starting at the boundary points described above. 
We present results for all integrals evaluated at specific phase space points in Appendix~\ref{appendix:PIII}.
In order to demonstrate the flexibility of our code we present the evaluation of the most complicated 7 point integrals along the arbitrary line $y = -2.6x - 1$ where we show results 
from $x=1$ to $x=3$ in Fig.\ref{fig:PII_plot}. We made this choice to compactly demonstrate the lack of issues around $x=1/2$ and to evaluate the integrals above and below threshold. 
Also shown in the figure is a comparison to the result obtained using the numerical package {\tt{AMFlow}}~\cite{Liu:2022chg} with a point taken in each region. We have checked all of our integrals with {\tt{AMFlow}}\footnote{and {\tt{pySecDec}}  for selected MIs} and we find perfect agreement within to the reported uncertainty from the numerical codes. 

%=====================================================================

%================================== NON  PLANAR RESULTS 

%=====================================================================

\section{Non-planar topologies} 
\label{sec:nonplan}

\subsection{Non-planar topology N$_\text{I}$}

%\subsubsection{Setup}
\begin{figure}[H]
	\centering
	\includegraphics[scale=1.5]{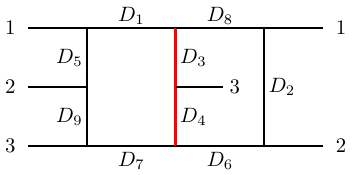}
	\caption{Auxiliary Topology diagram for N$_\text{I}$ as defined in Eq.~\ref{eq:topo_NI}. Propagators marked in red are massive.}
	\label{fig:aux_NI}
\end{figure}
The first non-planar topology which we encounter occurs when a single photon is radiated from the massive line and has the following propagator structure
\begin{eqnarray}
	I^{\text{N}_\text{I}}_{a_1\dots a_9} = \left( \frac{M_W^{2\epsilon}}{\Gamma(\epsilon)} \right)^2\int \frac{d^d \ell_1}{(2\pi)^d}\frac{d^d \ell_2}{(2\pi)^d} 
	\frac{1}{D_1^{a_1}D_2^{a_2}D_3^{a_3}D_4^{a_4}D_5^{a_5}D_6^{a_6}D_7^{a_7}D_8^{a_8}D_9^{a_9}},
\end{eqnarray}
with: 
\begin{align}
	\label{eq:topo_NI}
	&D_1=\ell_1^2,  
	&&D_{2}=(\ell_2+p_1)^2,
	&&D_{3}=(\ell_1 + \ell_2)^2-M_W^2, \nonumber\\  
	&D_{4}= (\ell_1 + \ell_2 - p_3)^2-M_W^2,
	&&D_{5}= (\ell_1 - p_1)^2,  
	&&D_{6}= (\ell_2 + p_1 + p_2)^2,\nonumber\\  
	&D_{7}= (\ell_1 - p_1 - p_2 - p_3)^2, 
	&&D_{8}= \ell_2^2,
	&&D_{9}= (\ell_1 - p_1 - p_2)^2.
\end{align}
These propagators form an auxiliary topology which is presented in Fig.~\ref{fig:aux_NI}. For physical scattering permutations of the external photons and gluons require us to evaluate the integral in all possible kinematic regions, e.g. with $p_1\cdot p_2$ and $p_2\cdot p_3$ having signs $(+-)$, $(-+)$ and $(--)$.  
In order to evaluate the Feynman diagrams occurring in this topology, 40 MI's are required. As a starting point we introduce the following 40 integral combinations (which are presented in Fig.~\ref{fig:MI_NI}): 
\begingroup
\allowdisplaybreaks
\begin{align}
	&\mathcal{J}_{1} = {\bf{D}}^-(I^{\text{N}_\text{I}}_{111000000}),\quad
	&&\mathcal{J}_{2} = {\bf{D}}^-(I^{\text{N}_\text{I}}_{110100000}),\quad
	&&\mathcal{J}_{3} = {\bf{D}}^-(I^{\text{N}_\text{I}}_{11(-1)100000}),\nonumber\\
	&\mathcal{J}_{4} = {\bf{D}}^-(I^{\text{N}_\text{I}}_{101001000}),\quad
	&&\mathcal{J}_{5} = {\bf{D}}^-(I^{\text{N}_\text{I}}_{1(-1)1001000}),\quad
	&&\mathcal{J}_{6} = {\bf{D}}^-(I^{\text{N}_\text{I}}_{000111000}),\nonumber\\
	&\mathcal{J}_{7} = {\bf{D}}^-(I^{\text{N}_\text{I}}_{(-1)00111000}),\quad
	&&\mathcal{J}_{8} = {\bf{D}}^-(I^{\text{N}_\text{I}}_{001010100}),\quad
	&&\mathcal{J}_{9} = I^{\text{N}_\text{I}}_{010210100},\nonumber\\
	&\mathcal{J}_{10} = I^{\text{N}_\text{I}}_{010220100},\quad
	&&\mathcal{J}_{11} = I^{\text{N}_\text{I}}_{111110000},\quad
	&&\mathcal{J}_{12} = I^{\text{N}_\text{I}}_{111101000},\nonumber\\
	&\mathcal{J}_{13} = I^{\text{N}_\text{I}}_{211101000},\quad
	&&\mathcal{J}_{14} = I^{\text{N}_\text{I}}_{111011000},\quad
	&&\mathcal{J}_{15} = I^{\text{N}_\text{I}}_{110111000},\nonumber\\
	&\mathcal{J}_{16} = I^{\text{N}_\text{I}}_{110211000},\quad
	&&\mathcal{J}_{17} = I^{\text{N}_\text{I}}_{101111000},\quad
	&&\mathcal{J}_{18} = I^{\text{N}_\text{I}}_{201111000},\nonumber\\
	&\mathcal{J}_{19} = I^{\text{N}_\text{I}}_{011111000},\quad
	&&\mathcal{J}_{20} = I^{\text{N}_\text{I}}_{111100100},\quad
	&&\mathcal{J}_{21} = I^{\text{N}_\text{I}}_{211100100},\nonumber\\
	&\mathcal{J}_{22} = I^{\text{N}_\text{I}}_{110110100},\quad
	&&\mathcal{J}_{23} = I^{\text{N}_\text{I}}_{110210100},\quad
	&&\mathcal{J}_{24} = I^{\text{N}_\text{I}}_{011110100},\nonumber\\
	&\mathcal{J}_{25} = I^{\text{N}_\text{I}}_{111001100},\quad
	&&\mathcal{J}_{26} = I^{\text{N}_\text{I}}_{112001100},\quad
	&&\mathcal{J}_{27} = I^{\text{N}_\text{I}}_{110101100},\nonumber\\
	&\mathcal{J}_{28} = I^{\text{N}_\text{I}}_{101101100},\quad
	&&\mathcal{J}_{29} = I^{\text{N}_\text{I}}_{101011100},\quad
	&&\mathcal{J}_{30} = I^{\text{N}_\text{I}}_{102011100},\nonumber\\
	&\mathcal{J}_{31} = I^{\text{N}_\text{I}}_{011011100},\quad
	&&\mathcal{J}_{32} = I^{\text{N}_\text{I}}_{111111000},\quad
	&&\mathcal{J}_{33} = I^{\text{N}_\text{I}}_{111110100},\nonumber\\
	&\mathcal{J}_{34} = I^{\text{N}_\text{I}}_{111101100},\quad
	&&\mathcal{J}_{35} = I^{\text{N}_\text{I}}_{101111100},\quad
	&&\mathcal{J}_{36} = I^{\text{N}_\text{I}}_{011111100},\nonumber\\
	&\mathcal{J}_{37} = I^{\text{N}_\text{I}}_{111111100},\quad
	&&\mathcal{J}_{38} = I^{\text{N}_\text{I}}_{1111111(-1)0},\quad
	&&\mathcal{J}_{39} = I^{\text{N}_\text{I}}_{11111110(-1)},\nonumber\\
	&\mathcal{J}_{40} = I^{\text{N}_\text{I}}_{1111111(-1)(-1)}.\quad
\end{align}
Here the dimension shifting operator is defined as
\begin{eqnarray}
	{\bf{D}}^-(I^{\text{N}_{\text{I}}}_{a_1\dots a_9})&=&\bigg((a_1 \mathbf{1}^+ + a_5  \mathbf{5}^++ a_7  \mathbf{7}^++ a_9 \mathbf{9}^+)(a_2 \mathbf{2}^+ + a_6 \mathbf{6}^+ + a_8 \mathbf{8}^+) 
	\nonumber\\&&+ (a_3 \mathbf{3}^++ a_4 \mathbf{4}^+) (\sum_{j \ne 3,4} \,a_j {\mathbf{j}^+}) \bigg)I^{\text{N}_{\text{I}}}_{a_1\dots a_9}.
\end{eqnarray}
As before we begin by rescaling the integrals to become dimensionless
\begin{eqnarray}
	\mathcal{F}_{X}= \epsilon^2 \; M_W^{2\alpha} \mathcal{J}_{X},
\end{eqnarray}
where again, $\alpha$ is defined for $ \mathcal{J}_{X}=  f(I^{\text{N}_\text{I}}_{a_1\dots a_9})$ as $(\sum {a_i}) -4$. Introducing the following dimensionless ratios
\begin{eqnarray}
	\label{eq:NII_ratios}
	x = \frac{2 p_1 \cdot p_2}{M_W^2},  \quad {\rm{and}}  \quad  y = \frac{2 p_1 \cdot p_3}{M_W^2}.
\end{eqnarray}
We can write a canonical basis as follows: 
\begingroup
\begin{align}
	&\mathcal{G}^{\text{N}_{\text{I}}}_{1}=\mathcal{F}_{1},
	&&\mathcal{G}^{\text{N}_{\text{I}}}_{2}=(1-y) \mathcal{F}_{2},
	&&\mathcal{G}^{\text{N}_{\text{I}}}_{3}=\mathcal{F}_{2}+\mathcal{F}_{3},\nonumber\\[5pt]
	&\mathcal{G}^{\text{N}_{\text{I}}}_{4}=(1-x) \mathcal{F}_{4},
	&&\mathcal{G}^{\text{N}_{\text{I}}}_{5}=\mathcal{F}_{5},
	&&\mathcal{G}^{\text{N}_{\text{I}}}_{6}=(1+x+y) \mathcal{F}_{6},\nonumber\\[5pt]
	&\mathcal{G}^{\text{N}_{\text{I}}}_{7}=\mathcal{F}_{7},
	&&\mathcal{G}^{\text{N}_{\text{I}}}_{8}=(x+y) \mathcal{F}_{8},
	&&\mathcal{G}^{\text{N}_{\text{I}}}_{9}=\epsilon(x+y)   \mathcal{F}_{9},\nonumber\\[5pt]
	&\mathcal{G}^{\text{N}_{\text{I}}}_{10}=(x+y) ((x+y)\mathcal{F}_{10}-\epsilon  \mathcal{F}_{1}),
	&&\mathcal{G}^{\text{N}_{\text{I}}}_{11}= \epsilon ^2 y \mathcal{F}_{11},
	&&\mathcal{G}^{\text{N}_{\text{I}}}_{12}=\epsilon ^2(x+y)  \mathcal{F}_{12},\nonumber\\[5pt]
	&\mathcal{G}^{\text{N}_{\text{I}}}_{13}=\epsilon (xy-x+y)  \mathcal{F}_{13},
	&&\mathcal{G}^{\text{N}_{\text{I}}}_{14}=x \epsilon ^2 \mathcal{F}_{14},
	&&\mathcal{G}^{\text{N}_{\text{I}}}_{15}=x \epsilon ^2 \mathcal{F}_{15},\nonumber\\[5pt]
	&\mathcal{G}^{\text{N}_{\text{I}}}_{16}=\epsilon \left(xy-x+y^2\right)\mathcal{F}_{16},
	&&\mathcal{G}^{\text{N}_{\text{I}}}_{17}=y \epsilon ^2 \mathcal{F}_{17},\nonumber\\[5pt]
	&\mathcal{G}^{\text{N}_{\text{I}}}_{18}=\mathrlap{\left(\frac{1}{2 x}-\frac{x (1+x+y)}{2 (x+y)^2}\right) \mathcal{F}_{1}+\frac{\left(x^2-y+x y \right) }{2 (x+y)^2}((1+x+y)\mathcal{F}_{6}-6 \mathcal{F}_{7})}\nonumber\\[5pt]
	&\hspace{30pt}\mathrlap{+\frac{y}{2 x (x+y)}\left((x-1)\mathcal{F}_{4}+6 \mathcal{F}_{5}+2(1+\epsilon ) x \left(x^2-y+x y \right) \mathcal{F}_{18}\right),}\nonumber \\[5pt]
	&\mathcal{G}^{\text{N}_{\text{I}}}_{19}= \epsilon ^2(x+y) \mathcal{F}_{19},
	&&\mathcal{G}^{\text{N}_{\text{I}}}_{20}= \epsilon ^2 x \mathcal{F}_{20},\nonumber\\[5pt]
	&\mathcal{G}^{\text{N}_{\text{I}}}_{21}=\mathrlap{\frac{x}{2 y (x+y)^2}((x (1+y)+y (2+y)) \mathcal{F}_{1}-y (1+x+y)((1+x+y) \mathcal{F}_{6}-6 \mathcal{F}_{7}))}\nonumber\\[5pt]
	&\hspace{30pt}\mathrlap{-\frac{x}{2 y(x+y)}\left((1+2 y) \mathcal{F}_{2}+3 \mathcal{F}_{3}-2 y \left(x(y-1)+y^2\right) (1+\epsilon ) \mathcal{F}_{21}\right),}\nonumber \\[5pt]
	&\mathcal{G}^{\text{N}_{\text{I}}}_{22}=(x+y) ((2 \epsilon^2-\epsilon  )\mathcal{F}_{22}+\mathcal{F}_{23}),
	&&\mathcal{G}^{\text{N}_{\text{I}}}_{23}=\epsilon y (x+y)   \mathcal{F}_{23},
	&&\mathcal{G}^{\text{N}_{\text{I}}}_{24}=\epsilon ^2 (x+y)  \mathcal{F}_{24},\nonumber\\[5pt]
	&\mathcal{G}^{\text{N}_{\text{I}}}_{25}= \epsilon ^2y \mathcal{F}_{25},
	&&\mathcal{G}^{\text{N}_{\text{I}}}_{26}=\epsilon \left(x^2-y+x y\right) \mathcal{F}_{26},
	&&\mathcal{G}^{\text{N}_{\text{I}}}_{27}=\epsilon ^2 y \mathcal{F}_{27},\nonumber\\[5pt]
	&\mathcal{G}^{\text{N}_{\text{I}}}_{29}= ((2 \epsilon^2-\epsilon  )(x+y)\mathcal{F}_{29}+\mathcal{F}_{30}),
	&&\mathcal{G}^{\text{N}_{\text{I}}}_{28}= \epsilon ^2 x \mathcal{F}_{28},
	&&\mathcal{G}^{\text{N}_{\text{I}}}_{30}=\epsilon x (x+y)   \mathcal{F}_{30},\nonumber\\[5pt]
	&\mathcal{G}^{\text{N}_{\text{I}}}_{31}=\epsilon ^2 (x+y)  \mathcal{F}_{31},
	&&\mathcal{G}^{\text{N}_{\text{I}}}_{32}=\epsilon ^2\sqrt{r_1}\mathcal{F}_{32},
	&&\mathcal{G}^{\text{N}_{\text{I}}}_{33}=\epsilon ^2 y (x+y)  \mathcal{F}_{33},\nonumber\\[5pt]
	&\mathcal{G}^{\text{N}_{\text{I}}}_{34}=\epsilon ^2\sqrt{r_1}\mathcal{F}_{34},
	&&\mathcal{G}^{\text{N}_{\text{I}}}_{35}=\epsilon ^2x (x+y)  \mathcal{F}_{35},
	&&\mathcal{G}^{\text{N}_{\text{I}}}_{36}=\epsilon ^2\sqrt{r_2}(x+y) \mathcal{F}_{36}.\nonumber\\[5pt]
\end{align}
Here, the 7-propagators family is defined as:
\begin{align}
	&\mathcal{G}^{\text{N}_{\text{I}}}_{37}=\epsilon ^2\frac{(x+y)}{2}   \left(x^2\mathcal{F}_{37}+x (\mathcal{F}_{32}+\mathcal{F}_{34}-2\mathcal{F}_{37}+y \mathcal{F}_{37}-\mathcal{F}_{39})+y(-2\mathcal{F}_{37}+\mathcal{F}_{39})\right),\nonumber\\[5pt]
	&\mathcal{G}^{\text{N}_{\text{I}}}_{38}=\frac{x}{2y}\bigg(\mathcal{F}_{1}-\mathcal{F}_{2}-\mathcal{F}_{3}-2\mathcal{F}_{7}-2\epsilon (x+y)   \mathcal{F}_{9}+6\epsilon ^2 x \mathcal{F}_{15}-2\epsilon \left(x (-1+y)+y^2\right)  \mathcal{F}_{16}\nonumber\\[5pt]
	&\hspace{30pt}+2\epsilon y (x+y) \mathcal{F}_{23}+2 \epsilon^2  y\mathcal{F}_{27}-2\epsilon ^2 (x+y)  \mathcal{F}_{31}\bigg)\nonumber\\[5pt]
	&\hspace{30pt}-\epsilon ^2\frac{x+y}{2} \bigg(x(\mathcal{F}_{32}-\mathcal{F}_{34})+(x+y) ((2+x) \mathcal{F}_{37}-2\mathcal{F}_{38}-\mathcal{F}_{39})\bigg),\nonumber\\[5pt]
	&\mathcal{G}^{\text{N}_{\text{I}}}_{39}= -\epsilon ^2\sqrt{r_2} \left(x^2\mathcal{F}_{37}+x (\mathcal{F}_{32}+\mathcal{F}_{34}+y\mathcal{F}_{37}-\mathcal{F}_{39})-y \mathcal{F}_{39}\right),\nonumber\\[5pt]
	&\mathcal{G}^{\text{N}_{\text{I}}}_{40}=-\epsilon ^2x (x+y)  \mathcal{F}_{37}+\epsilon^2(x+y)  \mathcal{F}_{40}+\frac{x}{16 y}\bigg(10 \mathcal{G}^{\text{N}_{\text{I}}}_{1}+2 \mathcal{G}^{\text{N}_{\text{I}}}_{2}-11\mathcal{G}^{\text{N}_{\text{I}}}_{3}+3 \mathcal{G}^{\text{N}_{\text{I}}}_{4}-10 \mathcal{G}^{\text{N}_{\text{I}}}_{5}\nonumber\\[5pt]
	&\hspace{30pt}-4\mathcal{G}^{\text{N}_{\text{I}}}_{6}-16 \mathcal{G}^{\text{N}_{\text{I}}}_{9}-12 \mathcal{G}^{\text{N}_{\text{I}}}_{12}-2\mathcal{G}^{\text{N}_{\text{I}}}_{13}+48 \mathcal{G}^{\text{N}_{\text{I}}}_{15}-16 \mathcal{G}^{\text{N}_{\text{I}}}_{16}-16\mathcal{G}^{\text{N}_{\text{I}}}_{17}+8 \mathcal{G}^{\text{N}_{\text{I}}}_{19}-12 \mathcal{G}^{\text{N}_{\text{I}}}_{20}\nonumber\\[5pt]
	&\hspace{30pt}-2\mathcal{G}^{\text{N}_{\text{I}}}_{21}+16 \mathcal{G}^{\text{N}_{\text{I}}}_{23}+8 \mathcal{G}^{\text{N}_{\text{I}}}_{25}+8\mathcal{G}^{\text{N}_{\text{I}}}_{27}+8 \mathcal{G}^{\text{N}_{\text{I}}}_{28}-16 \mathcal{G}^{\text{N}_{\text{I}}}_{31}\bigg).
\end{align}
\endgroup
The MIs above contain the roots of the following functions
\begin{align}
	\label{eq:roots_NI}
	& r_1 = xy(x+y),  &&r_2 = (4-x-y)(x+y),
\end{align}
and satisfy the usual canonical differential equations in $x$ and $y$. Given the simplicity of the non-rational terms which enter the differential equations, it is possible to simultaneously rationalize $r_1$ and $r_2$ to obtain purely GPL solutions. In all regions we find the initial mappings
\begin{eqnarray}
x\rightarrow \frac{u v}{1 + v} \quad {\rm{and}} \quad y \rightarrow \frac{u}{1 + v},
\label{eq:uvdef}
\end{eqnarray}
provides a convenient  set of variables to express the GPL solutions which only contain rational letters. 
In total there are 33 such integrals, $\mathcal{G}_{1-31}$, $\mathcal{G}_{33}$ and $\mathcal{G}_{35}$.
The alphabet for these integrals depends on the following letters
\begin{eqnarray}
\{ u ,\,  u \pm 1,\, v ,\, 1+v  ,\, 1-u v ,\, v-u  ,\,1 - u + v  ,\, 1 + v - uv ,\, 1 + 2 v - u v + v^2 \} .
\end{eqnarray}
Many of the integrals for this topology have been computed already in the previous sections, which allows for quick determination of the boundary constants for the GPL solutions, which includes all three propagator diagrams $\mathcal{G}_{1-8}$. 
The new integrals can be determined by requiring finiteness as $v\rightarrow 0$ $(\mathcal{G}_{12},\mathcal{G}_{16},\mathcal{G}_{18},\mathcal{G}_{21},\mathcal{G}_{22},\mathcal{G}_{26},\mathcal{G}_{30},\mathcal{G}_{33},\mathcal{G}_{35})$ or $u\rightarrow 0$ $(\mathcal{G}_{10})$. The remaining integrals can all be determined by noting that they vanish as $u\rightarrow 0$.

After fixing the GPL solutions in the $(u,v)$ variables we are left with seven integrals which contain roots in the initial $(x,y)$ variables.  As argued earlier these integrals can be rationalized, putting the entire solution in-terms of GPLs. Since the required variable change for the remaining integrals are sensitive to the sign of $x$ and $y$ we discuss each region separately.

%\subsubsection{Region  A - $x < 0$ and $y < 0$} 

We begin by considering the case where both $x$ and $y$ are less than zero, we introduce the following two variables
\begin{eqnarray}
u = \frac{4m_A^2}{m_A^2-1} \quad {\rm{and}} \quad v= \frac{1-m_A^2}{j_A^2},
\end{eqnarray}
then our requirements that $x <0$ and $y <0$ are satisfied by taking $j_A > 0$ and $m_A \in [0,1]$. 
%The alphabet for this region is 
%\begin{eqnarray}
%&&\{ j_A, \, m_A,  \, m_A \pm 1, 1 \pm i \sqrt{3} m_A  , \, 1 \pm \sqrt{5} m_A ,\, 1 + i j_A - m_A, \, \nonumber\\&& 1 - i j_A+  m_A , \,    1 - i j_A-   m_A , \,  1 + i j_A+  m_A , \,
%2m_A \pm ij_A ,\,   \nonumber\\ && 1 \pm  2 i\,  j_A m_A - m_A^2 ,\,  -j_A \pm \sqrt{-1 - 3 m_A^2} ,\, j_A \pm \sqrt{-1 + m_A^2}, \nonumber\\&& 1 - m_A^2  \pm i j_A \sqrt{1 + 3 m_A^2} \}.
%\end{eqnarray}
These variable changes rationalize the roots and result in a solution which be written in terms of GPLs.  The remaining boundary vectors for the seven unknown integrals can be quickly determined by noting their vanishing as $u \rightarrow 0$ $(m_A \rightarrow 0)$. 
%\subsubsection{Region  B - $x  > 0$ and $y < 0$} 

Next we consider the case where $x >0$ and $y < 0$, here the roots can be rationalized using the following variables: 
\begin{eqnarray}
u = \frac{4m_B^2}{1+m_B^2} \quad {\rm{and}} \quad v= -\frac{1+m_B^2}{j_B^2}.
\end{eqnarray}
Here $m_B >0$ and $j_B$ $\in [0,1]$ defines the region where $x$ is below the threshold at $4$, and the region $1 < j_B < \sqrt{1 + m_B^2}$ defines the region above threshold. 
The final region  $x  < 0$ and $y > 0$ can also be expressed in terms of Region B variables, here the domains are taken to be $m_B > 0$ and $j_B > 1+m_B^2$ for $y$ below the threshold at 4 and $\sqrt{1+m_B^2} < j_B < 1+m_B^2$ defining the region above $y >4$. 

To summarize, the kinematic regions for this topology are given by
\begin{align}
	\label{eq:NI_regionA}
	&{\rm{Region \; \; A}}: && (p_1\cdot p_2 < 0,\; p_1\cdot p_3 < 0) \qquad \to &&  x < 0,\; y < 0, \\[5pt]
	\label{eq:NI_regionB}
	&{\rm{Region \; \; B}}: && (p_1\cdot p_2 > 0,\; p_1\cdot p_3 < 0) \qquad \to &&  x > 0,\; y < 0,\; \lvert x \lvert>\lvert y \lvert,  \\[5pt]
	\label{eq:NI_regionC}
	&{\rm{Region \; \; C}}: && (p_1\cdot p_2 < 0,\; p_1\cdot p_3 > 0) \qquad \to &&  x < 0,\; y > 0 ,\; \lvert y \lvert>\lvert x \lvert.
\end{align}

We provide a table for all integrals for a test point in each region in Appendix~\ref{appendix:NI}. We have checked our results against {\tt{AMFlow}}~\cite{Liu:2022chg}, finding perfect agreement. 
\begingroup
\begin{figure}[H]
	\begin{center}
		\includegraphics[scale=0.75]{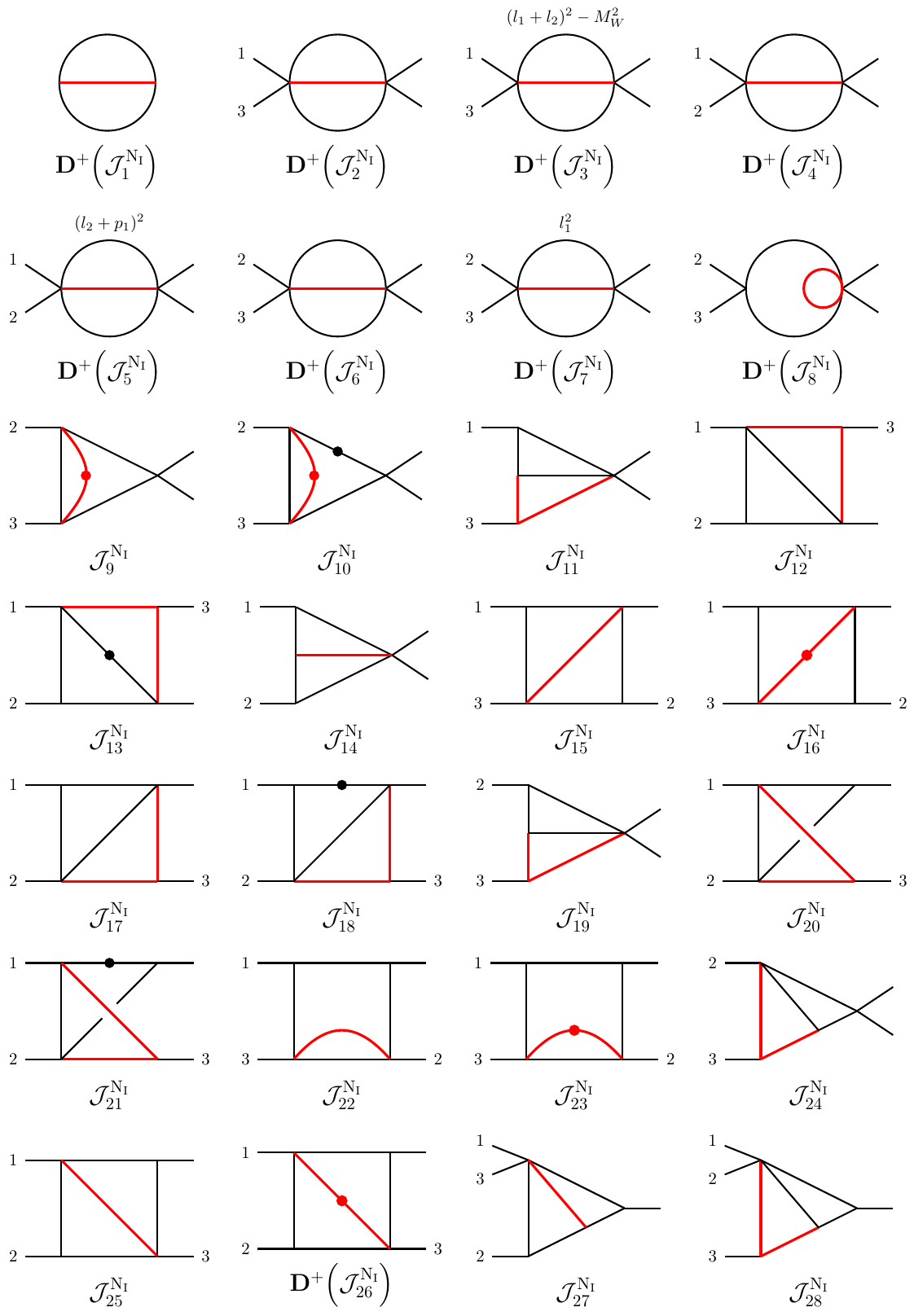}
	\end{center}
\end{figure}
\begin{figure}[H]
	\begin{center}
		\includegraphics[scale=0.75]{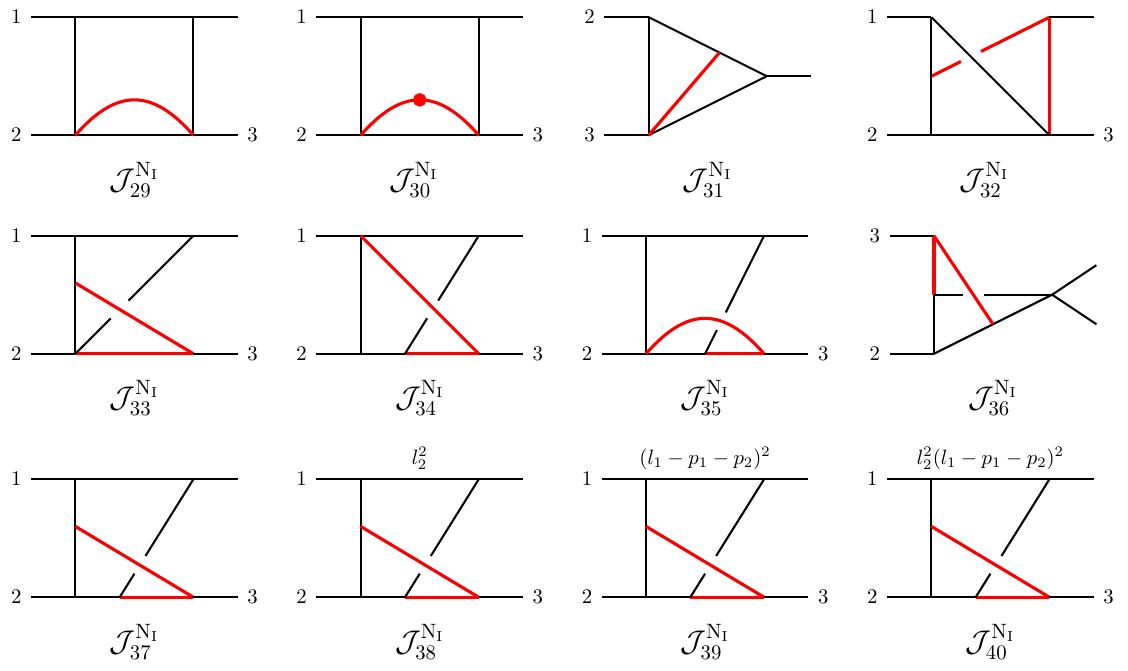}
		\caption{Master integrals for topology N$_\text{I}$. Dotted propagators have higher power of the respective denominator, while red propagators are massive}
		\label{fig:MI_NI}
	\end{center}
\end{figure}
\endgroup

\subsection{Non-planar topology N$_\text{II}$}

%\subsubsection{Setup}
\begin{figure}[H]
	\centering
	\includegraphics[scale=1.5]{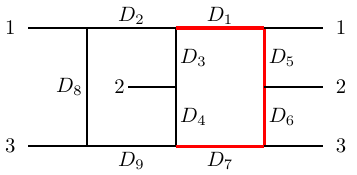}
	\caption{The auxiliary Topology diagram for N$_\text{II}$ as defined in Eq.~\ref{eq:topo_NII}. Propagators marked in red are massive.}
	\label{fig:aux_NII}
\end{figure}
The second non-planar topology, and most intricate one, corresponds to diagrams with three massive vector bosons and has the following propagator structure
\begin{eqnarray}
	I^{\text{N}_{\text{II}}}_{a_1\dots a_9} = \left( \frac{(-s^\epsilon)}{\Gamma(\epsilon)}\right)^2\int \frac{d^d \ell_1}{(2\pi)^d}\frac{d^d \ell_2}{(2\pi)^d} 
	\frac{1}{D_1^{a_1}D_2^{a_2}D_3^{a_3}D_4^{a_4}D_5^{a_5}D_6^{a_6}D_7^{a_7}D_8^{a_8}D_9^{a_9}},
\end{eqnarray}
\newpage\noindent with: 
\begin{align}
	\label{eq:topo_NII}
	&D_1=\ell_1^2- M_W^2,
	&&D_{2}=\ell_2^2,
	&&D_{3}=(\ell_1 + \ell_2)^2,\nonumber\\
	&D_{4}= (\ell_1 + \ell_2 - p_2)^2,
	&&D_{5}= (\ell_1 - p_1)^2-M_W^2, 
	&&D_{6}= (\ell_1 - p_1 - p_2)^2-M_W^2,\nonumber\\
	&D_{7}= (\ell_1 - p_1 - p_2 - p_3)^2-M_W^2, 
	&&D_{8}= (\ell_2 + p_1)^2,
	&&D_{9}= (\ell_2 + p_1 + p_3)^2.
\end{align}
These propagators form an auxiliary topology which is presented in Fig.~\ref{fig:aux_NII}.

After reduction the Feynman diagrams can be expressed in terms of 32 MI's. Our starting point is the following list of integrals (these are also presented in Fig.~\ref{fig:NII_MI}):
\begin{align}
	\mathcal{J}_1 &= {\bf{D}}^-(I^{\text{N}_{\text{II}}}_{111000000}), &
	\mathcal{J}_2 &= {\bf{D}}^-(I^{\text{N}_{\text{II}}}_{011001000}), &
	\mathcal{J}_3 &= {\bf{D}}^-(I^{\text{N}_{\text{II}}}_{(-1)11001000}),\nonumber\\
	\mathcal{J}_4 &= {\bf{D}}^-(I^{\text{N}_{\text{II}}}_{0 1 0 1 0 0 1 0 0}), & \mathcal{J}_5 &=  {\bf{D}}^-(I^{\text{N}_{\text{II}}}_{(-1) 1 0 1 0 0 1 0 0}), & \mathcal{J}_6 &= {\bf{D}}^-(I^{\text{N}_{\text{II}}}_ {0 0  1 0 0 0 1 1 0}),   \nonumber\\
	\mathcal{J}_7 &= {\bf{D}}^-(I^{\text{N}_{\text{II}}}_ {(-1) 0  1 0 0 0 1 1 0}), &
	\mathcal{J}_8 &= {\bf{D}}^- (I^{\text{N}_{\text{II}}}_{1 1 1 0 0 1 0 0 0}) ,& 
	\mathcal{J}_9&= I^{\text{N}_{\text{II}}}_{1 1  0  2  0  1  0  0 0}, \nonumber\\
	\mathcal{J}_{10}&= I^{\text{N}_{\text{II}}}_{2  1  0  2  0  1  0  0 0}   &
	\mathcal{J}_{11}&=I^{\text{N}_{\text{II}}}_ {1 1 1 1 0 1 0 0 0},&
	\mathcal{J}_{12}&=I^{\text{N}_{\text{II}}}_ {1 1 1 0 1 1 0 0 0},\nonumber\\
	\mathcal{J}_{13}&=I^{\text{N}_{\text{II}}}_ {1 1 1 1 0 0 1 0 0},  &
	\mathcal{J}_{14}&=I^{\text{N}_{\text{II}}}_ {1 1 0 1 0 1 1 0 0},&
	\mathcal{J}_{15}&=I^{\text{N}_{\text{II}}}_ {1 1 0 2 0 1 1 0 0},\nonumber\\
	\mathcal{J}_{16}&=I^{\text{N}_{\text{II}}}_ {0 1 1 1 0 1 1 0 0},&
	\mathcal{J}_{17}&=I^{\text{N}_{\text{II}}}_ {0 1 2 1 0 1 1 0 0},&
	\mathcal{J}_{18}&=I^{\text{N}_{\text{II}}}_ {1 1 1 1 0 0 0 1 0},\nonumber\\
	\mathcal{J}_{19}&=I^{\text{N}_{\text{II}}}_ {1 1 1 0 0 0 1 1 0},  &
	\mathcal{J}_{20}&=I^{\text{N}_{\text{II}}}_ {1 0 1 1 0 0 1 1 0},&
	\mathcal{J}_{21}&=I^{\text{N}_{\text{II}}}_ {1 0 2 1 0 0 1 1 0},\nonumber\\
	\mathcal{J}_{22}&=I^{\text{N}_{\text{II}}}_ {0 1 1 1 0 0 1 1 0},  &
	\mathcal{J}_{23}&=I^{\text{N}_{\text{II}}}_ {0 1 1 1 0 0 2 1 0},&
	\mathcal{J}_{24}&=I^{\text{N}_{\text{II}}}_ {1 0 1 0 0 1 1 1 0},\nonumber\\
	\mathcal{J}_{25}&=I^{\text{N}_{\text{II}}}_ {1 0 2 0 0 1 1 1 0},  &
	\mathcal{J}_{26}&= I^{\text{N}_{\text{II}}}_{1 1 1 1 0 1 1 0 0},&
	\mathcal{J}_{27}&= I^{\text{N}_{\text{II}}}_{1 1 1 1 0 1 0 1 0},\nonumber\\
	\mathcal{J}_{28}&= I^{\text{N}_{\text{II}}}_{1 1 1 1 0 0 1 1 0},  &
	\mathcal{J}_{29}&= I^{\text{N}_{\text{II}}}_{1 1 1 0 0 1 1 1 0},  &
	\mathcal{J}_{30}&= I^{\text{N}_{\text{II}}}_{1 1 1 1 0 1 1 1 0},\nonumber\\
	\mathcal{J}_{31}&= I^{\text{N}_{\text{II}}}_{1 1 1 1 (-1) 1 1 1 0},&
	\mathcal{J}_{32}&= I^{\text{N}_{\text{II}}}_{1 1 1 1 (-2) 1 1 1 0}.
\end{align}
Here the dimension shifting operator is defined as
\begin{eqnarray}
	{\bf{D}}^-(I^{\text{N}_{\text{II}}}_{a_1\dots a_9})&=&\bigg((a_1 \mathbf{1}^+ + a_5  \mathbf{5}^++ a_6  \mathbf{6}^++ a_7 \mathbf{7}^+)(a_2 \mathbf{2}^+ + a_8 \mathbf{8}^+ + a_9 \mathbf{9}^+) 
	\nonumber\\&&+ (a_3 \mathbf{3}^++ a_4 \mathbf{4}^+) (\sum_{j \ne 3,4} \,a_j {\mathbf{j}^+}) \bigg)I^{\text{N}_{\text{II}}}_{a_1\dots a_9}.
\end{eqnarray}
We proceed as in the earlier sections, first we render each integral dimensionless via the following scaling
\begin{eqnarray}
	\mathcal{F}_{X}= \epsilon^2 \; s^{\alpha} \mathcal{J}_{X}.
\end{eqnarray}
Where $\alpha$ is defined for $ \mathcal{J}_{X}=  f(I^{\text{N}_{\text{II}}}_{a_1\dots a_9})$ as $(\sum {a_i}) -4$. We then write a canonical basis in terms of our original integrals and the following dimensionless ratios
\begin{eqnarray}
	x = - \frac{p_1\cdot p_3 }{p_1\cdot p_2},  \quad {\rm{and}}  \quad  y = \frac{M_W^2}{2p_1\cdot p_2}.
\end{eqnarray}
Given the restriction that both photons must be radiated from the $W$ bosons all permutations of external photons and gluons result in $x > 0$ and $y > 0$.  Finally, in terms of these variables we construct the following canonical basis:
\newpage
\begingroup
\begin{align}
	&\mathcal{G}^{\text{N}_{\text{II}}}_1  = y\mathcal{F}_1, 
	&&\mathcal{G}^{\text{N}_{\text{II}}}_2 = \left(1 - y\right) \mathcal{F}_2, 
	&&\mathcal{G}^{\text{N}_{\text{II}}}_3 = \mathcal{F}_3 + \mathcal{F}_2, \nonumber\\[5pt]
	&\mathcal{G}^{\text{N}_{\text{II}}}_4 = \left(x + y\right) \mathcal{F}_4, 
	&&\mathcal{G}^{\text{N}_{\text{II}}}_5 = \mathcal{F}_5+y\mathcal{F}_4 , 
	&&\mathcal{G}^{\text{N}_{\text{II}}}_6 = \left(-1 + x - y\right) \mathcal{F}_6, \nonumber\\[5pt]
	&\mathcal{G}^{\text{N}_{\text{II}}}_7 = \mathcal{F}_7-\left(2-2x+y \right)\mathcal{F}_6 , 
	&&\mathcal{G}^{\text{N}_{\text{II}}}_8 = \sqrt{r_1}\left(y \mathcal{F}_8 + \mathcal{F}_2\right), &&\mathcal{G}^{\text{N}_{\text{II}}}_9 = \epsilon\,\mathcal{F}_9, \nonumber \\[5pt]
	&\mathcal{G}^{\text{N}_{\text{II}}}_{10} = \sqrt{r_1}\left(y \mathcal{F}_{10} + \frac{3\epsilon }{2}\mathcal{F}_9 \right), &&\mathcal{G}^{\text{N}_{\text{II}}}_{11} = \epsilon^2 \mathcal{F}_{11}, 
	&&\mathcal{G}^{\text{N}_{\text{II}}}_{12} = \epsilon\,(1 - 2 \epsilon) \mathcal{F}_{12}, \nonumber \\[5pt]
	&\mathcal{G}^{\text{N}_{\text{II}}}_{13} = \epsilon^2 x \;\mathcal{F}_{13}, 		
	&&\mathcal{G}^{\text{N}_{\text{II}}}_{14} = \epsilon(2 \epsilon -1) \,\mathcal{F}_{14}, 
	&&\mathcal{G}^{\text{N}_{\text{II}}}_{15} = \epsilon  \, \sqrt{r_2} \,\mathcal{F}_{15}, \nonumber \\[5pt]
	&\mathcal{G}^{\text{N}_{\text{II}}}_{16} =\epsilon^2 \,  (x-1)\;\mathcal{F}_{16},\nonumber\\[5pt]
	&\mathcal{G}^{\text{N}_{\text{II}}}_{17} = \mathrlap{
		(1+\epsilon)\frac{(1-x)(xy-x-y)}{x}\mathcal{F}_{17}-\frac{1}{4x^2}\bigg(2 (x+y-x^2 y)\mathcal{G}^{\text{N}_{\text{II}}}_{1}-2xy (-1+x)\mathcal{G}^{\text{N}_{\text{II}}}_{2}
	}\nonumber\\[5pt]
	&\hspace{30pt}\mathrlap{
		+3xy (x-1)\mathcal{G}^{\text{N}_{\text{II}}}_{3}+4 (x+y-x y) \mathcal{G}^{\text{N}_{\text{II}}}_{4}-6 (x+y-x y) \mathcal{G}^{\text{N}_{\text{II}}}_{5}\bigg),			
	} \nonumber \\[5pt]
	&\mathcal{G}^{\text{N}_{\text{II}}}_{18} = \epsilon^2 \mathcal{F}_{18}, 		
	&&\mathcal{G}^{\text{N}_{\text{II}}}_{19} = \epsilon^2 \,(1 - x)\; \mathcal{F}_{19} ,
	&&\mathcal{G}^{\text{N}_{\text{II}}}_{20} = \epsilon^2 x \mathcal{F}_{20},  \nonumber \\[5pt]
	&\mathcal{G}^{\text{N}_{\text{II}}}_{21} = \mathrlap{
	   (1+\epsilon )(1-x+x y)x \mathcal{F}_{21}+\frac{x}{4(1-x)}\bigg( 2(1+x (y-1)-2 y) \mathcal{G}_{1}} \nonumber\\[5pt]
	&\hspace{30pt}\mathrlap{
	+(x-1) (y-1) (2\mathcal{G}_{2}-3\mathcal{G}_{3})-8 y \mathcal{G}_{6}+6 y\mathcal{G}_{7}\bigg),
	}\nonumber\\[5pt]
	&\mathcal{G}^{\text{N}_{\text{II}}}_{22} = \epsilon^2\,\mathcal{F}_{22}, 
	&&\mathcal{G}^{\text{N}_{\text{II}}}_{23} = \epsilon\, (x - x^2 + y) \mathcal{F}_{23} ,
	&&\mathcal{G}^{\text{N}_{\text{II}}}_{24} =\epsilon(2\epsilon -1)\mathcal{F}_{24}, \nonumber\\[5pt]
	&\mathcal{G}^{\text{N}_{\text{II}}}_{25} =\epsilon\, \sqrt{r_3} \,\mathcal{F}_{25},
	&&\mathcal{G}^{\text{N}_{\text{II}}}_{26}=\epsilon^2 \, \sqrt{r_4}\,\mathcal{F}_{26},
	&&\mathcal{G}^{\text{N}_{\text{II}}}_{27}=  \epsilon^2\,\sqrt{r_1}\, \mathcal{F}_{27}, \nonumber\\[5pt]
	&\mathcal{G}^{\text{N}_{\text{II}}}_{28} = \epsilon^2\,\sqrt{r_5} \, \mathcal{F}_{28}, 
	&&\mathcal{G}^{\text{N}_{\text{II}}}_{29} = \epsilon^2 \,\sqrt{r_6}\, \mathcal{F}_{29} ,
\end{align}
\endgroup
the 7-propagator family is defined as follows:
\begin{align}
	\mathcal{G}^{\text{N}_{\text{II}}}_{30} = & -2 \epsilon^2 \bigg(2 (x-1) \mathcal{F}_{28} + (1 - x - 2 y + 4 x y) \mathcal{F}_{30} + \mathcal{F}_{31}\bigg),\nonumber\\[5pt]  
	\mathcal{G}^{\text{N}_{\text{II}}}_{31} = &-\epsilon^2 \,\frac{\sqrt{r_1}}{2} \bigg(2 (x-1) \mathcal{F}_{28} +(1 - x) \mathcal{F}_{30} + (1 - 2 x) \mathcal{F}_{31}\bigg),\nonumber\\[5pt]  
	\mathcal{G}^{\text{N}_{\text{II}}}_{32} = &\epsilon^2(y\mathcal{F}_{30}+\mathcal{F}_{31}+\mathcal{F}_{32})+\frac{1}{8x}(15\mathcal{G}^{\text{N}_{\text{II}}}_{1}+8\mathcal{G}^{\text{N}_{\text{II}}}_{2}-8\mathcal{G}^{\text{N}_{\text{II}}}_{3}+10\mathcal{G}^{\text{N}_{\text{II}}}_{4}-17\mathcal{G}^{\text{N}_{\text{II}}}_{5}\nonumber\\
	&+4\mathcal{G}^{\text{N}_{\text{II}}}_{6}-4\mathcal{G}^{\text{N}_{\text{II}}}_{7}-24\mathcal{G}^{\text{N}_{\text{II}}}_{13}-12\mathcal{G}^{\text{N}_{\text{II}}}_{16}-2\mathcal{G}^{\text{N}_{\text{II}}}_{17}+8\mathcal{G}^{\text{N}_{\text{II}}}_{18}-8\mathcal{G}^{\text{N}_{\text{II}}}_{19}+20\mathcal{G}^{\text{N}_{\text{II}}}_{20}\nonumber\\
	& +2\mathcal{G}^{\text{N}_{\text{II}}}_{21}-24\mathcal{G}^{\text{N}_{\text{II}}}_{22}-8\mathcal{G}^{\text{N}_{\text{II}}}_{23}).
\end{align}
Six non-rational roots appear in the integrals, defined via
\begin{align}
		\label{eq:rootsA}
		&r_1 = 1 - 4 y
		&&r_2 = x^2(1-4y)+2xy+y^2\nonumber\\
		&r_3= x^2(1-4 y)+x(6y-2)+(y-1)^2
		&&r_4=-x\left(4xy-x-4y\right)\nonumber\\
		&r_5=xy\left( x-1 \right)
		&&r_6=-(x-1)(x(4y-1)+1).
\end{align}
The basis integrals and resulting differential equations contain non-rational terms, several of which can be rationalized by an appropriate change of variables. 
In total there are 6 roots which appear for this topology, shown in Eq.~\ref{eq:rootsA} and not all of the roots can be simultaneously rationalized. 
One must therefore make a choice regarding which roots to rationalize to put the set of equations in a form most amenable for a solution in terms of Chen iterated integrals. 
Since it enters into many of the differential equations $\sqrt{r_1}$ is clearly of form which is desirable to rationalize.
Since we are interested in solving in the physical region $y > 0$ and the behavior of the MI's depends on whether one evaluates the integral above threshold $( s > 4 M_W^2 \implies y < 1/4$), or below  $( s \leq 4 M_W^2  \implies y \geq 1/4$), we divide the phase space into relevant regions and introduce the following variable transformations 
\begin{align}
	\label{eq:NII_regionA}
	& &&{\rm{Region \; \; A}}: && s  > 4 M_W^2 ,&&  x\in [0,1],   && y = \frac{z_A}{2}\left(1-\frac{z_A}{2}\right) && \\[5pt]
	\label{eq:NII_regionB}
	& &&{\rm{Region \; \; B}} : && s   \leq 4 M_W^2 ,&& x\in [0,1],  &&  y =  \frac{1-2z_B+z_B^2}{4z_B^2}. &&
\end{align}
With $z_{A} \in (0,1]$ and $z_{B} \in [0,1]$. 
After implementing these transformations the following integrals  $\mathcal{G}^{\text{N}_\text{III}}_{1-13}$, $\mathcal{G}^{\text{N}_\text{III}}_{16-23}$ and $\mathcal{G}^{\text{N}_\text{III}}_{27}$ are entirely rational and may therefore be written in terms of GPLs.
%\begin{eqnarray}
%	\{\mathcal{G}_{1},\mathcal{G}_{2},\mathcal{G}_{3},\mathcal{G}_{4},\mathcal{G}_{5},\mathcal{G}_{6},\mathcal{G}_{7},\mathcal{G}_{8},\mathcal{G}_{9},\mathcal{G}_{10},\mathcal{G}_{11},\mathcal{G}_{12},\mathcal{G}_{13},\mathcal{G}_{16},\mathcal{G}_{17},\mathcal{G}_{18},\mathcal{G}_{19},\mathcal{G}_{20},\mathcal{G}_{21},\mathcal{G}_{22},\mathcal{G}_{23},\mathcal{G}_{27}\}.\nonumber
%\end{eqnarray}
We evaluate the remaining MIs as Chen-iterated integrals, for which we need to determine boundary solutions. We do so by following the same procedure as for the planar topology discussed in section~\ref{sec:W3PA}, we solve the differential equation in $z_j$ at special points in $x$, namely
\begin{eqnarray}
\lim_{x\rightarrow 0 } \left( \frac {\pd \mathcal{G}^{\text{N}_\text{III}}_j}{\pd z_X }  \right) \quad {\rm{and}} \quad \lim_{x\rightarrow 1 } \left( \frac {\pd \mathcal{G}^{\text{N}_\text{III}}_j}{\pd z_X }  \right),
\end{eqnarray} 
with $X=A,B$ appropriately. Each of the roots presented in equation~\ref{eq:rootsA} ether vanish in these limits or are rationalized entirely by the variables change $y\rightarrow z_X$ and therefore the solution may be entirely determined in terms of GPLs. These boundary solutions can be further uniquely fixed by taking the decoupling limit $z_B\rightarrow 0$, and region A boundary points can then be obtained via analytic continuation and matching of the region B solutions. 

In order to demonstrate the flexibility of our solutions we present results for the most complicated 7-propagator integrals $\mathcal{G}^{\text{N}_{\text{II}}}_{30-32}$ in Figs.~\ref{fig:NII_A_plotj}-\ref{fig:NII_B_plotz}.
Specifically Figs.\ref{fig:NII_A_plotj},  and \ref{fig:NII_B_plotj} show the dependence of the loop integrals on $x$ as $z$ is held fixed in region $A$ and $B$ respectively. Figs.~\ref{fig:NII_A_plotz}) and \ref{fig:NII_B_plotz} alternatively show the dependence of the integrals on $z$ as $x$ is constant. On each figure a comparison point is shown which shows the result obtained using the numerical package {\tt{AMFLow}}, the two being in perfect agreement.

%by varying $x$ and $z$ is presented in Fig.(\ref{fig:NII_A_plotj},\ref{fig:NII_A_plotz}) and Fig.(\ref{fig:NII_B_plotj},\ref{fig:NII_B_plotz}) for region A and B respectively. In Appendix~\ref{appendix:NII} we present the numerical result for all the integrals in this topology. 
\begin{figure}[H]
	\centering	
	\includegraphics[width=4.9cm]{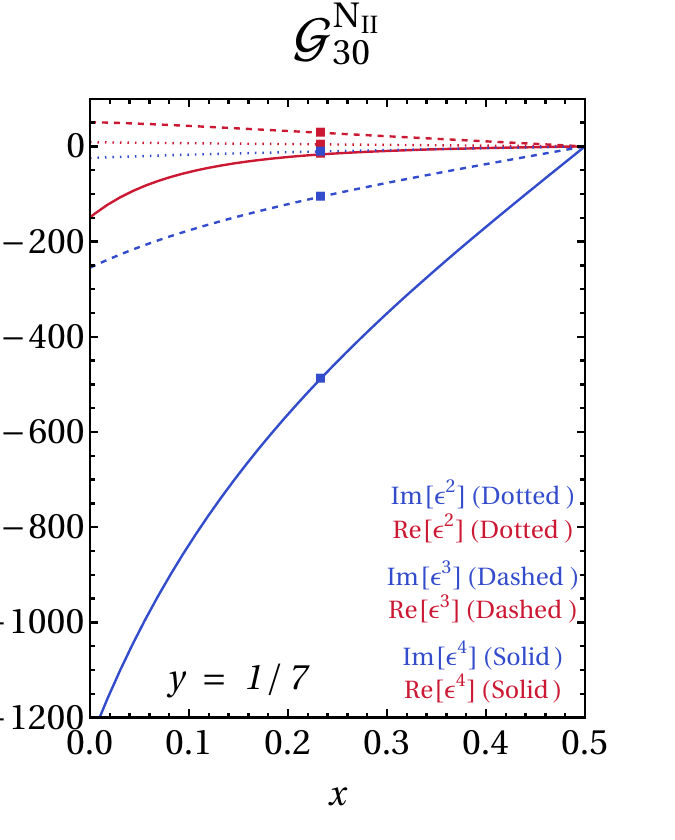}
	\includegraphics[width=4.9cm]{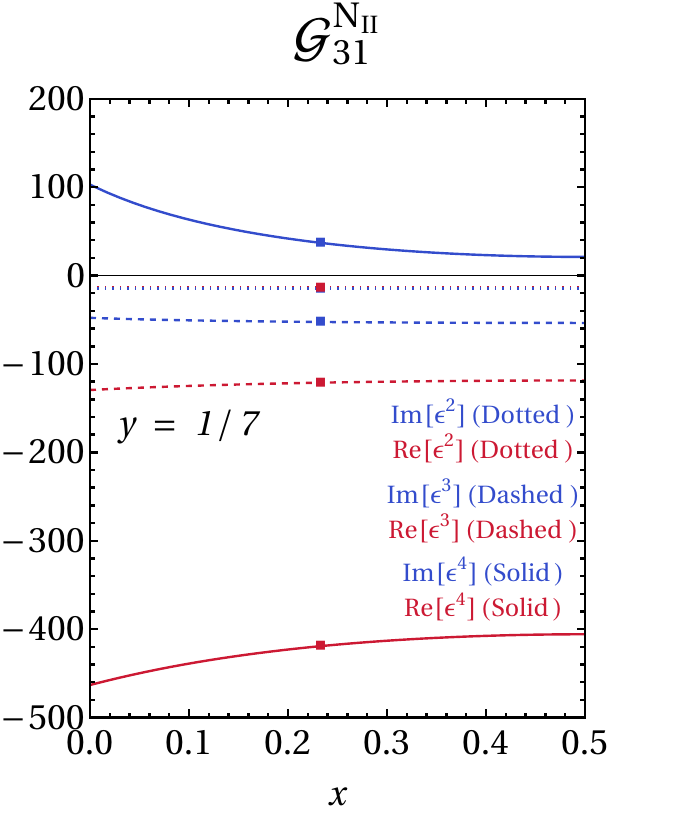}
	\includegraphics[width=4.9cm]{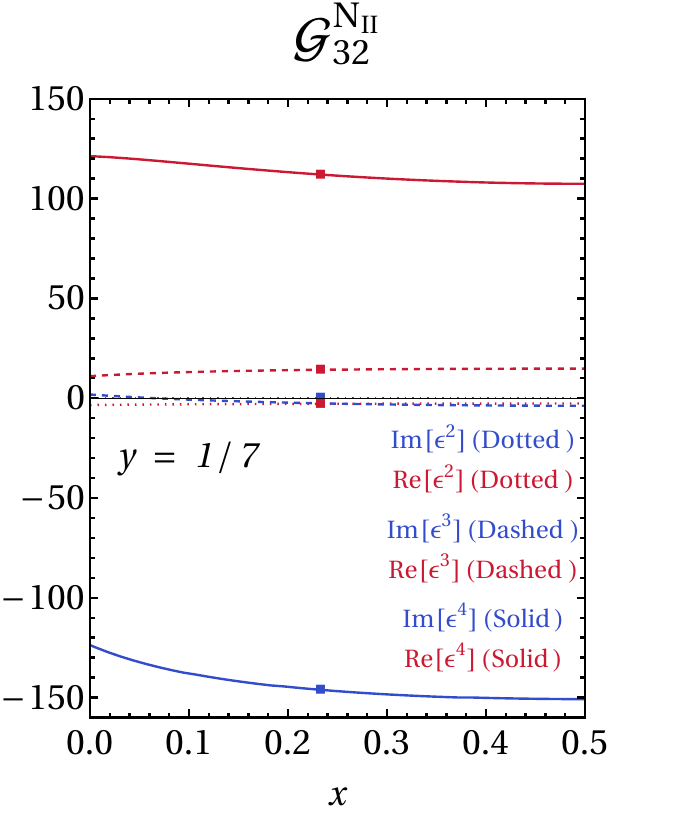}
	\caption{Coefficients of the $\epsilon^i$ terms (where $i=2,3,4$) in the expansion of the three 7 point integrals $\mathcal{G}^{\text{N}_{\text{II}}}_{30}$,	$\mathcal{G}^{\text{N}_{\text{II}}}_{31}$, and $\mathcal{G}^{\text{N}_{\text{II}}}_{32}$. Here $y = 1/7$ and $x$ is varied in the region $u < t$ (Region A, as defined in Eq.~\ref{eq:NII_regionA}). Red and blue squares represent the real and imaginary part, respectively, of independent evaluation using {\tt{AMFlow}}.}
	\label{fig:NII_A_plotj}
\end{figure}
\begin{figure}[H]
	\centering
	\includegraphics[width=4.9cm]{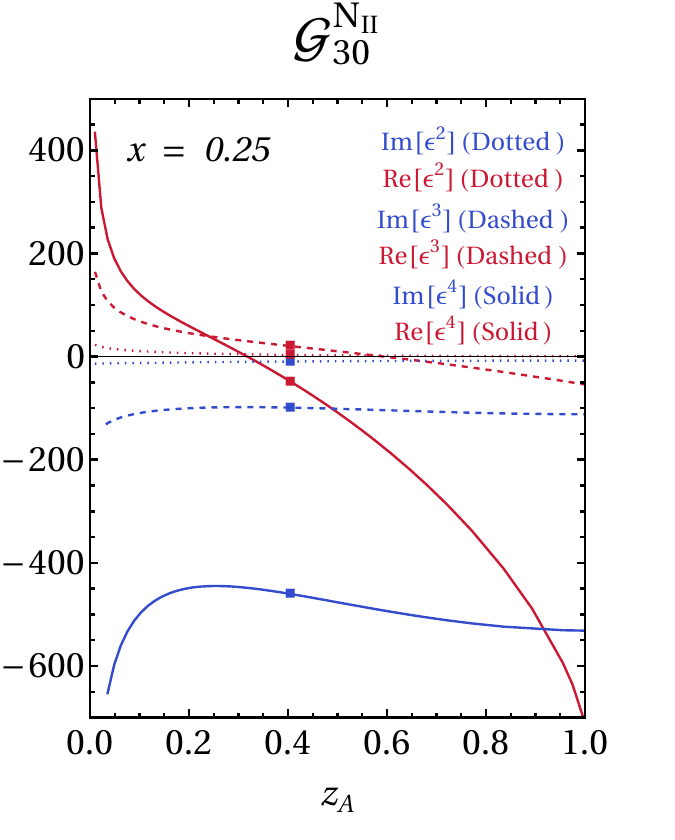}
	\includegraphics[width=4.9cm]{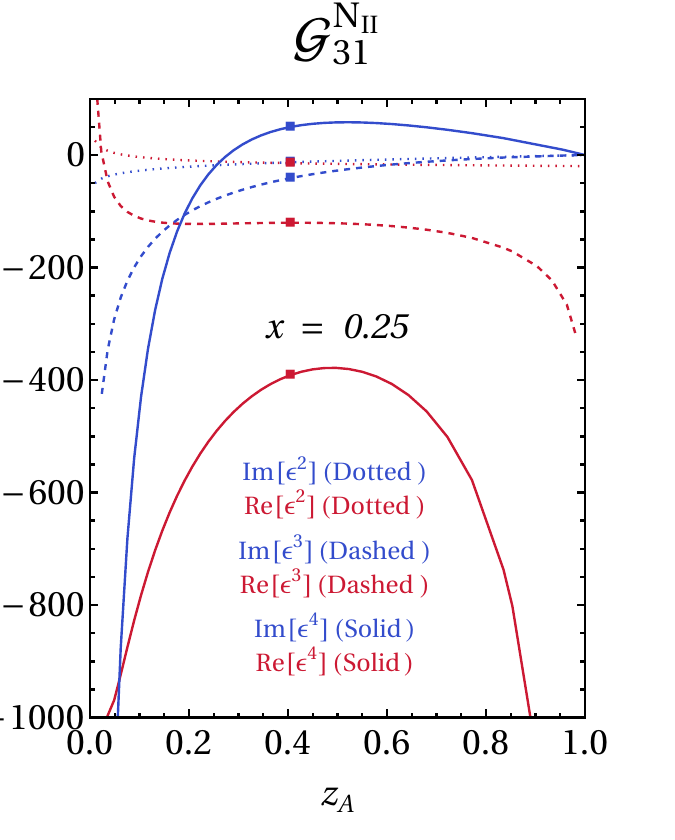}
	\includegraphics[width=4.9cm]{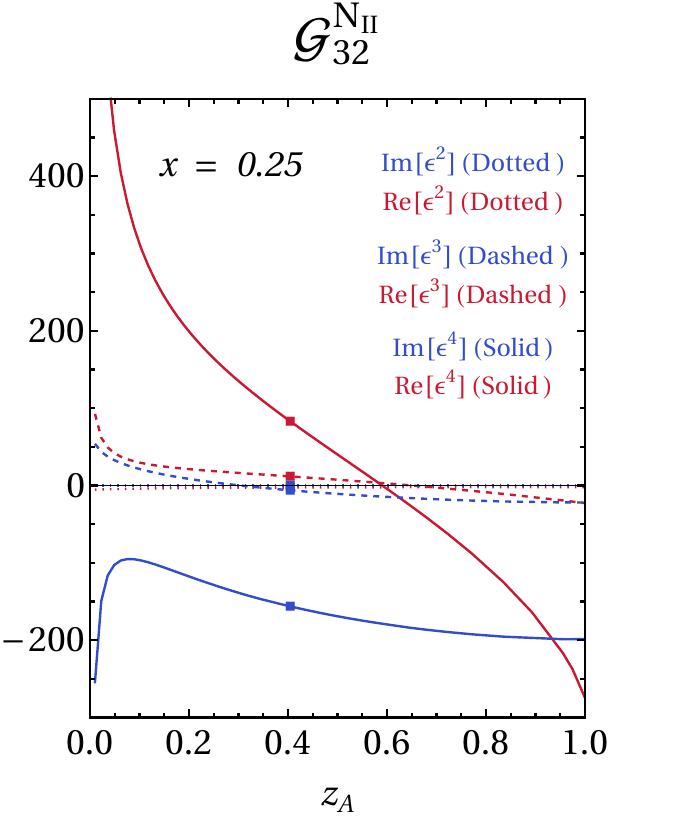}
	\caption{Coefficients of the $\epsilon^i$ terms (where $i=2,3,4$) in the expansion of the three 7 point integrals $\mathcal{G}^{\text{N}_{\text{II}}}_{30}$,	$\mathcal{G}^{\text{N}_{\text{II}}}_{31}$, and $\mathcal{G}^{\text{N}_{\text{II}}}_{32}$. Here $x = 0.25$ and $y$ is varied in the region $y < 1/4$ (Region A, as defined in Eq.~\ref{eq:NII_regionA}). Red and blue squares represent the real and imaginary part, respectively, of independent evaluation using {\tt{AMFlow}}.}
	\label{fig:NII_A_plotz}
\end{figure}
\raggedbottom
\begin{figure}[H]
	\centering
	\includegraphics[width=4.9cm]{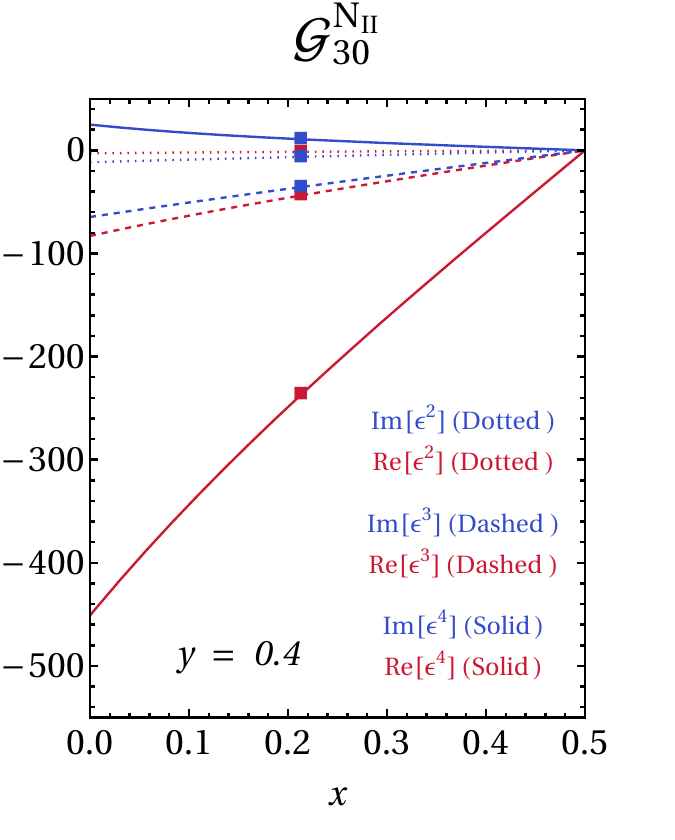}
	\includegraphics[width=4.9cm]{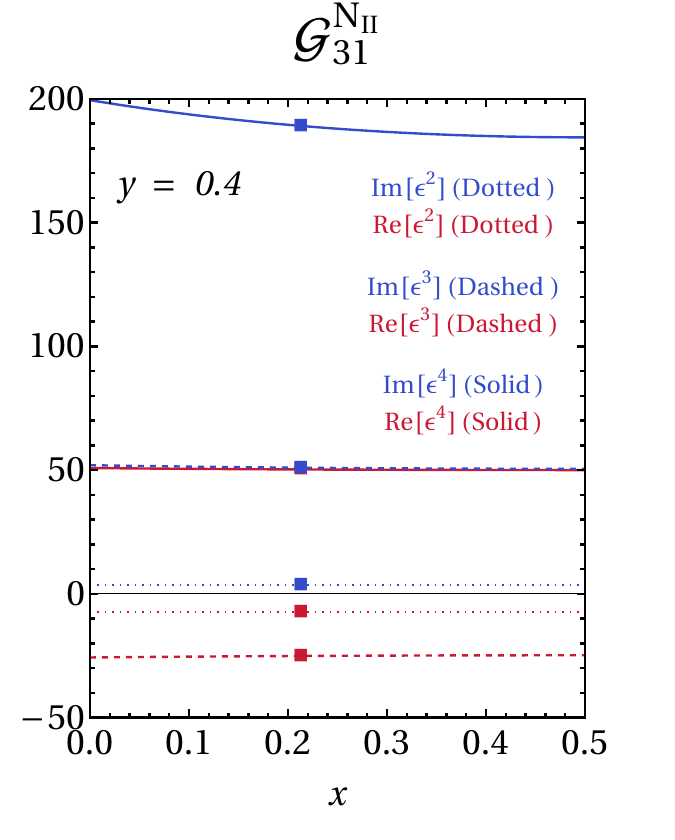}
	\includegraphics[width=4.9cm]{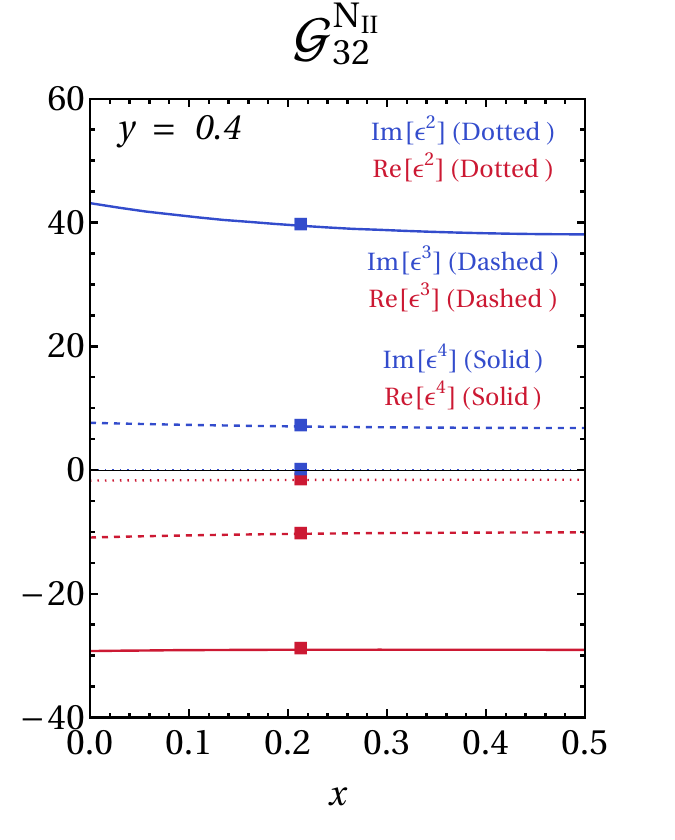}
	\caption{Coefficients of the $\epsilon^i$ terms (where $i=2,3,4$) in the expansion of the three 7 point integrals $\mathcal{G}^{\text{N}_{\text{II}}}_{30}$, $\mathcal{G}^{\text{N}_{\text{II}}}_{31}$, and $\mathcal{G}^{\text{N}_{\text{II}}}_{32}$. Here $y = 0.4$ and $x$ is varied in the region $u<t$ (Region B, as defined in Eq.~\ref{eq:NII_regionB}). Red and blue squares represent the real and imaginary part, respectively, of independent evaluation using {\tt{AMFlow}}.}
	\label{fig:NII_B_plotj}
\end{figure}

\begin{figure}[H]
	\centering
	\includegraphics[width=4.9cm]{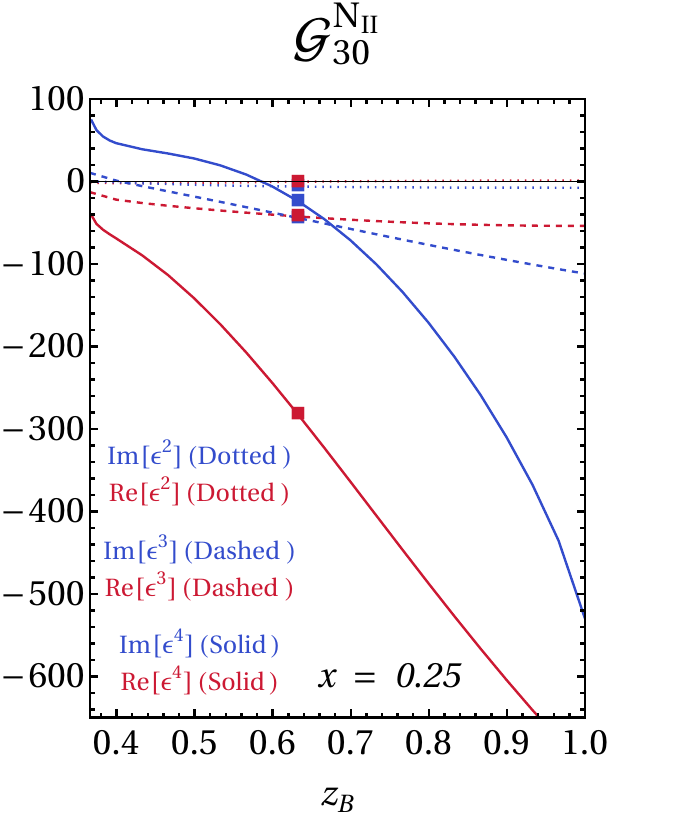}
	\includegraphics[width=4.9cm]{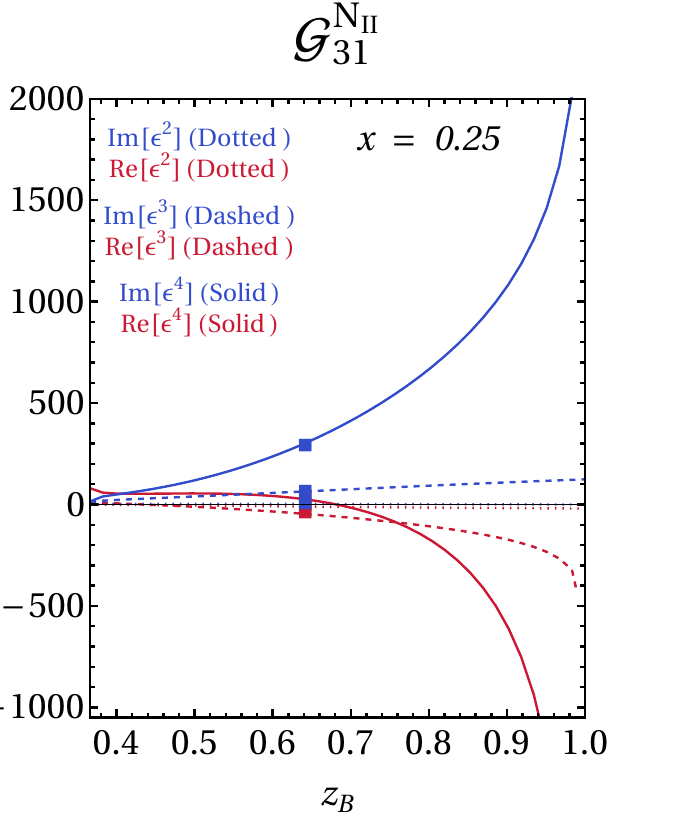}
	\includegraphics[width=4.9cm]{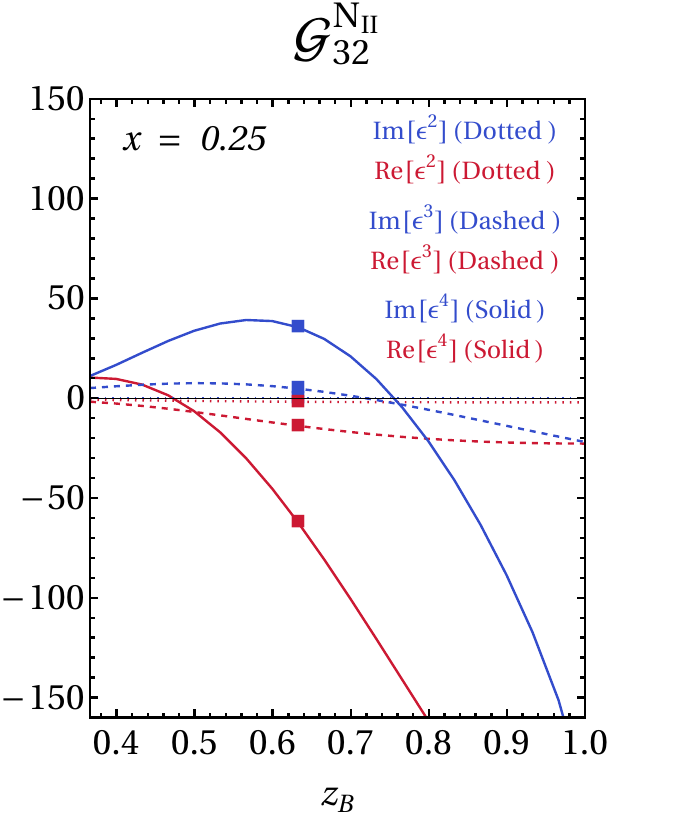}
	\caption{Coefficients of the $\epsilon^i$ terms (where $i=2,3,4$) in the expansion of the three 7 point integrals $\mathcal{G}^{\text{N}_{\text{II}}}_{30}$,$\mathcal{G}^{\text{N}_{\text{II}}}_{31}$, and $\mathcal{G}^{\text{N}_{\text{II}}}_{32}$. Here $x = 0.25$ and and $y$ is varied in the region $y > 1/4$ (Region B, as defined in Eq.~\ref{eq:NII_regionB}). Red and blue squares represent the real and imaginary part, respectively, of independent evaluation using {\tt{AMFlow}}.}
	\label{fig:NII_B_plotz}
\end{figure}
\begin{figure}[H]
	\centering
	\includegraphics[width=0.85\linewidth]{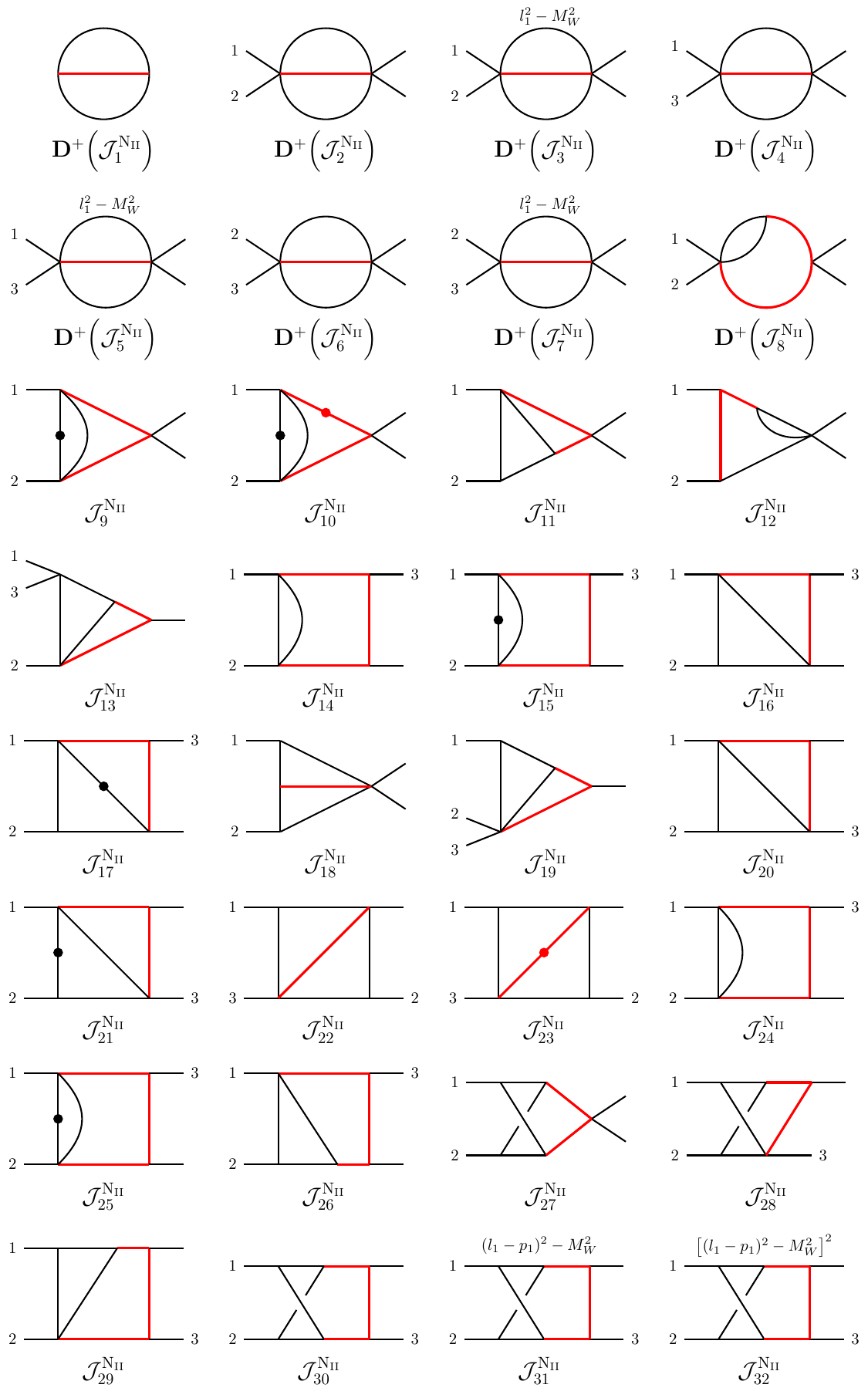}
	\caption{Master integrals for topology N$_{\text{II}}$. Dotted propagators have higher power of the respective denominator, while red propagators are massive.}
	\label{fig:NII_MI}
\end{figure}

%=====================================================================

%================================== CONCLUSIONS

%=====================================================================

\section{Conclusions}
\label{sec:conc}

In this paper we have presented the calculation of the Master Integrals required for the evaluation of electroweak corrections to the amplitude for $gg\rightarrow \gamma\gamma$ for the cases in which the closed loop of fermions corresponds to the first two generations. We solved for the MI's using a differential equation method, specifically by choosing a set of MI's which obeyed a differential equation in canonical form (in which the $\epsilon$ series factorizes from the kinematics in the differential equation), and we were able to solve our integrals using iterated integrals. 

We are able to classify our integrals based on the flow of the massive vector bosons through the diagrams. The situation in which there is no massive bosons (i.e. QED corrections) is well known and we did not discuss the answer in detail here. Topologies for which there is a single massive boson in the loop was also straightforward, there are only planar topologies with no issues regarding non-rational letters, such that the answer could be written entirely in terms of GPL solutions. For topologies with two massive vector bosons there are planar and non-planar topologies to consider. The planar topologies with two massive bosons are similar to those with one, in that they are readily written entirely in-terms of GPL solutions. For non-planar topologies non-rational letters initially appear in the alphabet, however by an appropriate variable change we were able to rationalize the roots such that a GPL solution was obtained for all two $W$ boson topologies. 

Finally the most complicated topologies we had to consider came from diagrams which contain three massive vector bosons. Again here there are both planar and non-planar topologies to consider. The increased number of internal massive lines lead to a proliferation of non-rational letters in the alphabet of the differential equations. In order to solve the differential equations in this instance we used a mixed Goncharov-Chen approach. By writing the non-rational contributions as Chen-iterated integrals we were able to define as solution as a two-parameter integral when expanded up to weight 4. 

We evaluated our integrals in all required regions of kinematic phase space for physical scattering. We checked all of our solutions against public numerical packages, finding prefect agreement, in general our analytic solutions were orders of magnitude quicker than the numerical checks, and should be amenable to implementation into a Monte Carlo generator for subsequent phenomenological analysis, which we will pursue in a subsequent publication. 

%=====================================================================

%==================================APPENDIX

%=====================================================================

\section*{Acknowledgments}
CW is indebted to Uli Schubert for many useful discussions and introducing him to the differential equation methodology. 
We thank Amlan Chakraborty for many useful discussions and collaboration at an early stage of this project. 
We thank Doreen Wackeroth for useful discussions. 
 The authors are supported by the National Science Foundation through awards NSF-PHY-1652066 and NSF-PHY-201402.

\appendix
\section{Numerical Tables}
\subsection{Topology P$_\text{I}$}
\label{appendix:PI}
\begin{table}[H]
	\scriptsize
	\begin{center}
		\begin{tabular}{|c| c c c c c| }
			\hline
			\rule{0pt}{10pt}
			Integral & $\epsilon^0$ &   $\epsilon^1$  &  $\epsilon^2$   & $\epsilon^3$  &  $\epsilon^4$  \\
			\hline
			\rule{0pt}{12pt}
			$\mathcal{G}^{\text{P}_{\text{I}}}_{1}$ &-1&0&-3.28987&2.40411&-9.74091  \\[3pt]
			$\mathcal{G}^{\text{P}_{\text{I}}}_{2}$ &2&-1.79943 + 6.28319$i$&-12.35 - 5.65307$i$&6.78877 - 18.1278$i$&18.5363 + 2.72968$i$  \\[3pt]
			$\mathcal{G}^{\text{P}_{\text{I}}}_{3}$ &-1&1.51073 - 12.5664$i$&59.9621 + 2.02519$i$&48.2216 + 169.018$i$&-300.438 + 290.282$i$  \\[3pt]
			$\mathcal{G}^{\text{P}_{\text{I}}}_{4}$ &1&-0.755365 + 6.28319$i$&-25.9088 + 1.81394$i$&-31.3687 - 73.4063$i$&129.896 - 137.865$i$  \\[3pt]
			$\mathcal{G}^{\text{P}_{\text{I}}}_{5}$ &2&-1.61875&-2.63478&-2.32223&-2.02061  \\[3pt]
			$\mathcal{G}^{\text{P}_{\text{I}}}_{6}$ &-1&4.71031&-4.96941&14.0924&-25.3592  \\[3pt]
			$\mathcal{G}^{\text{P}_{\text{I}}}_{7}$ &0&1.17758&-1.20438&3.00559&-6.20665  \\[3pt]
			$\mathcal{G}^{\text{P}_{\text{I}}}_{8}$ &0&0&0&2.46908 + 5.27231$i$&-10.8039 + 8.64876$i$  \\[3pt]
			$\mathcal{G}^{\text{P}_{\text{I}}}_{9}$ &0&0&-0.561726 - 3.89856$i$&8.7916 - 5.05245$i$&6.3134 + 0.0902883$i$  \\[3pt]
			$\mathcal{G}^{\text{P}_{\text{I}}}_{10}$ &0&0&0&-4.11017&-1.80935  \\[3pt]
			$\mathcal{G}^{\text{P}_{\text{I}}}_{11}$ &0&0&2.63271&2.30232&8.96751  \\[3pt]
			$\mathcal{G}^{\text{P}_{\text{I}}}_{12}$ &4&-6.47499&-7.91879&-0.758835&6.37784  \\[3pt]
			$\mathcal{G}^{\text{P}_{\text{I}}}_{13}$ &0&0&0&0.799872 - 4.00077$i$&14.1684 - 0.891901$i$  \\[3pt]
			$\mathcal{G}^{\text{P}_{\text{I}}}_{14}$ &0&0&-1.56899&-1.67582&-5.31164  \\[3pt]
			$\mathcal{G}^{\text{P}_{\text{I}}}_{15}$ &-4&3.41818 - 6.28319$i$&14.9929 + 5.08544$i$&-3.5778 + 18.0579$i$&-16.4178 - 1.15688$i$  \\[3pt]
			$\mathcal{G}^{\text{P}_{\text{I}}}_{16}$ &1&-1.18706 + 3.14159$i$&-13.1838 + 1.19078$i$&-18.7334 - 36.3309$i$&63.9675 - 71.8708$i$  \\[3pt]
			$\mathcal{G}^{\text{P}_{\text{I}}}_{17}$ &0&0.377683 - 3.14159$i$&16.721 + 4.46229$i$&10.4971 + 59.0473$i$&-114.05 + 88.9521$i$  \\[3pt]
			$\mathcal{G}^{\text{P}_{\text{I}}}_{18}$ &1&-2.07729 + 3.14159$i$&-7.04912 - 6.526$i$&11.2944 - 8.85682$i$&6.58424 + 9.52116$i$  \\[3pt]
			$\mathcal{G}^{\text{P}_{\text{I}}}_{19}$ &0&1.17758&-2.26387 + 3.69947$i$&-9.64506 - 4.79304$i$&2.65821 - 16.1321$i$  \\[3pt]
			$\mathcal{G}^{\text{P}_{\text{I}}}_{20}$ &0&0&0.134973 + 0.165281$i$&-0.397866 + 0.589612$i$&-1.62649 + 0.0748837$i$  \\[3pt]
			$\mathcal{G}^{\text{P}_{\text{I}}}_{21}$ &0&-1.55526 + 3.14159$i$&-16.1407 - 4.08498$i$&-7.58796 - 56.7198$i$&114.6 - 84.1576$i$  \\[3pt]
			$\mathcal{G}^{\text{P}_{\text{I}}}_{22}$ &0&0&0&-4.50119&7.32103  \\[3pt]
			$\mathcal{G}^{\text{P}_{\text{I}}}_{23}$ &-1&3.12948 - 12.5664$i$&55.1793 + 0.0852468$i$&63.663 + 169.72$i$&-291.552 + 326.643$i$  \\[3pt]
			$\mathcal{G}^{\text{P}_{\text{I}}}_{24}$ &1&-2.37411 + 6.28319$i$&-26.0172 + 2.94919$i$&-29.6303 - 71.2136$i$&131.079 - 143.964$i$  \\[3pt]
			$\mathcal{G}^{\text{P}_{\text{I}}}_{25}$ &0&0&0&0&13.308 + 2.2413$i$  \\[3pt]
			\hline 
		\end{tabular} 
		\caption{Numerical results for the 25 integrals of Topology $\text{P}_{\text{I}}$. The integrals are evaluated in Region A, as defined in Eq.\ref{eq:PI_regionA}, at the sample point $x=-2.4589$ and $y=2.2465$.}
		\label{table:PI_A}
	\end{center} 
\end{table} 
\raggedbottom
\begin{table}[H]
	\scriptsize
	\begin{center}
		\begin{tabular}{|c| c c c c c| }
			\hline
			\rule{0pt}{10pt}
			Integral & $\epsilon^0$ &   $\epsilon^1$  &  $\epsilon^2$   & $\epsilon^3$  &  $\epsilon^4$  \\
			\hline
			\rule{0pt}{12pt}
			$\mathcal{G}^{\text{P}_{\text{I}}}_{1}$ &-1&0&-3.28987&2.40411&-9.74091  \\[3pt]
			$\mathcal{G}^{\text{P}_{\text{I}}}_{2}$ &2&-2.90079&-1.18622&-1.05366&-0.988206  \\[3pt]
			$\mathcal{G}^{\text{P}_{\text{I}}}_{3}$ &-1&6.64417&-10.5138&20.7927&-42.9302  \\[3pt]
			$\mathcal{G}^{\text{P}_{\text{I}}}_{4}$ &1&-3.32209&4.4271&-10.1262&19.169  \\[3pt]
			$\mathcal{G}^{\text{P}_{\text{I}}}_{5}$ &2&-2.40949&-1.83846&-1.42764&-1.2897  \\[3pt]
			$\mathcal{G}^{\text{P}_{\text{I}}}_{6}$ &-1&5.86772&-7.96034&17.5021&-35.0958  \\[3pt]
			$\mathcal{G}^{\text{P}_{\text{I}}}_{7}$ &0&1.46693&-2.213&4.18951&-9.3411  \\[3pt]
			$\mathcal{G}^{\text{P}_{\text{I}}}_{8}$ &0&0&0&-6.12631&0.17907  \\[3pt]
			$\mathcal{G}^{\text{P}_{\text{I}}}_{9}$ &0&0&3.70427&1.46967&12.2132  \\[3pt]
			$\mathcal{G}^{\text{P}_{\text{I}}}_{10}$ &0&0&0&-5.27174&-0.802495 \\[3pt]
			$\mathcal{G}^{\text{P}_{\text{I}}}_{11}$ &0&0&3.26041&1.89421&10.7715  \\[3pt]
			$\mathcal{G}^{\text{P}_{\text{I}}}_{12}$ &4&-9.63794&-1.54824&3.14894&5.10092  \\[3pt]
			$\mathcal{G}^{\text{P}_{\text{I}}}_{13}$ &0&0&0&3.00395&-0.423949  \\[3pt]
			$\mathcal{G}^{\text{P}_{\text{I}}}_{14}$ &0&0&-2.09076&-1.73255&-6.52823  \\[3pt]
			$\mathcal{G}^{\text{P}_{\text{I}}}_{15}$ &-4&5.31028&12.9546&0.421382&-10.2771  \\[3pt]
			$\mathcal{G}^{\text{P}_{\text{I}}}_{16}$ &1&-2.86579&-0.824244&-1.5515&7.01  \\[3pt]
			$\mathcal{G}^{\text{P}_{\text{I}}}_{17}$ &0&1.66104&-5.0445&-8.01612&-4.02791  \\[3pt]
			$\mathcal{G}^{\text{P}_{\text{I}}}_{18}$ &1&-2.91732&-0.434013&-2.49897&7.90258  \\[3pt]
			$\mathcal{G}^{\text{P}_{\text{I}}}_{19}$ &0&1.46693&-4.34063&-6.67956&-3.65615  \\[3pt]
			$\mathcal{G}^{\text{P}_{\text{I}}}_{20}$ &0&0&-2.77842&2.09681&-4.45944  \\[3pt]
			$\mathcal{G}^{\text{P}_{\text{I}}}_{21}$ &0&-3.12797&-1.29177&5.28593&10.2361  \\[3pt]
			$\mathcal{G}^{\text{P}_{\text{I}}}_{22}$ &0&0&0&-4.90792&10.9452 \\[3pt]
			$\mathcal{G}^{\text{P}_{\text{I}}}_{23}$ &-1&9.05366&6.27915&-28.4251&-44.0975  \\[3pt]
			$\mathcal{G}^{\text{P}_{\text{I}}}_{24}$ &1&-5.73157&-8.58628&3.77865&20.6685  \\[3pt]
			$\mathcal{G}^{\text{P}_{\text{I}}}_{25}$ &0&0&0&0&-10.6725  \\[3pt]
			\hline 
		\end{tabular} 
		\caption{Numerical results for the 25 integrals of Topology $\text{P}_{\text{I}}$. The integrals are evaluated in Region B, as defined in Eq.\ref{eq:PI_regionB}, at the sample point $x=4.2648$ and $y=3.3359$.}
		\label{table:PI_B}
	\end{center} 
\end{table} 
\begin{table}[H]
	\scriptsize
	\begin{center}
		\begin{tabular}{|c| c c c c c| }
			\hline
			\rule{0pt}{10pt}
			Integral & $\epsilon^0$ &   $\epsilon^1$  &  $\epsilon^2$   & $\epsilon^3$  &  $\epsilon^4$  \\
			\hline
			\rule{0pt}{12pt}
			$\mathcal{G}^{\text{P}_{\text{I}}}_{1}$ &-1&0&-3.28987&2.40411&-9.74091  \\[3pt]
			$\mathcal{G}^{\text{P}_{\text{I}}}_{2}$ &2&2.11744&-2.16898&-7.89571&-11.7001  \\[3pt]
			$\mathcal{G}^{\text{P}_{\text{I}}}_{3}$ &-1&1.19122&-2.07511&6.89876&-8.20121  \\[3pt]
			$\mathcal{G}^{\text{P}_{\text{I}}}_{4}$ &1&-0.595611&2.36178&-4.88683&7.75393  \\[3pt]
			$\mathcal{G}^{\text{P}_{\text{I}}}_{5}$ &2&-0.272381 + 6.28319$i$&-13.1409 - 0.855709$i$&-3.01688 - 20.6126$i$&20.0146 - 12.293$i$  \\[3pt]
			$\mathcal{G}^{\text{P}_{\text{I}}}_{6}$ &-1&-7.69934 - 12.5664$i$&33.7392 - 99.3198$i$&466.546 - 227.039$i$&1909.94 + 576.108$i$  \\[3pt]
			$\mathcal{G}^{\text{P}_{\text{I}}}_{7}$ &0&-1.92483 - 3.14159$i$&10.2813 - 24.616$i$&119.158 - 54.2128$i$&487.629 + 158.085$i$  \\[3pt]
			$\mathcal{G}^{\text{P}_{\text{I}}}_{8}$ &0&0&0&-1.1328&-1.95545  \\[3pt]
			$\mathcal{G}^{\text{P}_{\text{I}}}_{9}$ &0&0&0.833835&1.83071&4.38072  \\[3pt]
			$\mathcal{G}^{\text{P}_{\text{I}}}_{10}$ &0&0&0&1.91488 + 2.89532$i$&-4.23565 + 7.46431$i$  \\[3pt]
			$\mathcal{G}^{\text{P}_{\text{I}}}_{11}$ &0&0&-0.81762 - 2.39879$i$&4.23739 - 5.61184$i$&7.33633 - 4.34692$i$  \\[3pt]
			$\mathcal{G}^{\text{P}_{\text{I}}}_{12}$ &4&-1.08952 + 25.1327$i$&-91.9679 - 6.84567$i$&5.84432 - 247.118$i$&512.679 - 53.3645$i$  \\[3pt]
			$\mathcal{G}^{\text{P}_{\text{I}}}_{13}$ &0&0&0&0.799592&1.66788  \\[3pt]
			$\mathcal{G}^{\text{P}_{\text{I}}}_{14}$ &0&0&2.04798 + 0.427855$i$&7.70385 + 5.12304$i$&18.4844 + 28.5217$i$  \\[3pt]
			$\mathcal{G}^{\text{P}_{\text{I}}}_{15}$ &-4&-1.84506 -	6.28319$i$&16.7377 - 6.65213$i$&24.1202 + 11.5908$i$&4.44028 + 26.4637$i$  \\[3pt]
			$\mathcal{G}^{\text{P}_{\text{I}}}_{16}$ &1&-0.433996 + 3.14159$i$&-7.31466 - 1.36344$i$&0.632443 - 10.1336$i$&8.23075 - 2.35015$i$  \\[3pt]
			$\mathcal{G}^{\text{P}_{\text{I}}}_{17}$ &0&0.297806&0.102776 + 0.935584$i$&-1.94227 + 0.707591$i$&-2.28442 - 2.17101$i$  \\[3pt]
			$\mathcal{G}^{\text{P}_{\text{I}}}_{18}$ &1&2.98355 + 3.14159$i$&-5.23195 + 28.7978$i$&-120.856 + 94.254$i$&-615.173 - 13.1543$i$  \\[3pt]
			$\mathcal{G}^{\text{P}_{\text{I}}}_{19}$ &0&-1.92483 - 3.14159$i$&8.24341 - 27.9421$i$&138.436 - 83.0437$i$&669.262 + 82.9275$i$  \\[3pt]
			$\mathcal{G}^{\text{P}_{\text{I}}}_{20}$ &0&0&1.33007 + 0.292072$i$&4.78264 + 3.48455$i$&10.9039 + 19.2933$i$  \\[3pt]
			$\mathcal{G}^{\text{P}_{\text{I}}}_{21}$ &0&1.62703 + 3.14159$i$&-6.83158 + 25.3565$i$&-108.996 + 63.017$i$&-461.856 - 109.585$i$  \\[3pt]
			$\mathcal{G}^{\text{P}_{\text{I}}}_{22}$ &0&0&0&-4.051 + 5.1399$i$&-23.0107 - 17.535$i$  \\[3pt]
			$\mathcal{G}^{\text{P}_{\text{I}}}_{23}$ &-1&1.4636 - 6.28319$i$&19.281 + 3.27123$i$&-22.1225 + 47.5912$i$&-161.918 - 57.2343$i$  \\[3pt]
			$\mathcal{G}^{\text{P}_{\text{I}}}_{24}$ &1&-0.867992 + 6.28319$i$&-20.1271 - 2.19925$i$&16.3575 - 49.4622$i$&147.355 + 46.4819$i$  \\[3pt]
			$\mathcal{G}^{\text{P}_{\text{I}}}_{25}$ &0&0&0&0&2.21507 - 3.01982$i$  \\[3pt]
			\hline 
		\end{tabular} 
		\caption{Numerical results for the 25 integrals of Topology $\text{P}_{\text{I}}$. The integrals are evaluated in Region C, as defined in Eq.\ref{eq:PI_regionC}, at the sample point $x=0.3469$ and $y=-1.1459$.}
		\label{table:PI_C}
	\end{center} 
\end{table}
\subsection{Topology P$_\text{II}$}
\label{appendix:PII}
\begin{table}[H]
	\scriptsize
	\begin{center}
		\begin{tabular}{|c| c c c c c| }
			\hline
			\rule{0pt}{10pt}
			Integral & $\epsilon^0$ &   $\epsilon^1$  &  $\epsilon^2$   & $\epsilon^3$  &  $\epsilon^4$  \\
			\hline
			\rule{0pt}{12pt}
			$\mathcal{G}^{\text{P}_{\text{II}}}_{1}$ &-1&0&-3.28987&2.40411&-9.74091  \\[3pt]
			$\mathcal{G}^{\text{P}_{\text{II}}}_{2}$ &2&-2.28832 + 6.28319$i$&-11.8504 - 7.18896$i$&9.74903 - 16.5582$i$&16.5125 + 6.97674$i$  \\[3pt]
			$\mathcal{G}^{\text{P}_{\text{II}}}_{3}$ &-1&3.04285 - 12.5664$i$&57.3151 + 16.6707$i$&-17.3517 + 160.478$i$&-324.915 + 104.559$i$  \\[3pt]
			$\mathcal{G}^{\text{P}_{\text{II}}}_{4}$ &1&-1.52142 + 6.28319$i$&-24.7256 - 4.74086$i$&-2.66226 - 70.1134$i$&142.408 - 56.4524$i$  \\[3pt]
			$\mathcal{G}^{\text{P}_{\text{II}}}_{5}$ &-1&3.57274&-2.9985&11.8091&-17.5852  \\[3pt]
			$\mathcal{G}^{\text{P}_{\text{II}}}_{6}$ &0&0.893186&-0.483293&2.25504&-3.68858  \\[3pt]
			$\mathcal{G}^{\text{P}_{\text{II}}}_{7}$ &2&-0.73331&-3.15543&-3.61841&-3.32713  \\[3pt]
			$\mathcal{G}^{\text{P}_{\text{II}}}_{8}$ &-1&1.14416 - 3.14159$i$&6.30304 + 8.05758$i$&-17.9535 + 6.24897$i$&1.00202 - 13.4629$i$  \\[3pt]
			$\mathcal{G}^{\text{P}_{\text{II}}}_{9}$ &0&0&0&2.58505 + 6.29354$i$&-13.7782 + 8.38369$i$  \\[3pt]
			$\mathcal{G}^{\text{P}_{\text{II}}}_{10}$ &0&0&0&-3.07869&-2.32866 \\[3pt]
			$\mathcal{G}^{\text{P}_{\text{II}}}_{11}$ &0&0&2.04657&2.45204&7.46877  \\[3pt]
			$\mathcal{G}^{\text{P}_{\text{II}}}_{12}$ &0&0&0&1.25399&0.690107  \\[3pt]
			$\mathcal{G}^{\text{P}_{\text{II}}}_{13}$ &0&0&0&0&1.47748 + 6.87675$i$  \\[3pt]
			$\mathcal{G}^{\text{P}_{\text{II}}}_{14}$ &0&0&-1.90921 + 0.868614$i$&-5.46692 - 3.68129$i$&3.85792 - 9.7775$i$  \\[3pt]
			$\mathcal{G}^{\text{P}_{\text{II}}}_{15}$ &1&-2.03734 + 3.14159$i$&-6.64448 - 6.40051$i$&11.301 - 6.40934$i$&1.11419 + 9.26507$i$  \\[3pt]
			$\mathcal{G}^{\text{P}_{\text{II}}}_{16}$ &0&0.893186&-1.50524 + 2.80603$i$&-6.69697 - 2.5648$i$&0.724511 - 9.41922$i$  \\[3pt]
			$\mathcal{G}^{\text{P}_{\text{II}}}_{17}$ &0&0&0&-1.78682 - 0.918304$i$&0.191481 - 4.36$i$  \\[3pt]
			$\mathcal{G}^{\text{P}_{\text{II}}}_{18}$ &0&1.6539 - 3.14159$i$&13.3872 + 3.78675$i$&0.717181 + 41.4859$i$&-81.4212 + 38.7063$i$  \\[3pt]
			$\mathcal{G}^{\text{P}_{\text{II}}}_{19}$ &-4&3.02163 - 6.28319$i$&15.6103 + 2.30376$i$&1.39621 + 16.2313$i$&-13.5634 + 6.1045$i$  \\[3pt]
			$\mathcal{G}^{\text{P}_{\text{II}}}_{20}$ &1&-1.12737 + 3.14159$i$&-13.0194 + 0.0721714$i$&-13.1936 - 33.9416$i$&59.5314 - 54.4339$i$  \\[3pt]
			$\mathcal{G}^{\text{P}_{\text{II}}}_{21}$ &0&0.760712 - 3.14159$i$&16.0159 + 7.11679$i$&1.74473 + 59.7822$i$&-117.48 + 68.6029$i$  \\[3pt]
			$\mathcal{G}^{\text{P}_{\text{II}}}_{22}$ &0&0&0&0&-2.90866  \\[3pt]
			$\mathcal{G}^{\text{P}_{\text{II}}}_{23}$ &0&0&1.11227&-1.88646 + 2.35627$i$&-7.16027 - 7.04561$i$  \\[3pt]
			$\mathcal{G}^{\text{P}_{\text{II}}}_{24}$ &0&0&-2.28707 - 3.59448$i$&10.1684 - 9.94568$i$&42.9927 + 10.1081$i$  \\[3pt]
			\hline 
		\end{tabular} 
		\caption{Numerical results for the 24 integrals of topology P$_\text{II}$. The integrals are evaluated in Region A, as defined in Eq.~\ref{eq:PII_regionA}, at the sample point $x=-3.1398$ and $y=1.4429$.}
		\label{table:PII_A}
	\end{center} 
\end{table} 
\begin{table}[H]
	\scriptsize
	\begin{center}
		\begin{tabular}{|c| c c c c c| }
			\hline
			\rule{0pt}{10pt}
			Integral & $\epsilon^0$ &   $\epsilon^1$  &  $\epsilon^2$   & $\epsilon^3$  &  $\epsilon^4$  \\
			\hline
			\rule{0pt}{12pt}
			$\mathcal{G}^{\text{P}_{\text{II}}}_{1}$ &-1&0&-3.28987&2.40411&-9.74091  \\[3pt]
			$\mathcal{G}^{\text{P}_{\text{II}}}_{2}$ &2&-2.11967&-2.16661&-1.71833&-1.51706  \\[3pt]
			$\mathcal{G}^{\text{P}_{\text{II}}}_{3}$ &-1&5.42942&-6.71206&16.0318&-31.168  \\[3pt]
			$\mathcal{G}^{\text{P}_{\text{II}}}_{4}$ &1&-2.71471&3.11478&-8.67794&15.0184  \\[3pt]
			$\mathcal{G}^{\text{P}_{\text{II}}}_{5}$ &-1&5.79661&-7.7483&17.2464&-34.4366  \\[3pt]
			$\mathcal{G}^{\text{P}_{\text{II}}}_{6}$ &0&1.44915&-2.1431&4.10035&-9.12827  \\[3pt]
			$\mathcal{G}^{\text{P}_{\text{II}}}_{7}$ &2&-2.36315&-1.89375&-1.47088&-1.32327  \\[3pt]
			$\mathcal{G}^{\text{P}_{\text{II}}}_{8}$ &-1&1.05984&-1.93498&3.59655&-8.04274  \\[3pt]
			$\mathcal{G}^{\text{P}_{\text{II}}}_{9}$ &0&0&0&-4.81701&-1.24457 \\[3pt]
			$\mathcal{G}^{\text{P}_{\text{II}}}_{10}$ &0&0&0&-5.19666&-0.879496  \\[3pt]
			$\mathcal{G}^{\text{P}_{\text{II}}}_{11}$ &0&0&3.22073&1.9269&10.6498  \\[3pt]
			$\mathcal{G}^{\text{P}_{\text{II}}}_{12}$ &0&0&0&2.51453&0.542128 \\[3pt]
			$\mathcal{G}^{\text{P}_{\text{II}}}_{13}$ &0&0&0&0&-4.41432  \\[3pt]
			$\mathcal{G}^{\text{P}_{\text{II}}}_{14}$ &0&0&-1.13211&-2.88739&-4.61031  \\[3pt]
			$\mathcal{G}^{\text{P}_{\text{II}}}_{15}$ &1&-2.50899&-1.51831&-2.22401&5.87141  \\[3pt]
			$\mathcal{G}^{\text{P}_{\text{II}}}_{16}$ &0&1.44915&-3.67896&-7.591&-5.79339  \\[3pt]
			$\mathcal{G}^{\text{P}_{\text{II}}}_{17}$ &0&0&0&3.79314&-1.96722\\[3pt]
			$\mathcal{G}^{\text{P}_{\text{II}}}_{18}$ &0&2.80651&2.27273&-2.99283&-8.27529  \\[3pt]
			$\mathcal{G}^{\text{P}_{\text{II}}}_{19}$ &-4&4.48282&13.9448&3.17825&-9.88202  \\[3pt]
			$\mathcal{G}^{\text{P}_{\text{II}}}_{20}$ &1&-2.53893&-1.31679&-2.59823&6.14537  \\[3pt]
			$\mathcal{G}^{\text{P}_{\text{II}}}_{21}$ &0&1.35735&-3.40245&-6.895&-5.37232  \\[3pt]
			$\mathcal{G}^{\text{P}_{\text{II}}}_{22}$ &0&0&0&0&-4.73229  \\[3pt]
			$\mathcal{G}^{\text{P}_{\text{II}}}_{23}$ &0&0&2.05698&-1.96362&-6.23  \\[3pt]
			$\mathcal{G}^{\text{P}_{\text{II}}}_{24}$ &0&0&1.88618&-1.77268&-5.29583  \\[3pt]
			\hline 
		\end{tabular} 
		\caption{Numerical results for the 24 integrals of topology P$_\text{II}$. The integrals are evaluated in Region B, as defined in Eq.~\ref{eq:PII_regionB}, at the sample point $x=2.8859$ and $y=3.2595$.}
		\label{table:PII_B}
	\end{center} 
\end{table} 
\subsection{Topology P$_\text{III}$}
\label{appendix:PIII}
\begin{table}[H]
	\scriptsize
	\begin{center}
		\begin{tabular}{|c| c c c c c| }
			\hline
			\rule{0pt}{10pt}
			Integral & $\epsilon^0$ &   $\epsilon^1$  &  $\epsilon^2$   & $\epsilon^3$  &  $\epsilon^4$  \\
			\hline
			\rule{0pt}{12pt}
			$\mathcal{G}^{\text{P}_{\text{III}}}_{1}$ &-1&0&-3.28987&2.40411&-9.74091  \\[3pt]
			$\mathcal{G}^{\text{P}_{\text{III}}}_{2}$ &2&-2.44431 + 6.28319$i$&-11.6658 - 7.67904$i$&10.6662 - 15.9784$i$&15.7162 + 8.24587$i$  \\[3pt]
			$\mathcal{G}^{\text{P}_{\text{III}}}_{3}$ &-1&3.4927 - 12.5664$i$&56.2276 + 20.8534$i$&-35.3395 + 155.103$i$&-318.936 + 54.3136$i$  \\[3pt]
			$\mathcal{G}^{\text{P}_{\text{III}}}_{4}$ &-1&1.74635 - 6.28319$i$&24.2455 + 6.5872$i$&-5.07237 + 67.9369$i$&-140.401 + 34.7934$i$  \\[3pt]
			$\mathcal{G}^{\text{P}_{\text{III}}}_{5}$ &-1&3.96632&-3.57047&12.5279&-20.0835  \\[3pt]
			$\mathcal{G}^{\text{P}_{\text{III}}}_{6}$ &0&0.99158&-0.702254&2.48225&-4.50394  \\[3pt]
			$\mathcal{G}^{\text{P}_{\text{III}}}_{7}$ &0&0&-0.307202 - 4.65067$i$&11.0493 - 3.93605$i$&3.7623 + 2.02169$i$  \\[3pt]
			$\mathcal{G}^{\text{P}_{\text{III}}}_{8}$ &0&0&0&2.61181 + 6.64895$i$&-14.8211 + 8.19906$i$  \\[3pt]
			$\mathcal{G}^{\text{P}_{\text{III}}}_{9}$ &0&0. + 2.34235$i$&-6.10326 + 4.37516$i$&-17.7599 - 7.33235$i$&2.81167 - 46.0371$i$  \\[3pt]
			$\mathcal{G}^{\text{P}_{\text{III}}}_{10}$ &0&0. - 9.3694$i$&29.4348 - 1.13343$i$&3.56079 + 60.2645$i$&-92.4897 + 30.0229$i$  \\[3pt]
			$\mathcal{G}^{\text{P}_{\text{III}}}_{11}$ &0&0&-5.4866&-7.32639&-16.7408  \\[3pt]
			$\mathcal{G}^{\text{P}_{\text{III}}}_{12}$ &0&0. + 2.34235$i$&0. + 3.14608$i$&0. + 3.72183$i$&0. - 7.11178$i$  \\[3pt]
			$\mathcal{G}^{\text{P}_{\text{III}}}_{13}$ &0&0&0&0&1.37323 + 7.20982$i$  \\[3pt]
			$\mathcal{G}^{\text{P}_{\text{III}}}_{14}$ &0&0&-1.91612 + 0.811151$i$&-5.08816 - 3.72197$i$&4.26994 - 8.47007$i$  \\[3pt]
			$\mathcal{G}^{\text{P}_{\text{III}}}_{15}$ &1&-2.21374 + 3.14159$i$&-6.4472 - 6.95466$i$&12.457 - 5.75855$i$&-0.0376254 + 10.2821$i$  \\[3pt]
			$\mathcal{G}^{\text{P}_{\text{III}}}_{16}$ &0&0.99158&-1.91412 + 3.11514$i$&-7.39861 - 3.52493$i$&2.01986 - 10.6336$i$  \\[3pt]
			$\mathcal{G}^{\text{P}_{\text{III}}}_{17}$ &0&0&0&-1.70036 - 0.943653$i$&0.64075 - 4.1616$i$  \\[3pt]
			$\mathcal{G}^{\text{P}_{\text{III}}}_{18}$ &0&1.86475 - 3.14159$i$&13.1082 + 5.49979$i$&-5.19227 + 42.2605$i$&-86.9637 + 25.7847$i$  \\[3pt]
			$\mathcal{G}^{\text{P}_{\text{III}}}_{19}$ &0&0&0&1.44546&0.713956 \\[3pt]
			$\mathcal{G}^{\text{P}_{\text{III}}}_{20}$ &0&0&-2.7433&-1.87382 - 2.77921$i$&-3.45475 - 10.6005$i$  \\[3pt]
			$\mathcal{G}^{\text{P}_{\text{III}}}_{21}$ &0&0&-2.7433&1.76327 - 8.61834$i$&17.1739 + 5.53946$i$  \\[3pt]
			$\mathcal{G}^{\text{P}_{\text{III}}}_{22}$ &0&0&0&3.12674 - 11.6783$i$&41.65 + 19.2504$i$  \\[3pt]
			$\mathcal{G}^{\text{P}_{\text{III}}}_{23}$ &0&0&0&-6.38107 - 5.12545$i$&-0.302294 - 21.7425$i$  \\[3pt]
			$\mathcal{G}^{\text{P}_{\text{III}}}_{24}$ &0&0&-2.7433&1.03354&-2.99786  \\[3pt]
			$\mathcal{G}^{\text{P}_{\text{III}}}_{25}$ &0&0&3.96819&8.78464&10.3528  \\[3pt]
			$\mathcal{G}^{\text{P}_{\text{III}}}_{26}$ &0&0&1.26421&-2.32486 + 2.56903$i$&-7.56341 - 8.62886$i$  \\[3pt]
			$\mathcal{G}^{\text{P}_{\text{III}}}_{27}$ &0&0&0&0&5.1483 + 4.05094$i$  \\[3pt]
			$\mathcal{G}^{\text{P}_{\text{III}}}_{28}$ &0&0&0&5.88157 + 6.62362$i$&1.72768 + 28.0861$i$  \\[3pt]
			$\mathcal{G}^{\text{P}_{\text{III}}}_{29}$ &0&0&3.96819&-5.76835 - 3.55034$i$&-16.7338 - 57.9513$i$  \\[3pt]
			$\mathcal{G}^{\text{P}_{\text{III}}}_{30}$ &0&0&0&0&-0.575238 - 0.729338$i$  \\[3pt]
			$\mathcal{G}^{\text{P}_{\text{III}}}_{31}$ &0&0&-2.7433&7.57798 - 3.05993$i$&14.1357 + 32.1012$i$  \\[3pt]
			$\mathcal{G}^{\text{P}_{\text{III}}}_{32}$ &0&0&0&3.40071 + 1.88731$i$&-4.87662 + 4.78924$i$  \\[3pt]
			\hline 
		\end{tabular} 
		\caption{Numerical results for the 32 integrals of topology P$_\text{III}$. The integrals are evaluated in Region A, as defined in Eq.~\ref{eq:PIII_regionA}, at the sample	point $x=1.69549$ and $y=-3.3945$.}
		\label{table:PIII_A}
	\end{center} 
\end{table} 
\begin{table}[H]
	\scriptsize
	\begin{center}
		\begin{tabular}{|c| c c c c c| }
			\hline
			\rule{0pt}{10pt}
			Integral & $\epsilon^0$ &   $\epsilon^1$  &  $\epsilon^2$   & $\epsilon^3$  &  $\epsilon^4$  \\
			\hline
			\rule{0pt}{12pt}
			$\mathcal{G}^{\text{P}_{\text{III}}}_{1}$ &-1&0&-3.28987&2.40411&-9.74091  \\[3pt]
			$\mathcal{G}^{\text{P}_{\text{III}}}_{2}$ &2&-2.99455 + 6.28319$i$&-10.9176 - 9.40767$i$&13.7763 - 13.6279$i$&12.3492 + 12.3294$i$  \\[3pt]
			$\mathcal{G}^{\text{P}_{\text{III}}}_{3}$ &-1&4.97604 - 12.5664$i$&51.6989 + 34.3078$i$&-90.1785 + 129.099$i$&-263.462 - 92.0311$i$  \\[3pt]
			$\mathcal{G}^{\text{P}_{\text{III}}}_{4}$ &-1&2.48802 - 6.28319$i$&22.2732 + 12.4501$i$&-28.2614 + 57.4298$i$&-119.079 - 27.3863$i$  \\[3pt]
			$\mathcal{G}^{\text{P}_{\text{III}}}_{5}$ &-1&4.92312&-5.44537&14.6137&-27.0022  \\[3pt]
			$\mathcal{G}^{\text{P}_{\text{III}}}_{6}$ &0&1.23078&-1.36913&3.18477&-6.73497  \\[3pt]
			$\mathcal{G}^{\text{P}_{\text{III}}}_{7}$ &0&0&-0.00936825 - 5.33815$i$&13.0223 - 2.46789$i$&0.329742 + 3.20266$i$  \\[3pt]
			$\mathcal{G}^{\text{P}_{\text{III}}}_{8}$ &0&0&0&2.65712 + 8.02212$i$&-18.859 + 7.07761$i$  \\[3pt]
			$\mathcal{G}^{\text{P}_{\text{III}}}_{9}$ &0&0.672459 - 3.14159$i$&17.2197 + 0.793569$i$&8.97323 + 49.1711$i$&-100.877 + 54.7732$i$  \\[3pt]
			$\mathcal{G}^{\text{P}_{\text{III}}}_{10}$ &0&-2.68983 + 12.5664$i$&-69.4825 - 17.7644$i$&54.1099 - 214.834$i$&487.856 + 80.9159$i$  \\[3pt]
			$\mathcal{G}^{\text{P}_{\text{III}}}_{11}$ &0&0&-9.4174 - 4.22518$i$&-10.8032 - 12.0621$i$&-14.9631 - 6.05697$i$  \\[3pt]
			$\mathcal{G}^{\text{P}_{\text{III}}}_{12}$ &0&0.672459 - 3.14159$i$&8.50787 - 2.37532$i$&2.60623 + 12.4742$i$&-21.1972 + 20.622$i$  \\[3pt]
			$\mathcal{G}^{\text{P}_{\text{III}}}_{13}$ &0&0&0&0&0.912314 + 8.42193$i$  \\[3pt]
			$\mathcal{G}^{\text{P}_{\text{III}}}_{14}$ &0&0&-1.9219 + 0.634321$i$&-3.9331 - 3.67257$i$&4.69996 - 4.43151$i$  \\[3pt]
			$\mathcal{G}^{\text{P}_{\text{III}}}_{15}$ &1&-2.72806 + 3.14159$i$&-5.56789 - 8.57044$i$&15.7439 - 2.66663$i$&-5.84004 + 12.6208$i$  \\[3pt]
			$\mathcal{G}^{\text{P}_{\text{III}}}_{16}$ &0&1.23078&-3.21195 + 3.86661$i$&-8.55908 - 6.55849$i$&6.66695 - 12.2735$i$  \\[3pt]
			$\mathcal{G}^{\text{P}_{\text{III}}}_{17}$ &0&0&0&-1.71668 - 1.20926$i$&2.18469 - 4.07839$i$  \\[3pt]
			$\mathcal{G}^{\text{P}_{\text{III}}}_{18}$ &0&2.47479 - 3.14159$i$&11.2287 + 10.1168$i$&-21.705 + 38.5434$i$&-85.3521 - 15.0574$i$  \\[3pt]
			$\mathcal{G}^{\text{P}_{\text{III}}}_{19}$ &0&0&0&1.96557&0.691136 \\[3pt]
			$\mathcal{G}^{\text{P}_{\text{III}}}_{20}$ &0&0&-4.7087 - 2.11259$i$&3.51701 - 13.3845$i$&26.3234 - 8.84441$i$  \\[3pt]
			$\mathcal{G}^{\text{P}_{\text{III}}}_{21}$ &0&0&-4.7087 - 2.11259$i$&11.4546 - 17.505$i$&40.7022 + 29.9841$i$  \\[3pt]
			$\mathcal{G}^{\text{P}_{\text{III}}}_{22}$ &0&0&0&9.53699 - 8.55241$i$&31.2663 + 57.1203$i$  \\[3pt]
			$\mathcal{G}^{\text{P}_{\text{III}}}_{23}$ &0&0&0&-2.01641 - 11.1864$i$&29.0282 - 20.3748$i$  \\[3pt]
			$\mathcal{G}^{\text{P}_{\text{III}}}_{24}$ &0&0&-4.7087 - 2.11259$i$&2.23036 - 5.55141$i$&2.25801 - 1.35728$i$  \\[3pt]
			$\mathcal{G}^{\text{P}_{\text{III}}}_{25}$ &0&0&8.11481 + 4.94635$i$&13.2002 + 14.1604$i$&4.97228 + 7.20444$i$  \\[3pt]
			$\mathcal{G}^{\text{P}_{\text{III}}}_{26}$ &0&0&1.66051&-3.66561 + 2.90444$i$&-7.09043 - 13.0777$i$  \\[3pt]
			$\mathcal{G}^{\text{P}_{\text{III}}}_{27}$ &0&0&0&0&-7.67426 + 1.63051$i$  \\[3pt]
			$\mathcal{G}^{\text{P}_{\text{III}}}_{28}$ &0&0&0&-0.9002 + 15.5455$i$&-39.7278 + 27.0031$i$  \\[3pt]
			$\mathcal{G}^{\text{P}_{\text{III}}}_{29}$ &0&0&8.11481 + 4.94635$i$&2.1968 - 7.84989$i$&59.9548 - 88.166$i$  \\[3pt]
			$\mathcal{G}^{\text{P}_{\text{III}}}_{30}$ &0&0&0&0&-1.95754 + 0.379603$i$  \\[3pt]
			$\mathcal{G}^{\text{P}_{\text{III}}}_{31}$ &0&0&-4.7087 - 2.11259$i$&8.88131 - 1.83875$i$&-16.5427 + 50.1519$i$  \\[3pt]
			$\mathcal{G}^{\text{P}_{\text{III}}}_{32}$ &0&0&0&3.43336 + 2.41853$i$&-11.6037 + 3.3266$i$  \\[3pt]
			\hline 
		\end{tabular} 
		\caption{Numerical results for the 32 integrals of topology P$_\text{III}$. The integrals are evaluated in Region B, as defined in Eq.~\ref{eq:PIII_regionB}, at the sample point $x=2.4239$ and $y=-4.4695$.}
		\label{table:PIII_B}
	\end{center} 
\end{table}
\subsection{Topology N$_\text{I}$}
\label{appendix:NI}
\begin{table}[H]
	\scriptsize
	\begin{center}
		\begin{tabular}{|c| c c c c c| }
			\hline
			\rule{0pt}{10pt}
			Integral & $\epsilon^0$ &   $\epsilon^1$  &  $\epsilon^2$   & $\epsilon^3$  &  $\epsilon^4$  \\
			\hline
			\rule{0pt}{12pt}
			$\mathcal{G}^{\text{N}_{\text{I}}}_{1}$ &-1&0&-3.28987&2.40411&-9.74091  \\[2pt]
			$\mathcal{G}^{\text{N}_{\text{I}}}_{2}$ &-1&3.25987&-2.62756&11.2623&-15.7512  \\[2pt]
			$\mathcal{G}^{\text{N}_{\text{I}}}_{3}$ &-1&1.62993&-1.96277&7.08085&-9.58688  \\[2pt]
			$\mathcal{G}^{\text{N}_{\text{I}}}_{4}$ &-1&4.94979&-5.50737&14.6818&-27.2124  \\[2pt]
			$\mathcal{G}^{\text{N}_{\text{I}}}_{5}$ &0&1.23745&-1.39043&3.2083&-6.80256  \\[2pt]
			$\mathcal{G}^{\text{N}_{\text{I}}}_{6}$ &-1&3.98174 - 12.5664$i$&54.8921 + 25.3445$i$&-54.193 + 147.895$i$&-306.422 + 2.57015$i$  \\[2pt]
			$\mathcal{G}^{\text{N}_{\text{I}}}_{7}$ &0&0.995435 - 3.14159$i$&15.6161 + 8.39375$i$&-20.5406 + 41.4401$i$&-85.3548 - 5.04259$i$  \\[2pt]
			$\mathcal{G}^{\text{N}_{\text{I}}}_{8}$ &2&-2.61985 + 6.28319$i$&-11.4436 - 8.23051$i$&11.6805 - 15.2802$i$&14.7354 + 9.618$i$  \\[2pt]
			$\mathcal{G}^{\text{N}_{\text{I}}}_{9}$ &0&0&0&2.63502 + 7.06654$i$&-16.0487 + 7.92493$i$  \\[2pt]
			$\mathcal{G}^{\text{N}_{\text{I}}}_{10}$ &0&0&0.220515 + 4.86575$i$&-16.948 - 10.6115$i$&29.3001 - 18.3132$i$  \\[2pt]
			$\mathcal{G}^{\text{N}_{\text{I}}}_{11}$ &0&0&0&-1.11048&-0.659894 \\[2pt]
			$\mathcal{G}^{\text{N}_{\text{I}}}_{12}$ &0&0&0&-2.66297&0.171531  \\[2pt]
			$\mathcal{G}^{\text{N}_{\text{I}}}_{13}$ &0&2.05241&3.60871&2.17674&-1.15127  \\[2pt]
			$\mathcal{G}^{\text{N}_{\text{I}}}_{14}$ &0&0&1.67211&1.7035&5.55794  \\[2pt]
			$\mathcal{G}^{\text{N}_{\text{I}}}_{15}$ &0&0&1.07122 + 2.06699$i$&-5.77731 + 4.49239$i$&-12.0055 - 5.92941$i$  \\[2pt]
			$\mathcal{G}^{\text{N}_{\text{I}}}_{16}$ &0&-1.8104 + 3.14159$i$&-12.1442 - 4.24105$i$&3.487 - 35.3989$i$&68.5213 - 19.7514$i$  \\[2pt]
			$\mathcal{G}^{\text{N}_{\text{I}}}_{17}$ &0&0&0&-1.15186 - 0.684574$i$&0.86833 - 2.73677$i$  \\[2pt]
			$\mathcal{G}^{\text{N}_{\text{I}}}_{18}$ &0&-0.995435 + 3.14159$i$&-15.8644 - 8.92608$i$&19.7524 - 45.146$i$&89.7615 - 4.06789$i$  \\[2pt]
			$\mathcal{G}^{\text{N}_{\text{I}}}_{19}$ &0&0&0&-4.21463 - 2.69534$i$&0.519373 - 11.9567$i$  \\[2pt]
			$\mathcal{G}^{\text{N}_{\text{I}}}_{20}$ &0&0&0&-2.44196 - 1.49785$i$&1.28909 - 6.22332$i$  \\[2pt]
			$\mathcal{G}^{\text{N}_{\text{I}}}_{21}$ &0&0&4.2849 + 8.26795$i$&-27.9932 + 14.9739$i$&-45.4438 - 36.1643$i$  \\[2pt]
			$\mathcal{G}^{\text{N}_{\text{I}}}_{22}$ &1&-2.12489 + 3.14159$i$&-6.31373 - 6.67555$i$&11.9904 - 4.65145$i$&-2.55735 + 9.7102$i$  \\[2pt]
			$\mathcal{G}^{\text{N}_{\text{I}}}_{23}$ &0&-0.814967&1.39994 - 2.56029$i$&5.84311 + 2.17981$i$&-0.627868 + 7.41755$i$  \\[2pt]
			$\mathcal{G}^{\text{N}_{\text{I}}}_{24}$ &0&0&0&0&1.24199 + 7.59051$i$  \\[2pt]
			$\mathcal{G}^{\text{N}_{\text{I}}}_{25}$ &0&0&0.473251 + 0.895731$i$&-2.81679 + 1.64898$i$&-4.9247 - 3.7518$i$  \\[2pt]
			$\mathcal{G}^{\text{N}_{\text{I}}}_{26}$ &0&-2.23288 + 3.14159$i$&-12.8019 - 8.92608$i$&16.2664 - 45.146$i$&101.433 - 4.06789$i$  \\[2pt]
			$\mathcal{G}^{\text{N}_{\text{I}}}_{27}$ &0&0&0.995946&1.35812&3.81909  \\[2pt]
			$\mathcal{G}^{\text{N}_{\text{I}}}_{28}$ &0&0&0&-1.98124&-0.688585 \\[2pt]
			$\mathcal{G}^{\text{N}_{\text{I}}}_{29}$ &1&-2.54737 + 3.14159$i$&-6.05462 - 8.00281$i$&14.6104 - 4.78095$i$&-1.47663 + 12.4471$i$  \\[2pt]
			$\mathcal{G}^{\text{N}_{\text{I}}}_{30}$ &0&-1.23745&3.0114 - 3.88755$i$&9.12397 + 6.2988$i$&-5.94287 + 13.9332$i$  \\[2pt]
			$\mathcal{G}^{\text{N}_{\text{I}}}_{31}$ &0&0&0&1.57961 - 4.3712$i$&15.5294 + 4.03181$i$  \\[2pt]
			$\mathcal{G}^{\text{N}_{\text{I}}}_{32}$ &0&0&1.84523 - 0.923495$i$&5.14416 + 2.17246$i$&-0.43744 + 10.0508$i$  \\[2pt]
			$\mathcal{G}^{\text{N}_{\text{I}}}_{33}$ &0&0&-0.995946&1.71917 - 1.93136$i$&5.41507 + 6.50726$i$  \\[2pt]
			$\mathcal{G}^{\text{N}_{\text{I}}}_{34}$ &0&0&1.84523 - 0.923495$i$&5.14416 + 2.17246$i$&-0.43744 + 10.0508$i$  \\[2pt]
			$\mathcal{G}^{\text{N}_{\text{I}}}_{35}$ &0&0&-1.67211&3.54962 - 3.24234$i$&9.0686 + 12.8483$i$  \\[2pt]
			$\mathcal{G}^{\text{N}_{\text{I}}}_{36}$ &0&0&0&-3.94759 - 15.6279$i$&39.9957 - 50.7967$i$  \\[2pt]
			$\mathcal{G}^{\text{N}_{\text{I}}}_{37}$ &0&2.05241&-3.29102 - 2.9352$i$&3.86669 - 28.3584$i$&67.4776 - 14.988$i$  \\[2pt]
			$\mathcal{G}^{\text{N}_{\text{I}}}_{38}$ &0&0.814967&-1.61544 - 3.60323$i$&9.43714 - 18.9226$i$&43.8465 + 4.13818$i$  \\[2pt]
			$\mathcal{G}^{\text{N}_{\text{I}}}_{39}$ &0&0&0&0.99494 + 4.45703$i$&-13.3029 + 12.9023$i$  \\[2pt]
			$\mathcal{G}^{\text{N}_{\text{I}}}_{40}$ &0&0&0&-2.44196 - 1.49785$i$&-0.597434 - 4.52663$i$  \\[2pt]
			\hline 
		\end{tabular} 
		\caption{Numerical results for the 40 integrals of Topology $\text{N}_{\text{I}}$. The integrals are evaluated in Region A, as defined in Eq.\ref{eq:NI_regionA}, at the sample point $x=-2.4468$ and $y=-1.2591$.}
		\label{table:NI_A}
	\end{center} 
\end{table}
\begin{table}[H]
	\scriptsize
	\begin{center}
		\begin{tabular}{|c| c c c c c| }
			\hline
			\rule{0pt}{10pt}
			Integral & $\epsilon^0$ &   $\epsilon^1$  &  $\epsilon^2$   & $\epsilon^3$  &  $\epsilon^4$  \\
			\hline
			\rule{0pt}{12pt}
			$\mathcal{G}^{\text{N}_{\text{I}}}_{1}$ &-1&0&-3.28987&2.40411&-9.74091  \\[3pt]
			$\mathcal{G}^{\text{N}_{\text{I}}}_{2}$ &-1&5.7937&-7.73972&17.2361&-34.4098  \\[3pt]
			$\mathcal{G}^{\text{N}_{\text{I}}}_{3}$ &-1&2.89685&-3.45918&9.04257&-16.1705  \\[3pt]
			$\mathcal{G}^{\text{N}_{\text{I}}}_{4}$ &-1&4.54587 - 12.5664$i$&53.1579 + 30.4564$i$&-75.0035 + 137.892$i$&-284.775 - 52.9579$i$  \\[3pt]
			$\mathcal{G}^{\text{N}_{\text{I}}}_{5}$ &0&1.13647 - 3.14159$i$&15.1267 + 9.83648$i$&-26.5496 + 38.461$i$&-78.7061 - 21.0217$i$  \\[3pt]
			$\mathcal{G}^{\text{N}_{\text{I}}}_{6}$ &-1&2.48087&-2.0293&9.8781&-11.8496  \\[3pt]
			$\mathcal{G}^{\text{N}_{\text{I}}}_{7}$ &0&0.620216&-0.0463515&1.7114&-1.74511  \\[3pt]
			$\mathcal{G}^{\text{N}_{\text{I}}}_{8}$ &2&0.303203&-3.26689&-5.30582&-5.63715  \\[3pt]
			$\mathcal{G}^{\text{N}_{\text{I}}}_{9}$ &0&0&0&-2.16594&-2.42064  \\[3pt]
			$\mathcal{G}^{\text{N}_{\text{I}}}_{10}$ &0&0&-1.49898&1.97982&-1.29796  \\[3pt]
			$\mathcal{G}^{\text{N}_{\text{I}}}_{11}$ &0&0&0&-2.51258&-0.542866 \\[3pt]
			$\mathcal{G}^{\text{N}_{\text{I}}}_{12}$ &0&0&0&0.715101 + 0.459292$i$&-0.8256 + 1.63381$i$  \\[3pt]
			$\mathcal{G}^{\text{N}_{\text{I}}}_{13}$ &0&2.58489 - 3.14159$i$&11.918 + 12.1192$i$&-27.3245 + 44.4824$i$&-106.273 - 21.9359$i$  \\[3pt]
			$\mathcal{G}^{\text{N}_{\text{I}}}_{14}$ &0&0&-2.02943 - 4.44478$i$&9.96187 - 11.1203$i$&21.8403 + 2.71965$i$  \\[3pt]
			$\mathcal{G}^{\text{N}_{\text{I}}}_{15}$ &0&0&-2.18264&-0.579008&-4.66848  \\[3pt]
			$\mathcal{G}^{\text{N}_{\text{I}}}_{16}$ &0&-2.06864&-3.76536&-2.75711&0.0101042  \\[3pt]
			$\mathcal{G}^{\text{N}_{\text{I}}}_{17}$ &0&0&0&-3.19913 - 2.19602$i$&2.34156 - 8.49664$i$  \\[3pt]
			$\mathcal{G}^{\text{N}_{\text{I}}}_{18}$ &0&-0.620216&6.01498 + 11.4821$i$&-39.3295 + 18.4562$i$&-48.4268 - 49.9119$i$  \\[3pt]
			$\mathcal{G}^{\text{N}_{\text{I}}}_{19}$ &0&0&0&0.784536&0.549118  \\[3pt]
			$\mathcal{G}^{\text{N}_{\text{I}}}_{20}$ &0&0&0&2.91521&-0.260722 \\[3pt]
			$\mathcal{G}^{\text{N}_{\text{I}}}_{21}$ &0&0&-8.73058&3.51438&-19.1953  \\[3pt]
			$\mathcal{G}^{\text{N}_{\text{I}}}_{22}$ &1&-1.29682&-3.82398&-4.18162&1.05095  \\[3pt]
			$\mathcal{G}^{\text{N}_{\text{I}}}_{23}$ &0&-1.44843&1.92068&9.75574&14.3738  \\[3pt]
			$\mathcal{G}^{\text{N}_{\text{I}}}_{24}$ &0&0&0&0&-2.08142 \\[3pt]
			$\mathcal{G}^{\text{N}_{\text{I}}}_{25}$ &0&0&1.31141 + 2.87053$i$&-7.88351 + 5.71205$i$&-14.3227 - 8.22965$i$  \\[3pt]
			$\mathcal{G}^{\text{N}_{\text{I}}}_{26}$ &0&-1.75668 + 3.14159$i$&-11.1411 - 2.79916$i$&1.61544 - 27.9809$i$&50.1745 - 13.3256$i$  \\[3pt]
			$\mathcal{G}^{\text{N}_{\text{I}}}_{27}$ &0&0&2.05561&1.73504&6.44769  \\[3pt]
			$\mathcal{G}^{\text{N}_{\text{I}}}_{28}$ &0&0&0&4.4335 + 3.14428$i$&-1.94521 + 12.8449$i$  \\[3pt]
			$\mathcal{G}^{\text{N}_{\text{I}}}_{29}$ &1&-0.984865 + 3.14159$i$&-12.8735 - 0.47064$i$&-10.6309 - 31.755$i$&53.8606 - 45.684$i$  \\[3pt]
			$\mathcal{G}^{\text{N}_{\text{I}}}_{30}$ &0&-1.13647 + 3.14159$i$&-15.299 - 9.36021$i$&4.62937 - 61.6318$i$&121.361 - 59.8471$i$  \\[3pt]
			$\mathcal{G}^{\text{N}_{\text{I}}}_{31}$ &0&0&0&1.38141&1.87152  \\[3pt]
			$\mathcal{G}^{\text{N}_{\text{I}}}_{32}$ &0&0&-1.75313 + 0.805685$i$&-4.49178 - 2.10017$i$&0.968643 - 8.17649$i$  \\[3pt]
			$\mathcal{G}^{\text{N}_{\text{I}}}_{33}$ &0&0&-2.05561&0.0839943&8.18846  \\[3pt]
			$\mathcal{G}^{\text{N}_{\text{I}}}_{34}$ &0&0&-1.75313 + 0.805685$i$&-4.49178 - 2.10017$i$&0.968643 - 8.17649$i$  \\[3pt]
			$\mathcal{G}^{\text{N}_{\text{I}}}_{35}$ &0&0&2.02943 + 4.44478$i$&-13.6379 + 9.59812$i$&-49.7803 - 21.9644$i$  \\[3pt]
			$\mathcal{G}^{\text{N}_{\text{I}}}_{36}$ &0&0&0&-1.25516&-4.20361  \\[3pt]
			$\mathcal{G}^{\text{N}_{\text{I}}}_{37}$ &0&2.58489 - 3.14159$i$&13.8943 + 10.6092$i$&-14.4297 + 66.4803$i$&-146.82 + 59.5817$i$  \\[3pt]
			$\mathcal{G}^{\text{N}_{\text{I}}}_{38}$ &0&1.44843&-0.862301 - 0.540505$i$&-4.91763 - 1.02503$i$&-4.10496 + 3.75441$i$  \\[3pt]
			$\mathcal{G}^{\text{N}_{\text{I}}}_{39}$ &0&0&0&2.80166 + 5.29482$i$&-15.1373 + 9.22942$i$  \\[3pt]
			$\mathcal{G}^{\text{N}_{\text{I}}}_{40}$ &0&0&0&2.91521&22.4086 + 8.96852$i$  \\[3pt]
			\hline 
		\end{tabular} 
		\caption{Numerical results for the 40 integrals of Topology $\text{N}_{\text{I}}$. The integrals are evaluated in Region B, as defined in Eq.\ref{eq:NI_regionB}, at the sample point $x=4.11574$ and $y=-3.25641$.}
		\label{table:NI_B}
	\end{center} 
\end{table}
\begin{table}[H]
	\scriptsize
	\begin{center}
		\begin{tabular}{|c| c c c c c| }
			\hline
			\rule{0pt}{10pt}
			Integral & $\epsilon^0$ &   $\epsilon^1$  &  $\epsilon^2$   & $\epsilon^3$  &  $\epsilon^4$  \\
			\hline
			\rule{0pt}{12pt}
			$\mathcal{G}^{\text{N}_{\text{I}}}_{1}$ &-1&0&-3.28987&2.40411&-9.74091  \\[3pt]
			$\mathcal{G}^{\text{N}_{\text{I}}}_{2}$ &-1&4.54587 - 12.5664$i$&53.1579 + 30.4564$i$&-75.0035 + 137.892$i$&-284.775 - 52.9579$i$  \\[3pt]
			$\mathcal{G}^{\text{N}_{\text{I}}}_{3}$ &-1&2.27293 - 6.28319$i$&22.9046 + 10.7834$i$&-21.9043 + 60.97$i$&-127.363 - 10.9144$i$  \\[3pt]
			$\mathcal{G}^{\text{N}_{\text{I}}}_{4}$ &-1&5.7937&-7.73972&17.2361&-34.4098  \\[3pt]
			$\mathcal{G}^{\text{N}_{\text{I}}}_{5}$ &0&1.44843&-2.14027&4.09676&-9.11964  \\[3pt]
			$\mathcal{G}^{\text{N}_{\text{I}}}_{6}$ &-1&2.48087&-2.0293&9.8781&-11.8496  \\[3pt]
			$\mathcal{G}^{\text{N}_{\text{I}}}_{7}$ &0&0.620216&-0.0463515&1.7114&-1.74511  \\[3pt]
			$\mathcal{G}^{\text{N}_{\text{I}}}_{8}$ &2&0.303203&-3.26689&-5.30582&-5.63715  \\[3pt]
			$\mathcal{G}^{\text{N}_{\text{I}}}_{9}$ &0&0&0&-2.16594&-2.42064  \\[3pt]
			$\mathcal{G}^{\text{N}_{\text{I}}}_{10}$ &0&0&-1.49898&1.97982&-1.29796  \\[3pt]
			$\mathcal{G}^{\text{N}_{\text{I}}}_{11}$ &0&0&0&4.4335 + 3.14428$i$&-1.94521 + 12.8449$i$  \\[3pt]
			$\mathcal{G}^{\text{N}_{\text{I}}}_{12}$ &0&0&0&0.715101 + 0.459292$i$&-0.8256 + 1.63381$i$  \\[3pt]
			$\mathcal{G}^{\text{N}_{\text{I}}}_{13}$ &0&2.58489 - 3.14159$i$&11.918 + 12.1192$i$&-27.3245 + 44.4824$i$&-106.273 - 21.9359$i$  \\[3pt]
			$\mathcal{G}^{\text{N}_{\text{I}}}_{14}$ &0&0&2.05561&1.73504&6.44769  \\[3pt]
			$\mathcal{G}^{\text{N}_{\text{I}}}_{15}$ &0&0&1.31141 + 2.87053$i$&-7.88351 + 5.71205$i$&-14.3227 - 8.22965$i$  \\[3pt]
			$\mathcal{G}^{\text{N}_{\text{I}}}_{16}$ &0&-1.75668 + 3.14159$i$&-11.1411 - 2.79916$i$&1.61544 - 27.9809$i$&50.1745 - 13.3256$i$  \\[3pt]
			$\mathcal{G}^{\text{N}_{\text{I}}}_{17}$ &0&0&0&2.91521&-0.260722  \\[3pt]
			$\mathcal{G}^{\text{N}_{\text{I}}}_{18}$ &0&-0.620216&-7.96124&2.11719&-15.0144  \\[3pt]
			$\mathcal{G}^{\text{N}_{\text{I}}}_{19}$ &0&0&0&0.784536&0.549118  \\[3pt]
			$\mathcal{G}^{\text{N}_{\text{I}}}_{20}$ &0&0&0&-3.19913 - 2.19602$i$&2.34156 - 8.49664$i$  \\[3pt]
			$\mathcal{G}^{\text{N}_{\text{I}}}_{21}$ &0&0&5.24564 + 11.4821$i$&-37.9323 + 18.4562$i$&-52.6077 - 49.9119$i$  \\[3pt]
			$\mathcal{G}^{\text{N}_{\text{I}}}_{22}$ &1&-0.984865 + 3.14159$i$&-12.8735 - 0.47064$i$&-10.6309 - 31.755$i$&53.8606 - 45.684$i$  \\[3pt]
			$\mathcal{G}^{\text{N}_{\text{I}}}_{23}$ &0&-1.13647 + 3.14159$i$&-15.299 - 9.36021$i$&4.62937 - 61.6318$i$&121.361 - 59.8471$i$  \\[3pt]
			$\mathcal{G}^{\text{N}_{\text{I}}}_{24}$ &0&0&0&0&-2.08142  \\[3pt]
			$\mathcal{G}^{\text{N}_{\text{I}}}_{25}$ &0&0&-2.18264&-0.579008&-4.66848  \\[3pt]
			$\mathcal{G}^{\text{N}_{\text{I}}}_{26}$ &0&-2.06864&-3.76536&-2.75711&0.0101042  \\[3pt]
			$\mathcal{G}^{\text{N}_{\text{I}}}_{27}$ &0&0&-2.02943 - 4.44478$i$&9.96187 - 11.1203$i$&21.8403 + 2.71965$i$  \\[3pt]
			$\mathcal{G}^{\text{N}_{\text{I}}}_{28}$ &0&0&0&-2.51258&-0.542866 \\[3pt]
			$\mathcal{G}^{\text{N}_{\text{I}}}_{29}$ &1&-1.29682&-3.82398&-4.18162&1.05095  \\[3pt]
			$\mathcal{G}^{\text{N}_{\text{I}}}_{30}$ &0&-1.44843&1.92068&9.75574&14.3738  \\[3pt]
			$\mathcal{G}^{\text{N}_{\text{I}}}_{31}$ &0&0&0&1.38141&1.87152  \\[3pt]
			$\mathcal{G}^{\text{N}_{\text{I}}}_{32}$ &0&0&-1.75313 + 0.805685$i$&-4.49178 - 2.10017$i$&0.968643 - 8.17649$i$  \\[3pt]
			$\mathcal{G}^{\text{N}_{\text{I}}}_{33}$ &0&0&2.02943 + 4.44478$i$&-13.6379 + 9.59812$i$&-49.7803 - 21.9644$i$  \\[3pt]
			$\mathcal{G}^{\text{N}_{\text{I}}}_{34}$ &0&0&-1.75313 + 0.805685$i$&-4.49178 - 2.10017$i$&0.968643 - 8.17649$i$  \\[3pt]
			$\mathcal{G}^{\text{N}_{\text{I}}}_{35}$ &0&0&-2.05561&0.0839943&8.18846  \\[3pt]
			$\mathcal{G}^{\text{N}_{\text{I}}}_{36}$ &0&0&0&-1.25516&-4.20361  \\[3pt]
			$\mathcal{G}^{\text{N}_{\text{I}}}_{37}$ &0&2.58489 - 3.14159$i$&13.8943 + 10.6092$i$&-14.4297 + 66.4803$i$&-146.82 + 59.5817$i$  \\[3pt]
			$\mathcal{G}^{\text{N}_{\text{I}}}_{38}$ &0&1.13647 - 3.14159$i$&15.7664 + 7.24545$i$&2.79455 + 57.0951$i$&-102.953 + 66.788$i$  \\[3pt]
			$\mathcal{G}^{\text{N}_{\text{I}}}_{39}$ &0&0&0&-2.80166 - 5.29482$i$&15.1373 - 9.22942$i$  \\[3pt]
			$\mathcal{G}^{\text{N}_{\text{I}}}_{40}$ &0&0&0&-3.19913 - 2.19602$i$&-13.7552 - 10.1177$i$  \\[3pt]
			\hline 
		\end{tabular} 
		\caption{Numerical results for the 40 integrals of Topology $\text{N}_{\text{I}}$. The integrals are evaluated in Region C, as defined in Eq.\ref{eq:NI_regionC}, at the sample point $x=-3.25641$ and $y=4.11574$.}
		\label{table:NI_C}
	\end{center} 
\end{table}
\subsection{Topology N$_\text{II}$}
\label{appendix:NII}
\begin{table}[H]
	\scriptsize
	\begin{center}
		\begin{tabular}{|c| c c c c c| }
			\hline
			\rule{0pt}{10pt}
			Integral & $\epsilon^0$ &   $\epsilon^1$  &  $\epsilon^2$   & $\epsilon^3$  &  $\epsilon^4$  \\
			\hline
			\rule{0pt}{12pt}
			$\mathcal{G}^{\text{N}_{\text{II}}}_{1}$ &-1&-3.89182 + 6.28319$i$&8.8762 + 24.453$i$&56.5976 + 26.9126$i$&114.63 - 33.824$i$ \\[3pt]
			$\mathcal{G}^{\text{N}_{\text{II}}}_{2}$ &1&-3.27522 + 6.28319$i$&-3.59618 + 16.1007$i$&-15.6054 + 5.54955$i$&-29.2737 - 31.491$i$  \\[3pt]
			$\mathcal{G}^{\text{N}_{\text{II}}}_{3}$ &0&-7.16704 + 12.5664$i$&2.78348 + 28.3273$i$&-2.23119 - 1.3617$i$&-28.9217 - 73.3651$i$  \\[3pt]
			$\mathcal{G}^{\text{N}_{\text{II}}}_{4}$ &-1&1.11923 + 6.28319$i$&26.0164 - 7.03235$i$&-1.17113 - 80.7823$i$&-161.929 - 85.3374$i$  \\[3pt]
			$\mathcal{G}^{\text{N}_{\text{II}}}_{5}$ &-1&-1.38629 + 6.28319$i$&19.1452 + 8.71034$i$&34.0088 - 37.6094$i$&-40.5868 - 99.2129$i$  \\[3pt]
			$\mathcal{G}^{\text{N}_{\text{II}}}_{6}$ &1&-2.92717 - 6.28319$i$&-27.5558 + 18.392$i$&37.6787 + 90.4545$i$&206.755 + 5.2864$i$  \\[3pt]
			$\mathcal{G}^{\text{N}_{\text{II}}}_{7}$ &1&-6.33667 - 6.28319$i$&-34.3304 + 39.8145$i$&93.1587 + 133.021$i$&337.315 - 61.3957$i$  \\[3pt]
			$\mathcal{G}^{\text{N}_{\text{II}}}_{8}$ &0&-6.2672 + 12.5664$i$&1.61621 + 44.945$i$&36.8713 + 52.4398$i$&97.0382 - 30.0204$i$  \\[3pt]
			$\mathcal{G}^{\text{N}_{\text{II}}}_{9}$ &0&0&7.41474 + 9.8445$i$&91.8121 + 5.51568$i$&276.733 - 257.824$i$  \\[3pt]
			$\mathcal{G}^{\text{N}_{\text{II}}}_{10}$ &0&1.5668 - 3.14159$i$&-7.11414 - 18.6196$i$&-68.9133 - 25.6336$i$&-199.549 + 118.811$i$  \\[3pt]
			$\mathcal{G}^{\text{N}_{\text{II}}}_{11}$ &0&0&0&-2.68058 + 9.32835$i$&19.9747 + 49.5367$i$  \\[3pt]
			$\mathcal{G}^{\text{N}_{\text{II}}}_{12}$ &0&0&-3.70737 - 4.92225$i$&-27.7816 - 2.53207$i$&-39.5993 + 42.0348$i$  \\[3pt]
			$\mathcal{G}^{\text{N}_{\text{II}}}_{13}$ &0&0&0&2.0165&8.52852 - 12.6701$i$  \\[3pt]
			$\mathcal{G}^{\text{N}_{\text{II}}}_{14}$ &0&0&3.70737 + 4.92225$i$&34.1453 - 2.19733$i$&50.2066 - 83.7076$i$  \\[3pt]
			$\mathcal{G}^{\text{N}_{\text{II}}}_{15}$ &0&0&-4.83578 - 10.4684$i$&-89.6305 - 26.6207$i$&-341.838 + 216.188$i$  \\[3pt]
			$\mathcal{G}^{\text{N}_{\text{II}}}_{16}$ &0&0&0&-2.89792 - 2.89391$i$&-21.9237 + 0.861081$i$  \\[3pt]
			$\mathcal{G}^{\text{N}_{\text{II}}}_{17}$ &0&-1.25276&0.791858 + 17.9342$i$&57.6023 + 28.452$i$&86.2574 - 74.1604$i$  \\[3pt]
			$\mathcal{G}^{\text{N}_{\text{II}}}_{18}$ &0&0&-1.24827 - 6.11326$i$&-26.9309 - 22.8598$i$&-103.622 - 9.12414$i$  \\[3pt]
			$\mathcal{G}^{\text{N}_{\text{II}}}_{19}$ &0&0&0&3.25763&12.8395 - 20.4683$i$  \\[3pt]
			$\mathcal{G}^{\text{N}_{\text{II}}}_{20}$ &0&0&0&1.47701 + 1.3967$i$&10.4495 - 1.15469$i$  \\[3pt]
			$\mathcal{G}^{\text{N}_{\text{II}}}_{21}$ &0&0&1.2949 + 4.59251$i$&14.4176 + 5.31907$i$&6.40806 - 16.2246$i$  \\[3pt]
			$\mathcal{G}^{\text{N}_{\text{II}}}_{22}$ &0&0&-2.72003&-8.9814 + 17.0904$i$&34.8718 + 56.46$i$  \\[3pt]
			$\mathcal{G}^{\text{N}_{\text{II}}}_{23}$ &0&2.95751&13.439 - 18.5826$i$&-32.3841 - 84.44$i$&-245.579 - 41.3448$i$  \\[3pt]
			$\mathcal{G}^{\text{N}_{\text{II}}}_{24}$ &0&0&3.70737 + 4.92225$i$&33.2564 - 3.01541$i$&42.0008 - 82.4754$i$  \\[3pt]
			$\mathcal{G}^{\text{N}_{\text{II}}}_{25}$ &0&0&-4.66689 - 12.7573$i$&-98.4288 - 36.1927$i$&-376.043 + 217.457$i$  \\[3pt]
			$\mathcal{G}^{\text{N}_{\text{II}}}_{26}$ &0&0&0&-7.58332 + 10.9578$i$&8.53532 + 86.7239$i$  \\[3pt]
			$\mathcal{G}^{\text{N}_{\text{II}}}_{27}$ &0&0&-26.8404 - 29.5335$i$&-258.918 - 95.6322$i$&-926.406 + 206.019$i$  \\[3pt]
			$\mathcal{G}^{\text{N}_{\text{II}}}_{28}$ &0&0&-2.59775 + 0.717265$i$&-8.34851 + 15.0074$i$&20.9528 + 51.2066$i$  \\[3pt]
			$\mathcal{G}^{\text{N}_{\text{II}}}_{29}$ &0&0&0&-9.84295 + 13.4833$i$&8.37554 + 103.762$i$  \\[3pt]
			$\mathcal{G}^{\text{N}_{\text{II}}}_{30}$ &0&0&2.23115 - 5.47036$i$&15.1731 - 53.7497$i$&-5.51732 - 244.721$i$  \\[3pt]
			$\mathcal{G}^{\text{N}_{\text{II}}}_{31}$ &0&0&-13.4202 - 14.7667$i$&-119.345 - 53.2964$i$&-409.465 + 25.3067$i$  \\[3pt]
			$\mathcal{G}^{\text{N}_{\text{II}}}_{32}$ &0&0&-2.72003&14.6476 - 3.46334$i$&108.73 - 149.403$i$  \\[3pt]
			\hline 
		\end{tabular} 
		\caption{Numerical results for the 32 integrals of Topology $\text{N}_{\text{II}}$. The integrals are evaluated in Region A, as defined in Eq.~\ref{eq:NII_regionA} at the sample point $x=0.357143$ and $z_A=0.345346$.}
		\label{table:NII_A}
	\end{center} 
\end{table}
~
\begin{table}[H]
	\scriptsize
	\begin{center}
		\begin{tabular}{|c| c c c c c| }
			\hline
			\rule{0pt}{10pt}
			Integral & $\epsilon^0$ &   $\epsilon^1$  &  $\epsilon^2$   & $\epsilon^3$  &  $\epsilon^4$  \\
			\hline
			\rule{0pt}{12pt}
			$\mathcal{G}^{\text{N}_{\text{II}}}_{1}$ &-1&-1.11923 + 6.28319$i$&15.823 + 7.03234$i$&20.5811 - 16.7354$i$&3.18735 - 36.7728$i$  \\[3pt]
			$\mathcal{G}^{\text{N}_{\text{II}}}_{2}$ &1&2.26996 + 6.28319$i$&0.819137 + 24.8111$i$&-40.2869 + 54.1249$i$&-144.19 + 72.8715$i$  \\[3pt]
			$\mathcal{G}^{\text{N}_{\text{II}}}_{3}$ &0&1.15073 + 12.5664$i$&11.7448 + 28.3273$i$&-46.8686 + 43.0408$i$&-171.091 + 65.7482$i$  \\[3pt]
			$\mathcal{G}^{\text{N}_{\text{II}}}_{4}$ &-1&0.8228 + 6.28319$i$&19.3965 - 5.16981$i$&-8.59555 - 39.1884$i$&-58.9126 - 14.0245$i$  \\[3pt]
			$\mathcal{G}^{\text{N}_{\text{II}}}_{5}$ &-1&-0.148216 + 6.28319$i$&18.1574 + 0.931267$i$&6.91088 - 31.4026$i$&-36.048 - 31.1674$i$  \\[3pt]
			$\mathcal{G}^{\text{N}_{\text{II}}}_{6}$ &1&-1.89586 - 6.28319$i$&-20.0981 + 11.912$i$&27.6044 + 43.5965$i$&75.1678 - 16.6877$i$  \\[3pt]
			$\mathcal{G}^{\text{N}_{\text{II}}}_{7}$ &1&-3.4034 - 6.28319$i$&-21.328 + 21.3842$i$&52.992 + 51.3241$i$&100.241 - 51.5542$i$  \\[3pt]
			$\mathcal{G}^{\text{N}_{\text{II}}}_{8}$ &0&0. + 5.78187$i$&28.8214 + 15.9573$i$&46.4377 - 33.9443$i$&-37.9255 - 38.1249$i$  \\[3pt]
			$\mathcal{G}^{\text{N}_{\text{II}}}_{9}$ &0&0&2.08938&4.634 - 13.128$i$&-29.5328 - 29.1163$i$  \\[3pt]
			$\mathcal{G}^{\text{N}_{\text{II}}}_{10}$ &0&0. - 1.44547$i$&-9.08215 - 2.18317$i$&-13.7172 + 23.6178$i$&28.8785 + 42.1751$i$  \\[3pt]
			$\mathcal{G}^{\text{N}_{\text{II}}}_{11}$ &0&0&0&2.99152 + 0.619702$i$&13.0558 - 13.1654$i$  \\[3pt]
			$\mathcal{G}^{\text{N}_{\text{II}}}_{12}$ &0&0&-1.04469&-0.972001 + 6.4362$i$&16.0754 + 5.02476$i$  \\[3pt]
			$\mathcal{G}^{\text{N}_{\text{II}}}_{13}$ &0&0&0&0.582402&1.10349 - 3.65934$i$  \\[3pt]
			$\mathcal{G}^{\text{N}_{\text{II}}}_{14}$ &0&0&1.04469&0.324381 - 6.56398$i$&-19.1223 - 2.02409$i$  \\[3pt]
			$\mathcal{G}^{\text{N}_{\text{II}}}_{15}$ &0&0&-1.17377&-4.28188 + 7.37503$i$&14.0183 + 26.993$i$  \\[3pt]
			$\mathcal{G}^{\text{N}_{\text{II}}}_{16}$ &0&0&0&-1.4531 - 0.294467$i$&-5.76062 + 6.43684$i$  \\[3pt]
			$\mathcal{G}^{\text{N}_{\text{II}}}_{17}$ &0&-0.485508&5.30877 + 7.10052$i$&38.8962 - 4.84398$i$&34.2225 - 45.6425$i$  \\[3pt]
			$\mathcal{G}^{\text{N}_{\text{II}}}_{18}$ &0&0&-2.44869 - 1.75809$i$&-16.0148 + 2.33374$i$&-25.2889 + 25.566$i$  \\[3pt]
			$\mathcal{G}^{\text{N}_{\text{II}}}_{19}$ &0&0&0&1.00336&1.75312 - 6.30431$i$  \\[3pt]
			$\mathcal{G}^{\text{N}_{\text{II}}}_{20}$ &0&0&0&0.766099 + 0.15557$i$&2.93573 - 3.39733$i$  \\[3pt]
			$\mathcal{G}^{\text{N}_{\text{II}}}_{21}$ &0&0&2.70937 + 2.08761$i$&14.4698 - 2.64757$i$&11.6396 - 12.1452$i$  \\[3pt]
			$\mathcal{G}^{\text{N}_{\text{II}}}_{22}$ &0&0&-1.23519&-2.60287 + 7.76096$i$&18.4967 + 16.3825$i$  \\[3pt]
			$\mathcal{G}^{\text{N}_{\text{II}}}_{23}$ &0&1.23928&4.72004 - 7.78662$i$&-15.1555 - 29.6569$i$&-80.0958 - 7.29967$i$  \\[3pt]
			$\mathcal{G}^{\text{N}_{\text{II}}}_{24}$ &0&0&1.04469&0.252028 - 6.56398$i$&-19.2679 - 1.58355$i$  \\[3pt]
			$\mathcal{G}^{\text{N}_{\text{II}}}_{25}$ &0&0&-1.02245&-3.53026 + 6.42425$i$&13.1505 + 22.1813$i$  \\[3pt]
			$\mathcal{G}^{\text{N}_{\text{II}}}_{26}$ &0&0&0&2.10017 + 0.670635$i$&13.2738 - 7.05277$i$  \\[3pt]
			$\mathcal{G}^{\text{N}_{\text{II}}}_{27}$ &0&0&-7.50723 + 7.2246$i$&-11.0856 + 70.1054$i$&112.479 + 195.179$i$  \\[3pt]
			$\mathcal{G}^{\text{N}_{\text{II}}}_{28}$ &0&0&-0.833325 + 1.02401$i$&0.255864 + 8.12278$i$&15.5687 + 15.4661$i$  \\[3pt]
			$\mathcal{G}^{\text{N}_{\text{II}}}_{29}$ &0&0&0&2.37381 + 0.765373$i$&14.7512 - 7.97368$i$  \\[3pt]
			$\mathcal{G}^{\text{N}_{\text{II}}}_{30}$ &0&0&-1.23122 - 1.96237$i$&-16.9157 - 7.25388$i$&-65.3901 + 18.5533$i$  \\[3pt]
			$\mathcal{G}^{\text{N}_{\text{II}}}_{31}$ &0&0&-3.75362 + 3.6123$i$&-5.36915 + 34.3514$i$&53.6674 + 91.7747$i$  \\[3pt]
			$\mathcal{G}^{\text{N}_{\text{II}}}_{32}$ &0&0&-1.23519&-5.34999 + 7.24985$i$&0.923416 + 28.7166$i$  \\[3pt]
			\hline 
		\end{tabular} 
		\caption{Numerical results for the 32 integrals of Topology $\text{N}_{\text{II}}$. The integrals are evaluated in Region B, as defined in Eq.~\ref{eq:NII_regionB}, at the sample point $x=0.357143$ and $z_B=0.468627$.}
		\label{table:NII_B}
	\end{center} 
\end{table}

\bibliographystyle{JHEP}
\newpage
\bibliography{gg2gaga}

\providecommand{\href}[2]{#2}\begingroup\raggedright\begin{thebibliography}{10}

\bibitem{Bonvin:1988yu}
{\scshape WA70 Collaboration} collaboration, \emph{{DOUBLE PROMPT PHOTON
  PRODUCTION AT HIGH TRANSVERSE MOMENTUM BY pi- ON PROTONS AT 280-GeV/c}},
  \href{https://doi.org/10.1007/BF01564702}{\emph{Z.Phys.} {\bfseries C41}
  (1989) 591}.

\bibitem{Alitti:1992hn}
{\scshape UA2 Collaboration} collaboration, \emph{{A Measurement of single and
  double prompt photon production at the CERN $\bar{p} p$ collider}},
  \href{https://doi.org/10.1016/0370-2693(92)91118-S}{\emph{Phys.Lett.}
  {\bfseries B288} (1992) 386}.

\bibitem{Abe:1992cy}
{\scshape CDF Collaboration} collaboration, \emph{{Measurement of the
  cross-section for production of two isolated prompt photons in $\bar{p}p$
  collisions at $\sqrt{s} = 1.8$ TeV}},
  \href{https://doi.org/10.1103/PhysRevLett.70.2232}{\emph{Phys.Rev.Lett.}
  {\bfseries 70} (1993) 2232}.

\bibitem{Abazov:2010ah}
{\scshape D0 Collaboration} collaboration, \emph{{Measurement of direct photon
  pair production cross sections in $p\bar{p}$ collisions at $\sqrt{s}=1.96$
  TeV}},
  \href{https://doi.org/10.1016/j.physletb.2010.05.017}{\emph{Phys.Lett.}
  {\bfseries B690} (2010) 108}
  [\href{https://arxiv.org/abs/1002.4917}{{\ttfamily 1002.4917}}].

\bibitem{Aaltonen:2012jd}
{\scshape CDF Collaboration} collaboration, \emph{{Measurement of the cross
  section for prompt isolated diphoton production using the full CDF Run II
  data sample}},
  \href{https://doi.org/10.1103/PhysRevLett.110.101801}{\emph{Phys.Rev.Lett.}
  {\bfseries 110} (2013) 101801}
  [\href{https://arxiv.org/abs/1212.4204}{{\ttfamily 1212.4204}}].

\bibitem{Chatrchyan:2011qt}
{\scshape CMS Collaboration} collaboration, \emph{{Measurement of the
  Production Cross Section for Pairs of Isolated Photons in $pp$ collisions at
  $\sqrt{s}=7$ TeV}},
  \href{https://doi.org/10.1007/JHEP01(2012)133}{\emph{JHEP} {\bfseries 1201}
  (2012) 133} [\href{https://arxiv.org/abs/1110.6461}{{\ttfamily 1110.6461}}].

\bibitem{Aad:2012tba}
{\scshape ATLAS Collaboration} collaboration, \emph{{Measurement of
  isolated-photon pair production in $pp$ collisions at $\sqrt{s}=7$ TeV with
  the ATLAS detector}},
  \href{https://doi.org/10.1007/JHEP01(2013)086}{\emph{JHEP} {\bfseries 1301}
  (2013) 086} [\href{https://arxiv.org/abs/1211.1913}{{\ttfamily 1211.1913}}].

\bibitem{Aad:2013zba}
{\scshape ATLAS Collaboration} collaboration, \emph{{Measurement of the
  inclusive isolated prompt photon cross section in pp collisions at sqrt(s) =
  7 TeV with the ATLAS detector using 4.6 fb-1}},
  \href{https://doi.org/10.1103/PhysRevD.89.052004}{\emph{Phys.Rev.} {\bfseries
  D89} (2014) 052004} [\href{https://arxiv.org/abs/1311.1440}{{\ttfamily
  1311.1440}}].

\bibitem{Aaltonen:2011vk}
{\scshape CDF Collaboration} collaboration, \emph{{Measurement of the Cross
  Section for Prompt Isolated Diphoton Production in $p\bar{p}$ Collisions at
  $\sqrt{s} = 1.96$ TeV}},
  \href{https://doi.org/10.1103/PhysRevD.84.052006}{\emph{Phys.Rev.} {\bfseries
  D84} (2011) 052006} [\href{https://arxiv.org/abs/1106.5131}{{\ttfamily
  1106.5131}}].

\bibitem{Chatrchyan:2013mwa}
{\scshape CMS Collaboration} collaboration, \emph{{Measurement of the
  triple-differential cross section for photon+jets production in proton-proton
  collisions at $\sqrt{s}$=7 TeV}},
  \href{https://doi.org/10.1007/JHEP06(2014)009}{\emph{JHEP} {\bfseries 1406}
  (2014) 009} [\href{https://arxiv.org/abs/1311.6141}{{\ttfamily 1311.6141}}].

\bibitem{Chatrchyan:2014fsa}
{\scshape CMS} collaboration, \emph{{Measurement of differential cross sections
  for the production of a pair of isolated photons in pp collisions at
  $\sqrt{s}=7\,\text {TeV} $}},
  \href{https://doi.org/10.1140/epjc/s10052-014-3129-3}{\emph{Eur. Phys. J.}
  {\bfseries C74} (2014) 3129}
  [\href{https://arxiv.org/abs/1405.7225}{{\ttfamily 1405.7225}}].

\bibitem{Marini:2016zjr}
{\scshape CMS} collaboration, \emph{{Measurements of Photon and Diphoton
  production cross-sections at CMS}},
  \href{https://doi.org/10.1016/j.nuclphysbps.2015.09.319}{\emph{Nucl. Part.
  Phys. Proc.} {\bfseries 273-275} (2016) 1973}.

\bibitem{CMS:2018qao}
{\scshape CMS} collaboration, \emph{{Measurement of differential cross sections
  for inclusive isolated-photon and photon+jets production in proton-proton
  collisions at $\sqrt{s} =$ 13 TeV}},
  \href{https://doi.org/10.1140/epjc/s10052-018-6482-9}{\emph{Eur. Phys. J. C}
  {\bfseries 79} (2019) 20} [\href{https://arxiv.org/abs/1807.00782}{{\ttfamily
  1807.00782}}].

\bibitem{ALICE:2019rtd}
{\scshape ALICE} collaboration, \emph{{Measurement of the inclusive isolated
  photon production cross section in pp collisions at $\sqrt{s}=7$ TeV}},
  \href{https://doi.org/10.1140/epjc/s10052-019-7389-9}{\emph{Eur. Phys. J. C}
  {\bfseries 79} (2019) 896}
  [\href{https://arxiv.org/abs/1906.01371}{{\ttfamily 1906.01371}}].

\bibitem{ATLAS:2019buk}
{\scshape ATLAS} collaboration, \emph{{Measurement of the inclusive
  isolated-photon cross section in $pp$ collisions at $\sqrt{s}=13$ TeV using
  36 fb$^{-1}$ of ATLAS data}},
  \href{https://doi.org/10.1007/JHEP10(2019)203}{\emph{JHEP} {\bfseries 10}
  (2019) 203} [\href{https://arxiv.org/abs/1908.02746}{{\ttfamily
  1908.02746}}].

\bibitem{ATLAS:2019iaa}
{\scshape ATLAS} collaboration, \emph{{Measurement of isolated-photon plus
  two-jet production in $pp$ collisions at $\sqrt s=13$ TeV with the ATLAS
  detector}}, \href{https://doi.org/10.1007/JHEP03(2020)179}{\emph{JHEP}
  {\bfseries 03} (2020) 179}
  [\href{https://arxiv.org/abs/1912.09866}{{\ttfamily 1912.09866}}].

\bibitem{CMS:2019jlq}
{\scshape CMS} collaboration, \emph{{Measurements of triple-differential cross
  sections for inclusive isolated-photon+jet events in pp collisions at
  $\sqrt{s} = 8\,\text {TeV} $}},
  \href{https://doi.org/10.1140/epjc/s10052-019-7451-7}{\emph{Eur. Phys. J. C}
  {\bfseries 79} (2019) 969}
  [\href{https://arxiv.org/abs/1907.08155}{{\ttfamily 1907.08155}}].

\bibitem{ATLAS:2021mbt}
{\scshape ATLAS} collaboration, \emph{{Measurement of the production cross
  section of pairs of isolated photons in $pp$ collisions at 13 TeV with the
  ATLAS detector}}, \href{https://doi.org/10.1007/JHEP11(2021)169}{\emph{JHEP}
  {\bfseries 11} (2021) 169}
  [\href{https://arxiv.org/abs/2107.09330}{{\ttfamily 2107.09330}}].

\bibitem{ATLAS:2023yrt}
{\scshape ATLAS} collaboration, \emph{{Inclusive-photon production and its
  dependence on photon isolation in $pp$ collisions at $\sqrt s=13$ TeV using
  139 fb$^{-1}$ of ATLAS data}},
  \href{https://arxiv.org/abs/2302.00510}{{\ttfamily 2302.00510}}.

\bibitem{Aad:2012tfa}
{\scshape ATLAS Collaboration} collaboration, \emph{{Observation of a new
  particle in the search for the Standard Model Higgs boson with the ATLAS
  detector at the LHC}},
  \href{https://doi.org/10.1016/j.physletb.2012.08.020}{\emph{Phys.Lett.}
  {\bfseries B716} (2012) 1} [\href{https://arxiv.org/abs/1207.7214}{{\ttfamily
  1207.7214}}].

\bibitem{Chatrchyan:2012ufa}
{\scshape CMS Collaboration} collaboration, \emph{{Observation of a new boson
  at a mass of 125 GeV with the CMS experiment at the LHC}},
  \href{https://doi.org/10.1016/j.physletb.2012.08.021}{\emph{Phys.Lett.}
  {\bfseries B716} (2012) 30}
  [\href{https://arxiv.org/abs/1207.7235}{{\ttfamily 1207.7235}}].

\bibitem{Binoth:1999qq}
T.~Binoth, J.~Guillet, E.~Pilon and M.~Werlen, \emph{{A Full next-to-leading
  order study of direct photon pair production in hadronic collisions}},
  \href{https://doi.org/10.1007/s100520050024}{\emph{Eur.Phys.J.} {\bfseries
  C16} (2000) 311} [\href{https://arxiv.org/abs/hep-ph/9911340}{{\ttfamily
  hep-ph/9911340}}].

\bibitem{Campbell:2016yrh}
J.~M. Campbell, R.~K. Ellis, Y.~Li and C.~Williams, \emph{{Predictions for
  diphoton production at the LHC through NNLO in QCD}},
  \href{https://doi.org/10.1007/JHEP07(2016)148}{\emph{JHEP} {\bfseries 07}
  (2016) 148} [\href{https://arxiv.org/abs/1603.02663}{{\ttfamily
  1603.02663}}].

\bibitem{Catani:2011qz}
S.~Catani, L.~Cieri, D.~de~Florian, G.~Ferrera and M.~Grazzini, \emph{{Diphoton
  production at hadron colliders: a fully-differential QCD calculation at
  NNLO}}, \href{https://doi.org/10.1103/PhysRevLett.108.072001}{\emph{Phys.
  Rev. Lett.} {\bfseries 108} (2012) 072001}
  [\href{https://arxiv.org/abs/1110.2375}{{\ttfamily 1110.2375}}].

\bibitem{Neumann:2021zkb}
T.~Neumann, \emph{{The diphoton $q_T$ spectrum at N$^3$LL$^\prime $~+~NNLO}},
  \href{https://doi.org/10.1140/epjc/s10052-021-09687-4}{\emph{Eur. Phys. J. C}
  {\bfseries 81} (2021) 905}
  [\href{https://arxiv.org/abs/2107.12478}{{\ttfamily 2107.12478}}].

\bibitem{DelDuca:2003uz}
V.~Del~Duca, F.~Maltoni, Z.~Nagy and Z.~Trocsanyi, \emph{{QCD radiative
  corrections to prompt diphoton production in association with a jet at hadron
  colliders}}, \href{https://doi.org/10.1088/1126-6708/2003/04/059}{\emph{JHEP}
  {\bfseries 0304} (2003) 059}
  [\href{https://arxiv.org/abs/hep-ph/0303012}{{\ttfamily hep-ph/0303012}}].

\bibitem{Gehrmann:2013bga}
T.~Gehrmann, N.~Greiner and G.~Heinrich, \emph{{Precise QCD predictions for the
  production of a photon pair in association with two jets}},
  \href{https://doi.org/10.1103/PhysRevLett.111.222002}{\emph{Phys. Rev. Lett.}
  {\bfseries 111} (2013) 222002}
  [\href{https://arxiv.org/abs/1308.3660}{{\ttfamily 1308.3660}}].

\bibitem{Badger:2013ava}
S.~Badger, A.~Guffanti and V.~Yundin, \emph{{Next-to-leading order QCD
  corrections to di-photon production in association with up to three jets at
  the Large Hadron Collider}},
  \href{https://doi.org/10.1007/JHEP03(2014)122}{\emph{JHEP} {\bfseries 03}
  (2014) 122} [\href{https://arxiv.org/abs/1312.5927}{{\ttfamily 1312.5927}}].

\bibitem{Chawdhry:2021hkp}
H.~A. Chawdhry, M.~Czakon, A.~Mitov and R.~Poncelet, \emph{{NNLO QCD
  corrections to diphoton production with an additional jet at the LHC}},
  \href{https://doi.org/10.1007/JHEP09(2021)093}{\emph{JHEP} {\bfseries 09}
  (2021) 093} [\href{https://arxiv.org/abs/2105.06940}{{\ttfamily
  2105.06940}}].

\bibitem{Caola:2020dfu}
F.~Caola, A.~Von~Manteuffel and L.~Tancredi, \emph{{Diphoton Amplitudes in
  Three-Loop Quantum Chromodynamics}},
  \href{https://doi.org/10.1103/PhysRevLett.126.112004}{\emph{Phys. Rev. Lett.}
  {\bfseries 126} (2021) 112004}
  [\href{https://arxiv.org/abs/2011.13946}{{\ttfamily 2011.13946}}].

\bibitem{Bargiela:2022lxz}
P.~Bargiela, A.~Chakraborty and G.~Gambuti, \emph{{Three-loop helicity
  amplitudes for photon+jet production}},
  \href{https://doi.org/10.1103/PhysRevD.107.L051502}{\emph{Phys. Rev. D}
  {\bfseries 107} (2023) L051502}
  [\href{https://arxiv.org/abs/2212.14069}{{\ttfamily 2212.14069}}].

\bibitem{Glover:1988fe}
E.~N. Glover and J.~van~der Bij, \emph{{Vector Boson Pair Production via Gluon
  Fusion}},
  \href{https://doi.org/10.1016/0370-2693(89)91099-X}{\emph{Phys.Lett.}
  {\bfseries B219} (1989) 488}.

\bibitem{Bern:2001df}
Z.~Bern, A.~De~Freitas and L.~J. Dixon, \emph{{Two loop amplitudes for gluon
  fusion into two photons}},
  \href{https://doi.org/10.1088/1126-6708/2001/09/037}{\emph{JHEP} {\bfseries
  09} (2001) 037} [\href{https://arxiv.org/abs/hep-ph/0109078}{{\ttfamily
  hep-ph/0109078}}].

\bibitem{Bern:2002jx}
Z.~Bern, L.~J. Dixon and C.~Schmidt, \emph{{Isolating a light Higgs boson from
  the diphoton background at the CERN LHC}},
  \href{https://doi.org/10.1103/PhysRevD.66.074018}{\emph{Phys.Rev.} {\bfseries
  D66} (2002) 074018} [\href{https://arxiv.org/abs/hep-ph/0206194}{{\ttfamily
  hep-ph/0206194}}].

\bibitem{Chen:2019fla}
L.~Chen, G.~Heinrich, S.~Jahn, S.~P. Jones, M.~Kerner, J.~Schlenk et~al.,
  \emph{{Photon pair production in gluon fusion: Top quark effects at NLO with
  threshold matching}},
  \href{https://doi.org/10.1007/JHEP04(2020)115}{\emph{JHEP} {\bfseries 04}
  (2020) 115} [\href{https://arxiv.org/abs/1911.09314}{{\ttfamily
  1911.09314}}].

\bibitem{Maltoni:2018zvp}
F.~Maltoni, M.~K. Mandal and X.~Zhao, \emph{{Top-quark effects in diphoton
  production through gluon fusion at next-to-leading order in QCD}},
  \href{https://doi.org/10.1103/PhysRevD.100.071501}{\emph{Phys. Rev. D}
  {\bfseries 100} (2019) 071501}
  [\href{https://arxiv.org/abs/1812.08703}{{\ttfamily 1812.08703}}].

\bibitem{Bargiela:2021wuy}
P.~Bargiela, F.~Caola, A.~von Manteuffel and L.~Tancredi, \emph{{Three-loop
  helicity amplitudes for diphoton production in gluon fusion}},
  \href{https://doi.org/10.1007/JHEP02(2022)153}{\emph{JHEP} {\bfseries 02}
  (2022) 153} [\href{https://arxiv.org/abs/2111.13595}{{\ttfamily
  2111.13595}}].

\bibitem{Martin:2012xc}
S.~P. Martin, \emph{{Shift in the LHC Higgs Diphoton Mass Peak from
  Interference with Background}},
  \href{https://doi.org/10.1103/PhysRevD.86.073016}{\emph{Phys. Rev. D}
  {\bfseries 86} (2012) 073016}
  [\href{https://arxiv.org/abs/1208.1533}{{\ttfamily 1208.1533}}].

\bibitem{Dixon:2003yb}
L.~J. Dixon and M.~S. Siu, \emph{{Resonance continuum interference in the
  diphoton Higgs signal at the LHC}},
  \href{https://doi.org/10.1103/PhysRevLett.90.252001}{\emph{Phys. Rev. Lett.}
  {\bfseries 90} (2003) 252001}
  [\href{https://arxiv.org/abs/hep-ph/0302233}{{\ttfamily hep-ph/0302233}}].

\bibitem{Dixon:2013haa}
L.~J. Dixon and Y.~Li, \emph{{Bounding the Higgs Boson Width Through
  Interferometry}},
  \href{https://doi.org/10.1103/PhysRevLett.111.111802}{\emph{Phys. Rev. Lett.}
  {\bfseries 111} (2013) 111802}
  [\href{https://arxiv.org/abs/1305.3854}{{\ttfamily 1305.3854}}].

\bibitem{deFlorian:2013psa}
D.~de~Florian, N.~Fidanza, R.~J. Hern\'andez-Pinto, J.~Mazzitelli,
  Y.~Rotstein~Habarnau and G.~F.~R. Sborlini, \emph{{A complete $O(\alpha_S^2)$
  calculation of the signal-background interference for the Higgs diphoton
  decay channel}},
  \href{https://doi.org/10.1140/epjc/s10052-013-2387-9}{\emph{Eur. Phys. J. C}
  {\bfseries 73} (2013) 2387}
  [\href{https://arxiv.org/abs/1303.1397}{{\ttfamily 1303.1397}}].

\bibitem{Coradeschi:2015tna}
F.~Coradeschi, D.~de~Florian, L.~J. Dixon, N.~Fidanza, S.~H\"oche, H.~Ita
  et~al., \emph{{Interference effects in the $H(\rightarrow \gamma\gamma) + 2$
  jets channel at the LHC}},
  \href{https://doi.org/10.1103/PhysRevD.92.013004}{\emph{Phys. Rev. D}
  {\bfseries 92} (2015) 013004}
  [\href{https://arxiv.org/abs/1504.05215}{{\ttfamily 1504.05215}}].

\bibitem{Campbell:2017rke}
J.~Campbell, M.~Carena, R.~Harnik and Z.~Liu, \emph{{Interference in the
  $gg\rightarrow h \rightarrow \gamma\gamma$ On-Shell Rate and the Higgs Boson
  Total Width}},
  \href{https://doi.org/10.1103/PhysRevLett.119.181801}{\emph{Phys. Rev. Lett.}
  {\bfseries 119} (2017) 181801}
  [\href{https://arxiv.org/abs/1704.08259}{{\ttfamily 1704.08259}}].

\bibitem{Cieri:2017kpq}
L.~Cieri, F.~Coradeschi, D.~de~Florian and N.~Fidanza,
  \emph{{Transverse-momentum resummation for the signal-background interference
  in the H\textrightarrow{}\ensuremath{\gamma}\ensuremath{\gamma} channel at
  the LHC}}, \href{https://doi.org/10.1103/PhysRevD.96.054003}{\emph{Phys. Rev.
  D} {\bfseries 96} (2017) 054003}
  [\href{https://arxiv.org/abs/1706.07331}{{\ttfamily 1706.07331}}].

\bibitem{Bargiela:2022dla}
P.~Bargiela, F.~Buccioni, F.~Caola, F.~Devoto, A.~von Manteuffel and
  L.~Tancredi, \emph{{Signal-background interference effects in Higgs-mediated
  diphoton production beyond NLO}},
  \href{https://doi.org/10.1140/epjc/s10052-023-11337-w}{\emph{Eur. Phys. J. C}
  {\bfseries 83} (2023) 174}
  [\href{https://arxiv.org/abs/2212.06287}{{\ttfamily 2212.06287}}].

\bibitem{Kotikov:1990kg}
A.~Kotikov, \emph{{Differential equations method: New technique for massive
  Feynman diagrams calculation}},
  \href{https://doi.org/10.1016/0370-2693(91)90413-K}{\emph{Phys.Lett.}
  {\bfseries B254} (1991) 158}.

\bibitem{Remiddi:1997ny}
E.~Remiddi, \emph{{Differential equations for Feynman graph amplitudes}},
  {\emph{Nuovo Cim.} {\bfseries A110} (1997) 1435}
  [\href{https://arxiv.org/abs/hep-th/9711188}{{\ttfamily hep-th/9711188}}].

\bibitem{Gehrmann:1999as}
T.~Gehrmann and E.~Remiddi, \emph{{Differential equations for two loop four
  point functions}},
  \href{https://doi.org/10.1016/S0550-3213(00)00223-6}{\emph{Nucl.Phys.}
  {\bfseries B580} (2000) 485}
  [\href{https://arxiv.org/abs/hep-ph/9912329}{{\ttfamily hep-ph/9912329}}].

\bibitem{Gehrmann:2000xj}
T.~Gehrmann and E.~Remiddi, \emph{{Using differential equations to compute two
  loop box integrals}},
  \href{https://doi.org/10.1016/S0920-5632(00)00851-3}{\emph{Nucl. Phys. B
  Proc. Suppl.} {\bfseries 89} (2000) 251}
  [\href{https://arxiv.org/abs/hep-ph/0005232}{{\ttfamily hep-ph/0005232}}].

\bibitem{Argeri:2007up}
M.~Argeri and P.~Mastrolia, \emph{{Feynman Diagrams and Differential
  Equations}}, \href{https://doi.org/10.1142/S0217751X07037147}{\emph{Int. J.
  Mod. Phys.} {\bfseries A22} (2007) 4375}
  [\href{https://arxiv.org/abs/0707.4037}{{\ttfamily 0707.4037}}].

\bibitem{Henn:2013pwa}
J.~M. Henn, \emph{{Multiloop integrals in dimensional regularization made
  simple}},
  \href{https://doi.org/10.1103/PhysRevLett.110.251601}{\emph{Phys.Rev.Lett.}
  {\bfseries 110} (2013) 251601}
  [\href{https://arxiv.org/abs/1304.1806}{{\ttfamily 1304.1806}}].

\bibitem{Gehrmann:2001ck}
T.~Gehrmann and E.~Remiddi, \emph{{Two loop master integrals for $\gamma^*
  \rightarrow$ 3 jets: The Nonplanar topologies}},
  \href{https://doi.org/10.1016/S0550-3213(01)00074-8}{\emph{Nucl.Phys.}
  {\bfseries B601} (2001) 287}
  [\href{https://arxiv.org/abs/hep-ph/0101124}{{\ttfamily hep-ph/0101124}}].

\bibitem{Bonciani:2016ypc}
R.~Bonciani, S.~Di~Vita, P.~Mastrolia and U.~Schubert, \emph{{Two-Loop Master
  Integrals for the mixed EW-QCD virtual corrections to Drell-Yan scattering}},
  \href{https://doi.org/10.1007/JHEP09(2016)091}{\emph{JHEP} {\bfseries 09}
  (2016) 091} [\href{https://arxiv.org/abs/1604.08581}{{\ttfamily
  1604.08581}}].

\bibitem{DiVita:2017xlr}
S.~Di~Vita, P.~Mastrolia, A.~Primo and U.~Schubert, \emph{{Two-loop master
  integrals for the leading QCD corrections to the Higgs coupling to a $W$ pair
  and to the triple gauge couplings $ZWW$ and $\gamma^*WW$}},
  \href{https://doi.org/10.1007/JHEP04(2017)008}{\emph{JHEP} {\bfseries 04}
  (2017) 008} [\href{https://arxiv.org/abs/1702.07331}{{\ttfamily
  1702.07331}}].

\bibitem{Heller:2019gkq}
M.~Heller, A.~von Manteuffel and R.~M. Schabinger, \emph{{Multiple
  polylogarithms with algebraic arguments and the two-loop EW-QCD Drell-Yan
  master integrals}},
  \href{https://doi.org/10.1103/PhysRevD.102.016025}{\emph{Phys. Rev. D}
  {\bfseries 102} (2020) 016025}
  [\href{https://arxiv.org/abs/1907.00491}{{\ttfamily 1907.00491}}].

\bibitem{Hasan:2020vwn}
S.~M. Hasan and U.~Schubert, \emph{{Master Integrals for the mixed QCD-QED
  corrections to the Drell-Yan production of a massive lepton pair}},
  \href{https://doi.org/10.1007/JHEP11(2020)107}{\emph{JHEP} {\bfseries 11}
  (2020) 107} [\href{https://arxiv.org/abs/2004.14908}{{\ttfamily
  2004.14908}}].

\bibitem{Bonetti:2020hqh}
M.~Bonetti, E.~Panzer, V.~A. Smirnov and L.~Tancredi, \emph{{Two-loop mixed
  QCD-EW corrections to $gg \to Hg$}},
  \href{https://doi.org/10.1007/JHEP11(2020)045}{\emph{JHEP} {\bfseries 11}
  (2020) 045} [\href{https://arxiv.org/abs/2007.09813}{{\ttfamily
  2007.09813}}].

\bibitem{Behring:2020cqi}
A.~Behring, F.~Buccioni, F.~Caola, M.~Delto, M.~Jaquier, K.~Melnikov et~al.,
  \emph{{Mixed QCD-electroweak corrections to $W$-boson production in hadron
  collisions}}, \href{https://doi.org/10.1103/PhysRevD.103.013008}{\emph{Phys.
  Rev. D} {\bfseries 103} (2021) 013008}
  [\href{https://arxiv.org/abs/2009.10386}{{\ttfamily 2009.10386}}].

\bibitem{Canko:2020ylt}
D.~D. Canko, C.~G. Papadopoulos and N.~Syrrakos, \emph{{Analytic representation
  of all planar two-loop five-point Master Integrals with one off-shell leg}},
  \href{https://doi.org/10.1007/JHEP01(2021)199}{\emph{JHEP} {\bfseries 01}
  (2021) 199} [\href{https://arxiv.org/abs/2009.13917}{{\ttfamily
  2009.13917}}].

\bibitem{Duhr:2021fhk}
C.~Duhr, V.~A. Smirnov and L.~Tancredi, \emph{{Analytic results for two-loop
  planar master integrals for Bhabha scattering}},
  \href{https://doi.org/10.1007/JHEP09(2021)120}{\emph{JHEP} {\bfseries 09}
  (2021) 120} [\href{https://arxiv.org/abs/2108.03828}{{\ttfamily
  2108.03828}}].

\bibitem{Abreu:2021smk}
S.~Abreu, H.~Ita, B.~Page and W.~Tschernow, \emph{{Two-loop hexa-box integrals
  for non-planar five-point one-mass processes}},
  \href{https://doi.org/10.1007/JHEP03(2022)182}{\emph{JHEP} {\bfseries 03}
  (2022) 182} [\href{https://arxiv.org/abs/2107.14180}{{\ttfamily
  2107.14180}}].

\bibitem{Bonciani:2021zzf}
R.~Bonciani, L.~Buonocore, M.~Grazzini, S.~Kallweit, N.~Rana, F.~Tramontano
  et~al., \emph{{Mixed Strong-Electroweak Corrections to the Drell-Yan
  Process}}, \href{https://doi.org/10.1103/PhysRevLett.128.012002}{\emph{Phys.
  Rev. Lett.} {\bfseries 128} (2022) 012002}
  [\href{https://arxiv.org/abs/2106.11953}{{\ttfamily 2106.11953}}].

\bibitem{Kardos:2022tpo}
A.~Kardos, C.~G. Papadopoulos, A.~V. Smirnov, N.~Syrrakos and C.~Wever,
  \emph{{Two-loop non-planar hexa-box integrals with one massive leg}},
  \href{https://doi.org/10.1007/JHEP05(2022)033}{\emph{JHEP} {\bfseries 05}
  (2022) 033} [\href{https://arxiv.org/abs/2201.07509}{{\ttfamily
  2201.07509}}].

\bibitem{Abreu:2022vei}
S.~Abreu, M.~Becchetti, C.~Duhr and M.~A. Ozcelik, \emph{{Two-loop master
  integrals for pseudo-scalar quarkonium and leptonium production and decay}},
  \href{https://doi.org/10.1007/JHEP09(2022)194}{\emph{JHEP} {\bfseries 09}
  (2022) 194} [\href{https://arxiv.org/abs/2206.03848}{{\ttfamily
  2206.03848}}].

\bibitem{Badger:2022hno}
S.~Badger, M.~Becchetti, E.~Chaubey and R.~Marzucca, \emph{{Two-loop master
  integrals for a planar topology contributing to pp \textrightarrow{}$
  t\overline{t}j $}},
  \href{https://doi.org/10.1007/JHEP01(2023)156}{\emph{JHEP} {\bfseries 01}
  (2023) 156} [\href{https://arxiv.org/abs/2210.17477}{{\ttfamily
  2210.17477}}].

\bibitem{Armadillo:2022bgm}
T.~Armadillo, R.~Bonciani, S.~Devoto, N.~Rana and A.~Vicini, \emph{{Two-loop
  mixed QCD-EW corrections to neutral current Drell-Yan}},
  \href{https://doi.org/10.1007/JHEP05(2022)072}{\emph{JHEP} {\bfseries 05}
  (2022) 072} [\href{https://arxiv.org/abs/2201.01754}{{\ttfamily
  2201.01754}}].

\bibitem{Bonetti:2022lrk}
M.~Bonetti, E.~Panzer and L.~Tancredi, \emph{{Two-loop mixed QCD-EW corrections
  to $ q\overline{q} $ \textrightarrow{} Hg, qg \textrightarrow{} Hq, and $
  \overline{q}g $ \textrightarrow{} $ H\overline{q} $}},
  \href{https://doi.org/10.1007/JHEP06(2022)115}{\emph{JHEP} {\bfseries 06}
  (2022) 115} [\href{https://arxiv.org/abs/2203.17202}{{\ttfamily
  2203.17202}}].

\bibitem{Buccioni:2022kgy}
F.~Buccioni, F.~Caola, H.~A. Chawdhry, F.~Devoto, M.~Heller, A.~von Manteuffel
  et~al., \emph{{Mixed QCD-electroweak corrections to dilepton production at
  the LHC in the high invariant mass region}},
  \href{https://doi.org/10.1007/JHEP06(2022)022}{\emph{JHEP} {\bfseries 06}
  (2022) 022} [\href{https://arxiv.org/abs/2203.11237}{{\ttfamily
  2203.11237}}].

\bibitem{Peraro:2019cjj}
T.~Peraro and L.~Tancredi, \emph{{Physical projectors for multi-leg helicity
  amplitudes}}, \href{https://doi.org/10.1007/JHEP07(2019)114}{\emph{JHEP}
  {\bfseries 07} (2019) 114}
  [\href{https://arxiv.org/abs/1906.03298}{{\ttfamily 1906.03298}}].

\bibitem{Peraro:2020sfm}
T.~Peraro and L.~Tancredi, \emph{{Tensor decomposition for bosonic and
  fermionic scattering amplitudes}},
  \href{https://doi.org/10.1103/PhysRevD.103.054042}{\emph{Phys. Rev. D}
  {\bfseries 103} (2021) 054042}
  [\href{https://arxiv.org/abs/2012.00820}{{\ttfamily 2012.00820}}].

\bibitem{Chen:1977oja}
K.-T. Chen, \emph{{Iterated path integrals}},
  \href{https://doi.org/10.1090/S0002-9904-1977-14320-6}{\emph{Bull.Am.Math.Soc.}
  {\bfseries 83} (1977) 831}.

\bibitem{Caron-Huot:2014lda}
S.~Caron-Huot and J.~M. Henn, \emph{{Iterative structure of finite loop
  integrals}}, \href{https://doi.org/10.1007/JHEP06(2014)114}{\emph{JHEP}
  {\bfseries 06} (2014) 114} [\href{https://arxiv.org/abs/1404.2922}{{\ttfamily
  1404.2922}}].

\bibitem{Klappert:2020nbg}
J.~Klappert, F.~Lange, P.~Maierh\"ofer and J.~Usovitsch, \emph{{Integral
  reduction with Kira 2.0 and finite field methods}},
  \href{https://doi.org/10.1016/j.cpc.2021.108024}{\emph{Comput. Phys. Commun.}
  {\bfseries 266} (2021) 108024}
  [\href{https://arxiv.org/abs/2008.06494}{{\ttfamily 2008.06494}}].

\bibitem{Maierhofer:2017gsa}
P.~Maierh\"ofer, J.~Usovitsch and P.~Uwer, \emph{{Kira\textemdash{}A Feynman
  integral reduction program}},
  \href{https://doi.org/10.1016/j.cpc.2018.04.012}{\emph{Comput. Phys. Commun.}
  {\bfseries 230} (2018) 99}
  [\href{https://arxiv.org/abs/1705.05610}{{\ttfamily 1705.05610}}].

\bibitem{Argeri:2014qva}
M.~Argeri, S.~Di~Vita, P.~Mastrolia, E.~Mirabella, J.~Schlenk, U.~Schubert
  et~al., \emph{{Magnus and Dyson Series for Master Integrals}},
  \href{https://doi.org/10.1007/JHEP03(2014)082}{\emph{JHEP} {\bfseries 1403}
  (2014) 082} [\href{https://arxiv.org/abs/1401.2979}{{\ttfamily 1401.2979}}].

\bibitem{Liu:2022chg}
X.~Liu and Y.-Q. Ma, \emph{{AMFlow: A Mathematica package for Feynman integrals
  computation via auxiliary mass flow}},
  \href{https://doi.org/10.1016/j.cpc.2022.108565}{\emph{Comput. Phys. Commun.}
  {\bfseries 283} (2023) 108565}
  [\href{https://arxiv.org/abs/2201.11669}{{\ttfamily 2201.11669}}].

\bibitem{Borowka:2015mxa}
S.~Borowka, G.~Heinrich, S.~P. Jones, M.~Kerner, J.~Schlenk and T.~Zirke,
  \emph{{SecDec-3.0: numerical evaluation of multi-scale integrals beyond one
  loop}}, \href{https://doi.org/10.1016/j.cpc.2015.05.022}{\emph{Comput. Phys.
  Commun.} {\bfseries 196} (2015) 470}
  [\href{https://arxiv.org/abs/1502.06595}{{\ttfamily 1502.06595}}].

\bibitem{Duhr:2019tlz}
C.~Duhr and F.~Dulat, \emph{{PolyLogTools \textemdash{} polylogs for the
  masses}}, \href{https://doi.org/10.1007/JHEP08(2019)135}{\emph{JHEP}
  {\bfseries 08} (2019) 135}
  [\href{https://arxiv.org/abs/1904.07279}{{\ttfamily 1904.07279}}].

\bibitem{Naterop:2019xaf}
L.~Naterop, A.~Signer and Y.~Ulrich, \emph{{handyG \textemdash{}Rapid numerical
  evaluation of generalised polylogarithms in Fortran}},
  \href{https://doi.org/10.1016/j.cpc.2020.107165}{\emph{Comput. Phys. Commun.}
  {\bfseries 253} (2020) 107165}
  [\href{https://arxiv.org/abs/1909.01656}{{\ttfamily 1909.01656}}].

\end{thebibliography}\endgroup

\end{document}